\PassOptionsToPackage{unicode}{hyperref}
\PassOptionsToPackage{hyphens}{url}
\PassOptionsToPackage{dvipsnames,svgnames,x11names}{xcolor}
\documentclass[
]{article}

\usepackage{amsmath,amssymb}
\usepackage{lmodern}
\usepackage{iftex}
\ifPDFTeX
  \usepackage[T1]{fontenc}
  \usepackage[utf8]{inputenc}
  \usepackage{textcomp} 
\else 
  \usepackage{unicode-math}
  \defaultfontfeatures{Scale=MatchLowercase}
  \defaultfontfeatures[\rmfamily]{Ligatures=TeX,Scale=1}
\fi
\IfFileExists{upquote.sty}{\usepackage{upquote}}{}
\IfFileExists{microtype.sty}{
  \usepackage[]{microtype}
  \UseMicrotypeSet[protrusion]{basicmath} 
}{}
\makeatletter
\@ifundefined{KOMAClassName}{
  \IfFileExists{parskip.sty}{%
    \usepackage{parskip}
  }{
    \setlength{\parindent}{0pt}
    \setlength{\parskip}{6pt plus 2pt minus 1pt}}
}{
  \KOMAoptions{parskip=half}}
\makeatother
\usepackage{xcolor}
\ifx\paragraph\undefined\else
  \let\oldparagraph\paragraph
  \renewcommand{\paragraph}[1]{\oldparagraph{#1}\mbox{}}
\fi
\ifx\subparagraph\undefined\else
  \let\oldsubparagraph\subparagraph
  \renewcommand{\subparagraph}[1]{\oldsubparagraph{#1}\mbox{}}
\fi

\usepackage{color}
\usepackage{fancyvrb}

\DefineVerbatimEnvironment{Highlighting}{Verbatim}{commandchars=\\\{\}}
\usepackage{framed}
\definecolor{shadecolor}{RGB}{241,243,245}
\newenvironment{Shaded}{\begin{snugshade}}{\end{snugshade}}

\newcommand{\AttributeTok}[1]{\textcolor[rgb]{0.40,0.45,0.13}{#1}}

\newcommand{\CommentTok}[1]{\textcolor[rgb]{0.37,0.37,0.37}{#1}}

\newcommand{\ConstantTok}[1]{\textcolor[rgb]{0.56,0.35,0.01}{#1}}
\newcommand{\ControlFlowTok}[1]{\textcolor[rgb]{0.00,0.23,0.31}{#1}}

\newcommand{\DecValTok}[1]{\textcolor[rgb]{0.68,0.00,0.00}{#1}}
\newcommand{\DocumentationTok}[1]{\textcolor[rgb]{0.37,0.37,0.37}{\textit{#1}}}

\newcommand{\FloatTok}[1]{\textcolor[rgb]{0.68,0.00,0.00}{#1}}
\newcommand{\FunctionTok}[1]{\textcolor[rgb]{0.28,0.35,0.67}{#1}}

\newcommand{\NormalTok}[1]{\textcolor[rgb]{0.00,0.23,0.31}{#1}}

\newcommand{\OtherTok}[1]{\textcolor[rgb]{0.00,0.23,0.31}{#1}}

\newcommand{\SpecialCharTok}[1]{\textcolor[rgb]{0.37,0.37,0.37}{#1}}

\newcommand{\StringTok}[1]{\textcolor[rgb]{0.13,0.47,0.30}{#1}}

\providecommand{\tightlist}{%
  \setlength{\itemsep}{0pt}\setlength{\parskip}{0pt}}\usepackage{longtable,booktabs,array}
\usepackage{calc} 
\usepackage{etoolbox}
\makeatletter
\patchcmd\longtable{\par}{\if@noskipsec\mbox{}\fi\par}{}{}
\makeatother
\IfFileExists{footnotehyper.sty}{\usepackage{footnotehyper}}{\usepackage{footnote}}
\makesavenoteenv{longtable}
\usepackage{graphicx}
\makeatletter
\def\maxwidth{\ifdim\Gin@nat@width>\linewidth\linewidth\else\Gin@nat@width\fi}
\def\maxheight{\ifdim\Gin@nat@height>\textheight\textheight\else\Gin@nat@height\fi}
\makeatother
\setkeys{Gin}{width=\maxwidth,height=\maxheight,keepaspectratio}
\makeatletter
\def\fps@figure{htbp}
\makeatother
\newlength{\cslhangindent}
\newlength{\csllabelwidth}
\newlength{\cslentryspacingunit} 
\newenvironment{CSLReferences}[2] 
 {
  \setlength{\parindent}{0pt}
  \ifodd #1
  \let\oldpar\par
  \def\par{\hangindent=\cslhangindent\oldpar}
  \fi
  \setlength{\parskip}{#2\cslentryspacingunit}
 }%
 {}
\usepackage{calc}

\usepackage{arxiv}
\usepackage{orcidlink}
\usepackage{amsmath}
\usepackage[T1]{fontenc}
\makeatletter
\@ifpackageloaded{caption}{}{\usepackage{caption}}
\AtBeginDocument{%
\ifdefined\contentsname
  \renewcommand*\contentsname{Table of contents}
\else
  \newcommand\contentsname{Table of contents}
\fi
\ifdefined\listfigurename
  \renewcommand*\listfigurename{List of Figures}
\else
  \newcommand\listfigurename{List of Figures}
\fi
\ifdefined\listtablename
  \renewcommand*\listtablename{List of Tables}
\else
  \newcommand\listtablename{List of Tables}
\fi
\ifdefined\figurename
  \renewcommand*\figurename{Figure}
\else
  \newcommand\figurename{Figure}
\fi
\ifdefined\tablename
  \renewcommand*\tablename{Table}
\else
  \newcommand\tablename{Table}
\fi
}
\@ifpackageloaded{float}{}{\usepackage{float}}
\floatstyle{ruled}
\@ifundefined{c@chapter}{\newfloat{codelisting}{h}{lop}}{\newfloat{codelisting}{h}{lop}[chapter]}
\floatname{codelisting}{Listing}

\makeatother
\makeatletter
\@ifpackageloaded{caption}{}{\usepackage{caption}}
\@ifpackageloaded{subcaption}{}{\usepackage{subcaption}}
\makeatother
\makeatletter
\@ifpackageloaded{tcolorbox}{}{\usepackage[many]{tcolorbox}}
\makeatother
\makeatletter
\@ifundefined{shadecolor}{\definecolor{shadecolor}{HTML}{FAFAFA}}
\makeatother
\ifLuaTeX
  \usepackage{selnolig}  
\fi
\IfFileExists{bookmark.sty}{\usepackage{bookmark}}{\usepackage{hyperref}}
\IfFileExists{xurl.sty}{\usepackage{xurl}}{} 
\hypersetup{
  pdftitle={tmfast fits topic models fast},
  pdfauthor={Daniel J. Hicks},
  colorlinks=true,
  linkcolor={blue},
  filecolor={Maroon},
  citecolor={Blue},
  urlcolor={Blue},
  pdfcreator={LaTeX via pandoc}}

\renewcommand{\today}{2023-05-02}
\newcommand{\runninghead}{A Preprint }
\title{\texttt{tmfast} fits topic models fast}
\author{
\textbf{Daniel J. Hicks}~\orcidlink{0000-0001-7945-4416}\\Department of
Philosophy\\University of California, Merced\\Merced,
CA,\ 95343\\\href{mailto:dhicks4@ucmerced.edu}{dhicks4@ucmerced.edu}}
\date{2023-05-02}
\begin{document}
\maketitle
\begin{abstract}
\texttt{tmfast} is an R package for fitting topic models using a fast
algorithm based on partial PCA and the varimax rotation. After providing
mathematical background to the method, we present two examples, using a
simulated corpus and aggregated works of a selection of authors from the
long nineteenth century, and compare the quality of the fitted models to
a standard topic modeling package.
\end{abstract}
\ifdefined\Shaded\renewenvironment{Shaded}{\begin{tcolorbox}[breakable, enhanced, colback={shadecolor}, boxrule=0pt, frame hidden]}{\end{tcolorbox}}\fi

\renewcommand*\contentsname{Table of contents}
{
\hypersetup{linkcolor=}
\setcounter{tocdepth}{3}
\tableofcontents
}
\hypertarget{introduction}{%
\section{Introduction}\label{introduction}}

Topic modeling is a natural language processing (NLP) technique popular
among digital humanists, computational social scientists, and data
scientists working with textual data (eg, product reviews)
(\protect\hyperlink{ref-RobertsStmPackageStructural2019}{Roberts,
Stewart, and Tingley 2019}). Compared to methods such as vector space
embeddings or general-use clustering algorithms such as \(k\)-means, a
key advantage of topic modeling is that it simultaneously clusters both
text units (terms or phrases) and documents, enabling analysts to
provide human-meaningful, domain-specific labels to the clusters
(topics).

However, a major disadvantage of topic modeling is that the models are
relatively computationally intensive and slow to fit. This strongly
discourages analysts from fitting and comparing multiple models, which
is arguably the best way to determine to what extent results are
sensitive to researcher degrees of freedom
(\protect\hyperlink{ref-GelmanGardenForkingPaths2013}{Gelman and Loken
2013};
\protect\hyperlink{ref-SteegenIncreasingTransparencyMultiverse2016}{Steegen
et al. 2016}). Instead, typically analysts fit a few models to a given
corpus and focus interpretation on a single ``best'' model, often chosen
by informal assessments of ``interpretability'' of the fitted topics,
introducing additional researcher degrees of freedom.

This paper reports \texttt{tmfast}, an R package designed to facilitate
a multiple-model approach by using a significantly faster fitting
algorithm. After giving a brief mathematical background in
Section~\ref{sec-math}, we walk through two examples of \texttt{tmfast}
in action: generating and fitting models to a simulated text corpus
(Section~\ref{sec-sim}), and then fitting models to a collection of
books by different authors retrieved from Project Gutenberg
(Section~\ref{sec-realbooks}). Note that both of these examples are
supervised cases --- the true topics are known \emph{a priori} --- and
we use a method from Malaterre and Lareau
(\protect\hyperlink{ref-MalaterreEarlyDaysContemporary2022}{2022}) to
assess goodness of fit. In addition, we also fit models using the
\texttt{stm} package
(\protect\hyperlink{ref-RobertsStmPackageStructural2019}{Roberts,
Stewart, and Tingley 2019}) --- generally regarded as the state of the
art in topic modelling in R --- and compare the models fitted by the two
packages. \texttt{tmfast} is available at
\texttt{\textless{}https://github.com/dhicks/tmfast\textgreater{}}.

\hypertarget{sec-math}{%
\section{Mathematical background}\label{sec-math}}

Topic modeling is typically framed using a generative model. A corpus
\(C\) is defined by a fixed vocabulary or collection of terms \(T\); a
collection of \(k\) topics \(B\), where each topic \(\beta \in B\) is a
multinomial distribution over \(W\); and parameters \(\lambda > 0\) and
\(\alpha = (\alpha_1, \ldots, \alpha_k)\) with each \(\alpha_i > 0\).
Then a document \(d\) is generated as follows:

\begin{enumerate}
\def\labelenumi{\arabic{enumi}.}
\tightlist
\item
  Draw the total length \(N_d\) of \(d\) from a Poisson distribution,
  \(N_ds \sim \textrm{Poisson}(\lambda)\) (other distributions over the
  whole numbers might be used here, eg, negative binomial)
\item
  Draw a (\(k\)-element) topic distribution \(\theta_d\) from the
  Dirichlet distribution defined by \(\alpha\),
  \(\theta_d \sim \textrm{Dir}(\alpha)\)
\item
  For each token \(t_i\) (\(i = 1, \ldots, N\)),

  \begin{enumerate}
  \def\labelenumii{\alph{enumii}.}
  \tightlist
  \item
    Draw a topic \(b_i \sim \textrm{Multinomial}(\theta_d)\)
  \item
    Draw a term from the topic, \(t_i \sim b_i\)
    (\protect\hyperlink{ref-BleiLatentDirichletAllocation2003}{Blei, Ng,
    and Jordan 2003, 996}).
  \end{enumerate}
\end{enumerate}

This generative model is used to define a joint probability distribution
that is fit to the data (observed document lengths and token counts)
using numerical methods such as variational Bayes.

Rohe and Zeng
(\protect\hyperlink{ref-RoheVintageFactorAnalysis2020}{2020}) take a
different approach to topic modeling, viewing it through the lens of
principal component analysis (PCA) and the varimax rotation.\\
Consider a rectangular dataset \(X\) with \(n\) observations of \(p\)
variables (\(n \times p\)). In a statistics or data science context, PCA
is used for \emph{dimension reduction}, representing these data with
\(k < p\) dimensions while preserving as much of the original variance
as possible. Contemporary approaches to PCA use the singular value
decomposition

\[ X = U \Sigma V^t = U L^t \]

where \(U\) is a \(n \times n\) orthogonal matrix (the column vectors
are orthogonal and length 1), \(\Sigma\) is a \(n \times p\) diagonal
matrix (all non-diagonal entries are 0), and \(V\) is a \(p \times p\)
orthogonal matrix. \(L = V \Sigma^t\) is a \(p \times n\) matrix called
the \emph{loadings}. When \(p < n\) (that is, more observations than
variables) then columns \(p+1, p+2, \ldots, n\) of the loadings will be
zero, and columns \(1, 2, \ldots, p\) can be interpreted as a new set of
\(p\) variables constructed from the observed \(p\) variables. The rows
of \(U\) are called the \emph{scores}; they represent the values of the
observations in the new variables.

If \(X\) is centered (mean of each column/variable is 0) then the SVD is
related to the covariance of the original variables in such a way that
the new variables are ordered from greatest to least variance, and the
original and new variables have the same total variance. So if we
restrict our attention to the first \(k\) new variables we will have a
smaller representation of the original dataset that captures as much of
the original variance as possible. Formally, let \(U_k\) be the
\(n \times k\) matrix with columns \(1, \ldots, k\) of \(U\) and
\(L_k = V_k \Sigma_k^t\) the corresponding \(p \times k\) partial
loadings matrix. Then \(X \approx U_k \Sigma_k V_k^t\).

The loadings matrix is generally not easy to interpret, because the new
variables are arbitrary linear combinations of the original variables.
Such interpretations are essential in factor analysis, which attempts to
identify interpretable latent variables from the data, such as
psychological constructs corresponding to (weighted) sets of items in a
survey instrument.\footnote{Strictly speaking, PCA and factor analysis
  are two different analytical tasks. Factor analysis models are
  typically fit by optimizing a maximum likelihood model, rather than an
  algebraic method like SVD. And the rotation introduced in the next
  sentence means that the new variables are not orthogonal/uncorrelated
  and are not ordered from greatest to least variance, which are key
  desiderata of PCA. Nonetheless, the approach to topic modeling
  proposed by Rohe and Zeng
  (\protect\hyperlink{ref-RoheVintageFactorAnalysis2020}{2020}) combines
  PCA with varimax.} Psychometricians proposed to address this problem
by finding a \(k \times k\) orthogonal\footnote{Orthogonal matrices have
  the property that \(T^t = T^{-1}\).} matrix \(T\)

\[ U_k L_k^t = U_k T^t T L_k^t = U_k T^t (L_k T)^t \]

that (roughly) makes the ``rotated'' scores and loadings, \(U_k T^t\)
and \(L_k T\), as \emph{sparse} as possible, that is, have as few
non-zero entries as possible. This makes the new variables much more
interpretable, as generalizations or abstractions of a small collection
of observed variables. Because orthogonal matrices generalize rotations
and the method for finding this \(T\) involves maximizing a total
variance, this method is called the \emph{varimax rotation}.

Finally, to semi-formally motivate a connection between PCA and topic
modeling, consider \(r_{td}\), the occurrence rate of term \(t\) in
document \(d\). This rate estimates the conditional probability of \(t\)
given \(d\):

\[ r_{td} \approx \Pr(t | d) = \sum_i \Pr(t | b_i) \Pr(b_i | d) = \sum_i b_i \theta_d, \]

with a slight abuse of notation, where \(i \in 1, \ldots, k\) indexes
topics. In other words, topic modeling can be seen as factoring the
(more-or-less observed) term-document distribution into two sets of
latent distributions, term-topic and topic-document, much like PCA
factors a data matrix into scores and loadings in latent variables. See
Rohe and Zeng
(\protect\hyperlink{ref-RoheVintageFactorAnalysis2020}{2020}) lemma 5.2
for a formal development of this connection.

The upshot is that the latent variables constructed using PCA + varimax
can be interpreted as topics. Sparsity means that a given document will
have near-zero value for all but a few topics, and a given topic will
have near-zero value for all but a few documents.

The most obvious potential advantage of this approach is speed. Text
data is typically extremely sparse --- documents typically contain only
a small fraction of the words in the full vocabulary --- and efficient
algorithms have been developed for partial SVD of sparse matrices
(\protect\hyperlink{ref-BaglamaAugmentedImplicitlyRestarted2005}{James
Baglama and Reichel 2005}).

The \texttt{tmfast} package implements this PCA + varimax approach to
topic modeling in R, with specific support for the widely-used tidyverse
idiom. The \texttt{irlba} package
(\protect\hyperlink{ref-BaglamaIrlbaFastTruncated2022}{Jim Baglama,
Reichel, and Lewis 2022}) is used for efficient SVD (by default; users
can specify an alternative SVD method if they prefer). \texttt{tmfast}
is available at \url{https://github.com/dhicks/tmfast}.

\hypertarget{sec-sim}{%
\section{Example 1: A simulated corpus}\label{sec-sim}}

\texttt{tmfast} includes a collection of functions to generate a
simulated corpus according to the standard generative model. In this
section, we use these functions to generate a corpus, fit topic models
using \texttt{tmfast} and \texttt{stm}
(\protect\hyperlink{ref-RobertsStmPackageStructural2019}{Roberts,
Stewart, and Tingley 2019}) --- widely used for topic modeling in R ---
and compare their respective ability to identify the true topics used to
generate the corpus.

We first load the \texttt{tidyverse} suite, the \texttt{lpSolve} package
to match fitted and true topics, the \texttt{tictoc} package to
calculate wall compute times, and \texttt{tmfast} and \texttt{stm}. The
\texttt{tidytext} package is also loaded for its \texttt{stm} tidiers
(eg, functions to represent a fitted \texttt{stm} model as a dataframe).

\begin{Shaded}
\begin{Highlighting}[]
\FunctionTok{library}\NormalTok{(tidyverse)          }\CommentTok{\# infrastructure}
\FunctionTok{theme\_set}\NormalTok{(}\FunctionTok{theme\_minimal}\NormalTok{())  }\CommentTok{\# make plots not look bad}
\FunctionTok{library}\NormalTok{(lpSolve)            }\CommentTok{\# used to match fitted and true topics}
\FunctionTok{library}\NormalTok{(tictoc)             }\CommentTok{\# timing}
\FunctionTok{library}\NormalTok{(tmfast)             }\CommentTok{\# fit topic models fast! }
\FunctionTok{library}\NormalTok{(stm)                }\CommentTok{\# standard topic model package}
\FunctionTok{library}\NormalTok{(tidytext)           }\CommentTok{\# tidiers for stm models}
\end{Highlighting}
\end{Shaded}

\hypertarget{simulation-parameters}{%
\subsection{Simulation parameters}\label{simulation-parameters}}

We create simulated text data following the data-generating process
assumed by LDA. Specifically, each document will be generated from one
of several ``journals.'' Each journal corresponds to a topic, and vice
versa, in that documents from journal \(j\) will tend to have a much
greater probability for topic \(j\) than the other topics.

We first specify the number of topics/journals \texttt{k}, and the
number of documents to draw from each journal \texttt{Mj}, for a total
of \texttt{M\ =\ Mj\ *\ k} documents in the corpus. We also specify the
length of the vocabulary (total unique words) as a multiple of the total
number of documents \texttt{M}. Document lengths are generated using a
negative binomial distribution, using the size-mean parameterization.
Per \texttt{?NegBinomial}, the standard deviation of document lengths in
this parameterization is \(\sqrt{\mu + \frac{\mu^2}{\mathrm{size}}}\).

\begin{Shaded}
\begin{Highlighting}[]
\NormalTok{k }\OtherTok{=} \DecValTok{10}                \CommentTok{\# Num. topics / journals}
\NormalTok{Mj }\OtherTok{=} \DecValTok{100}              \CommentTok{\# Num. documents per journal}
\NormalTok{M }\OtherTok{=}\NormalTok{ Mj}\SpecialCharTok{*}\NormalTok{k              }\CommentTok{\# Total corpus size}
\NormalTok{vocab }\OtherTok{=}\NormalTok{ M             }\CommentTok{\# Vocabulary length}

\DocumentationTok{\#\# Negative binomial distribution of doc lengths}
\NormalTok{size }\OtherTok{=} \DecValTok{10}             \CommentTok{\# Size and mean}
\NormalTok{mu }\OtherTok{=} \DecValTok{300}
\FunctionTok{sqrt}\NormalTok{(mu }\SpecialCharTok{+}\NormalTok{ mu}\SpecialCharTok{\^{}}\DecValTok{2}\SpecialCharTok{/}\NormalTok{size)  }\CommentTok{\# Resulting SD of document sizes}
\end{Highlighting}
\end{Shaded}

\begin{verbatim}
[1] 96.43651
\end{verbatim}

Topic-document and word-topic distributions are both sampled from
Dirichlet distributions. For topic-docs, we use an asymmetric Dirichlet
distribution\footnote{The \(k\)-component Dirichlet distribution is
  parameterized by a \(k\)-component vector
  \(\mathbf{\alpha} = \alpha (n_1, \ldots, n_k)\), where \(\alpha\) is a
  scalar and \(\sum_i n_i = 1\). Using this parameterization, the
  expected value for component \(i\) is \(n_i\) with variance
  \(\frac{n_i(1-n_i)}{\alpha + 1}\). So increasing the scaling factor
  \(\alpha\) means samples from the Dirichlet distribution will be more
  likely to look like \((n_1, \ldots, n_k)\).} where one component will
have (in expectation) most of the probability mass (eg, 80\%) and the
remaining probability mass will be (in expectation) distributed evenly
over the remaining components (eg, \(0.2/(k-1)\)). For word-topics we
use a symmetric Dirichlet distribution (parametized only by the scaling
factor). \texttt{tmfast} includes utility functions for constructing and
drawing both kinds of Dirichlet distributions.

\begin{Shaded}
\begin{Highlighting}[]
\DocumentationTok{\#\# Dirichlet distributions for topic{-}docs and word{-}topics}
\NormalTok{topic\_peak }\OtherTok{=}\NormalTok{ .}\DecValTok{8}
\NormalTok{topic\_scale }\OtherTok{=} \DecValTok{10}
\FunctionTok{peak\_alpha}\NormalTok{(k, }\DecValTok{1}\NormalTok{, }\AttributeTok{peak =}\NormalTok{ topic\_peak, }\AttributeTok{scale =}\NormalTok{ topic\_scale)}
\end{Highlighting}
\end{Shaded}

\begin{verbatim}
 [1] 8.0000000 0.2222222 0.2222222 0.2222222 0.2222222 0.2222222 0.2222222
 [8] 0.2222222 0.2222222 0.2222222
\end{verbatim}

\begin{Shaded}
\begin{Highlighting}[]
\FunctionTok{peak\_alpha}\NormalTok{(k, }\DecValTok{2}\NormalTok{, }\AttributeTok{peak =}\NormalTok{ topic\_peak, }\AttributeTok{scale =}\NormalTok{ topic\_scale)}
\end{Highlighting}
\end{Shaded}

\begin{verbatim}
 [1] 0.2222222 8.0000000 0.2222222 0.2222222 0.2222222 0.2222222 0.2222222
 [8] 0.2222222 0.2222222 0.2222222
\end{verbatim}

\begin{Shaded}
\begin{Highlighting}[]
\NormalTok{word\_beta }\OtherTok{=} \FloatTok{0.1}
\end{Highlighting}
\end{Shaded}

Because the simulations involve drawing samples using a RNG, we set a
seed.

\begin{Shaded}
\begin{Highlighting}[]
\FunctionTok{set.seed}\NormalTok{(}\DecValTok{2022{-}06{-}19}\NormalTok{)}
\end{Highlighting}
\end{Shaded}

\hypertarget{draw-true-topic-distributions}{%
\subsection{Draw true topic
distributions}\label{draw-true-topic-distributions}}

We generate the true topic-document distributions
\(p(\theta = t | \mathrm{doc}_m)\), often simply notated \(\theta\) or
\(\gamma\). In this vignette we use \(\theta\) for the true distribution
and \(\gamma\) for the fitted distribution in the topic model. Each
document's \(\theta\) is sampled from a Dirichlet distribution
(\texttt{rdirichlet()}), with the parameter \(\mathbf{\alpha}\)
corresponding to the document's journal \(j\). The variable
\texttt{theta} is a \texttt{M} by \texttt{k} matrix; \texttt{theta\_df}
is a tidy representation with columns \texttt{doc}, \texttt{topic}, and
\texttt{prob}. The visualization confirms that documents are generally
most strongly associated with the corresponding topics, though with some
noise: in the median document, 82\% of its topic probability mass is
associated with the single dominant topic.

\begin{Shaded}
\begin{Highlighting}[]
\DocumentationTok{\#\# Journal{-}specific alpha, with a peak value (80\%) and uniform otherwise; }
\DocumentationTok{\#\# For each topic, draw Mj documents}
\NormalTok{theta }\OtherTok{=} \FunctionTok{map}\NormalTok{(}\DecValTok{1}\SpecialCharTok{:}\NormalTok{k, }
            \SpecialCharTok{\textasciitilde{}}\FunctionTok{rdirichlet}\NormalTok{(Mj, }
                        \FunctionTok{peak\_alpha}\NormalTok{(k, .x, }
                                   \AttributeTok{peak =}\NormalTok{ topic\_peak, }
                                   \AttributeTok{scale =}\NormalTok{ topic\_scale))) }\SpecialCharTok{|\textgreater{}} 
    \FunctionTok{reduce}\NormalTok{(rbind)}

\NormalTok{theta\_df }\OtherTok{=}\NormalTok{ theta }\SpecialCharTok{|\textgreater{}}
    \FunctionTok{as\_tibble}\NormalTok{(}\AttributeTok{rownames =} \StringTok{\textquotesingle{}doc\textquotesingle{}}\NormalTok{, }
              \AttributeTok{.name\_repair =}\NormalTok{ tmfast}\SpecialCharTok{:::}\NormalTok{make\_colnames) }\SpecialCharTok{|\textgreater{}}
    \FunctionTok{mutate}\NormalTok{(}\AttributeTok{doc =} \FunctionTok{as.integer}\NormalTok{(doc)) }\SpecialCharTok{|\textgreater{}}
    \FunctionTok{pivot\_longer}\NormalTok{(}\FunctionTok{starts\_with}\NormalTok{(}\StringTok{\textquotesingle{}V\textquotesingle{}}\NormalTok{),}
                 \AttributeTok{names\_to =} \StringTok{\textquotesingle{}topic\textquotesingle{}}\NormalTok{,}
                 \AttributeTok{values\_to =} \StringTok{\textquotesingle{}theta\textquotesingle{}}\NormalTok{)}
\NormalTok{theta\_df}
\end{Highlighting}
\end{Shaded}

\begin{verbatim}
# A tibble: 10,000 x 3
     doc topic      theta
   <int> <chr>      <dbl>
 1     1 V01   0.872     
 2     1 V02   0.0644    
 3     1 V03   0.0278    
 4     1 V04   0.00169   
 5     1 V05   0.00249   
 6     1 V06   0.00137   
 7     1 V07   0.00693   
 8     1 V08   0.000124  
 9     1 V09   0.00000643
10     1 V10   0.0235    
# ... with 9,990 more rows
\end{verbatim}

\begin{Shaded}
\begin{Highlighting}[]
\FunctionTok{ggplot}\NormalTok{(theta\_df, }\FunctionTok{aes}\NormalTok{(doc, topic, }\AttributeTok{fill =}\NormalTok{ theta)) }\SpecialCharTok{+}
    \FunctionTok{geom\_tile}\NormalTok{()}
\end{Highlighting}
\end{Shaded}

\begin{figure}[H]

{\centering \includegraphics{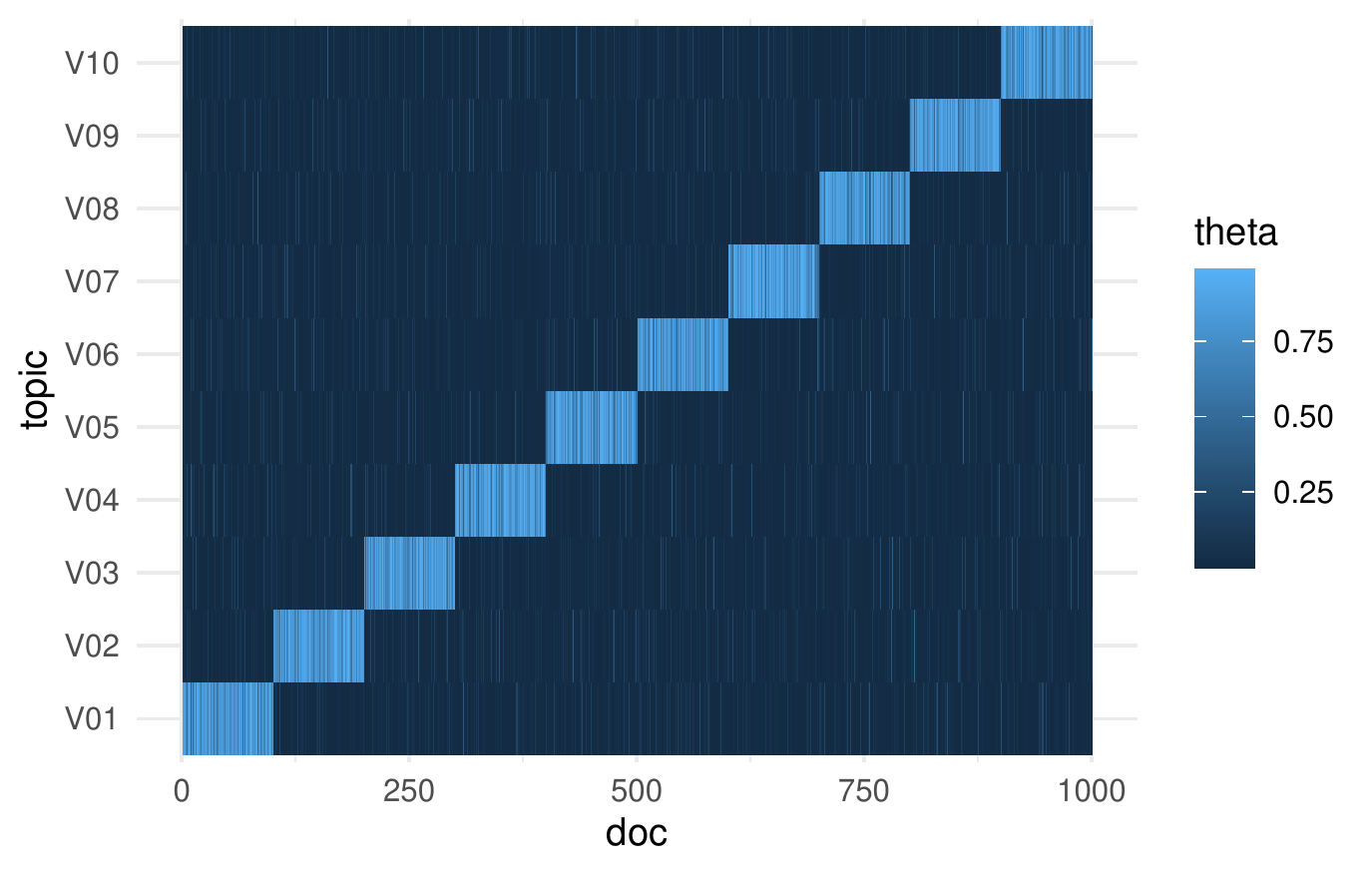}

}

\end{figure}

\begin{Shaded}
\begin{Highlighting}[]
\NormalTok{theta\_df }\SpecialCharTok{|\textgreater{}} 
    \FunctionTok{group\_by}\NormalTok{(doc) }\SpecialCharTok{|\textgreater{}} 
    \FunctionTok{summarize}\NormalTok{(}\AttributeTok{max =} \FunctionTok{max}\NormalTok{(theta)) }\SpecialCharTok{|\textgreater{}} 
    \FunctionTok{pull}\NormalTok{(max) }\SpecialCharTok{|\textgreater{}} 
    \FunctionTok{summary}\NormalTok{()}
\end{Highlighting}
\end{Shaded}

\begin{verbatim}
   Min. 1st Qu.  Median    Mean 3rd Qu.    Max. 
 0.3567  0.7148  0.8192  0.7958  0.8963  0.9901 
\end{verbatim}

\hypertarget{draw-true-word-distributions}{%
\subsection{Draw true word
distributions}\label{draw-true-word-distributions}}

Next we generate the true word-topic distributions
\(p(\phi = w | \theta = t)\), often designed as either \(\phi\) or
\(\beta\). We use \(\phi\) for the true distribution and \(\beta\) for
the fitted distribution. We sample these distributions from a symmetric
Dirichlet distribution over the length of the vocabulary with
\(\alpha = .01\). Tile and Zipfian (probability vs.~rank on a log-log
scale) plots confirm these distributions are working correctly.

\begin{Shaded}
\begin{Highlighting}[]
\DocumentationTok{\#\# phi\_j:  Word distribution for topic j}
\NormalTok{phi }\OtherTok{=} \FunctionTok{rdirichlet}\NormalTok{(k, word\_beta, }\AttributeTok{k =}\NormalTok{ vocab)}

\NormalTok{phi\_df }\OtherTok{=}\NormalTok{ phi }\SpecialCharTok{|\textgreater{}}
    \FunctionTok{t}\NormalTok{() }\SpecialCharTok{|\textgreater{}} 
    \FunctionTok{as\_tibble}\NormalTok{(}\AttributeTok{rownames =} \StringTok{\textquotesingle{}token\textquotesingle{}}\NormalTok{, }
              \AttributeTok{.name\_repair =}\NormalTok{ tmfast}\SpecialCharTok{:::}\NormalTok{make\_colnames) }\SpecialCharTok{|\textgreater{}}
    \FunctionTok{pivot\_longer}\NormalTok{(}\FunctionTok{starts\_with}\NormalTok{(}\StringTok{\textquotesingle{}V\textquotesingle{}}\NormalTok{),}
                 \AttributeTok{names\_to =} \StringTok{\textquotesingle{}topic\textquotesingle{}}\NormalTok{,}
                 \AttributeTok{values\_to =} \StringTok{\textquotesingle{}phi\textquotesingle{}}\NormalTok{)}
\NormalTok{phi\_df}
\end{Highlighting}
\end{Shaded}

\begin{verbatim}
# A tibble: 10,000 x 3
   token topic      phi
   <chr> <chr>    <dbl>
 1 1     V01   2.28e- 4
 2 1     V02   2.51e-18
 3 1     V03   4.06e- 4
 4 1     V04   5.86e- 4
 5 1     V05   5.73e- 8
 6 1     V06   1.04e- 5
 7 1     V07   2.67e- 4
 8 1     V08   9.02e- 6
 9 1     V09   1.00e-25
10 1     V10   2.03e- 4
# ... with 9,990 more rows
\end{verbatim}

\begin{Shaded}
\begin{Highlighting}[]
\DocumentationTok{\#\# Word distributions}
\FunctionTok{ggplot}\NormalTok{(phi\_df, }\FunctionTok{aes}\NormalTok{(topic, token, }\AttributeTok{fill =}\NormalTok{ (phi))) }\SpecialCharTok{+}
    \FunctionTok{geom\_tile}\NormalTok{() }\SpecialCharTok{+}
    \FunctionTok{scale\_y\_discrete}\NormalTok{(}\AttributeTok{breaks =} \ConstantTok{NULL}\NormalTok{)}
\end{Highlighting}
\end{Shaded}

\begin{figure}[H]

{\centering \includegraphics{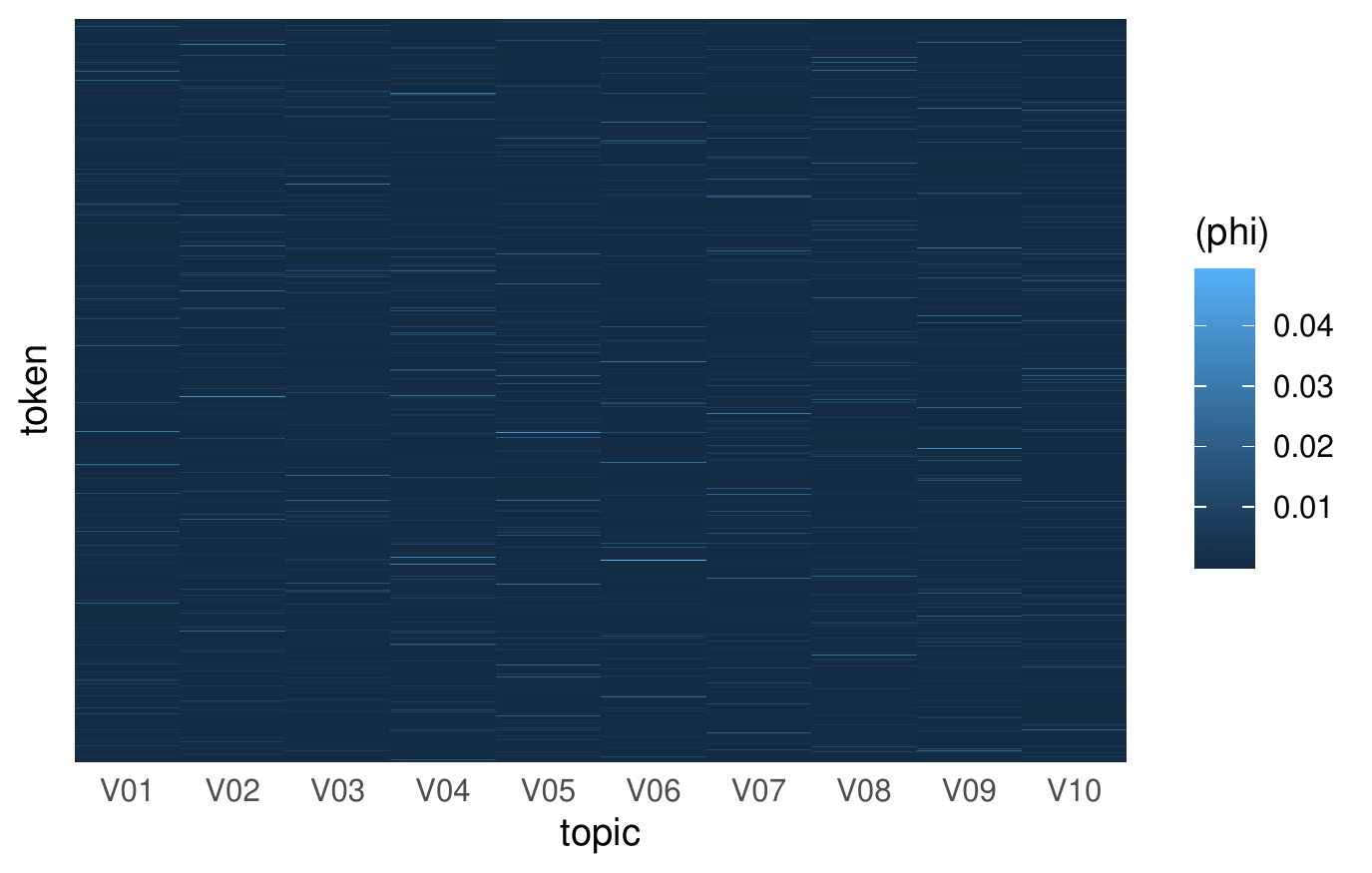}

}

\end{figure}

\begin{Shaded}
\begin{Highlighting}[]
\DocumentationTok{\#\# Zipfian plot}
\NormalTok{phi\_df }\SpecialCharTok{|\textgreater{}}
    \FunctionTok{group\_by}\NormalTok{(topic) }\SpecialCharTok{|\textgreater{}}
    \FunctionTok{mutate}\NormalTok{(}\AttributeTok{rank =} \FunctionTok{rank}\NormalTok{(}\FunctionTok{desc}\NormalTok{(phi))) }\SpecialCharTok{|\textgreater{}}
    \FunctionTok{arrange}\NormalTok{(topic, rank) }\SpecialCharTok{|\textgreater{}}
    \FunctionTok{filter}\NormalTok{(rank }\SpecialCharTok{\textless{}}\NormalTok{ vocab}\SpecialCharTok{/}\DecValTok{2}\NormalTok{) }\SpecialCharTok{|\textgreater{}}
    \FunctionTok{ggplot}\NormalTok{(}\FunctionTok{aes}\NormalTok{(rank, phi, }\AttributeTok{color =}\NormalTok{ topic)) }\SpecialCharTok{+}
    \FunctionTok{geom\_line}\NormalTok{() }\SpecialCharTok{+}
    \FunctionTok{scale\_x\_log10}\NormalTok{() }\SpecialCharTok{+}
    \FunctionTok{scale\_y\_log10}\NormalTok{()}
\end{Highlighting}
\end{Shaded}

\begin{figure}[H]

{\centering \includegraphics{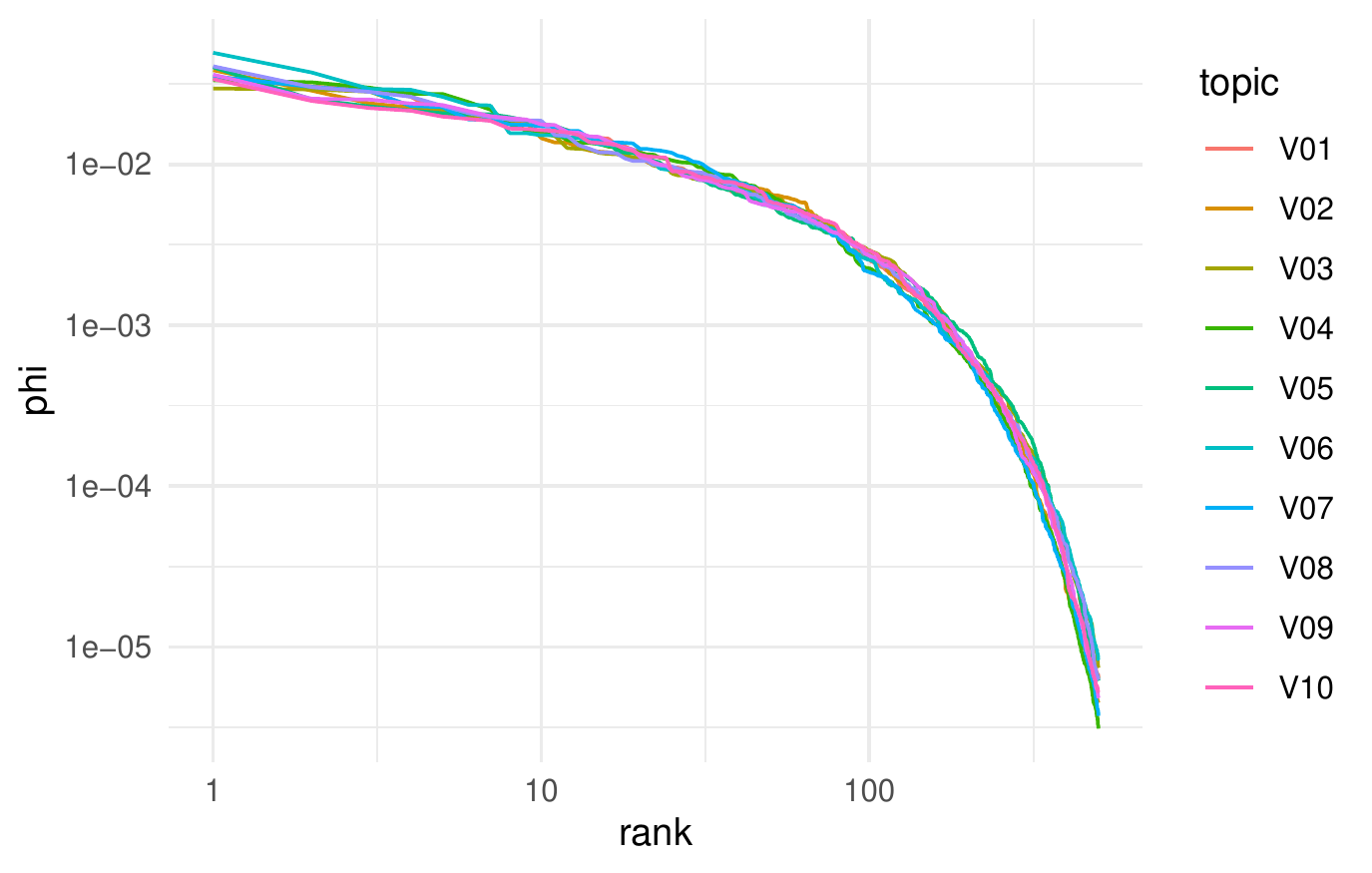}

}

\end{figure}

\hypertarget{document-lengths}{%
\subsection{Document lengths}\label{document-lengths}}

Again, document lengths are drawn from a negative binomial distribution.

\begin{Shaded}
\begin{Highlighting}[]
\DocumentationTok{\#\# N\_i:  Length of document i}
\NormalTok{N }\OtherTok{=} \FunctionTok{rnbinom}\NormalTok{(M, }\AttributeTok{size =}\NormalTok{ size, }\AttributeTok{mu =}\NormalTok{ mu)}
\FunctionTok{summary}\NormalTok{(N)}
\end{Highlighting}
\end{Shaded}

\begin{verbatim}
   Min. 1st Qu.  Median    Mean 3rd Qu.    Max. 
   93.0   240.8   300.5   308.6   364.5   774.0 
\end{verbatim}

\begin{Shaded}
\begin{Highlighting}[]
\FunctionTok{sd}\NormalTok{(N)}
\end{Highlighting}
\end{Shaded}

\begin{verbatim}
[1] 95.1555
\end{verbatim}

\begin{Shaded}
\begin{Highlighting}[]
\FunctionTok{hist}\NormalTok{(N)}
\end{Highlighting}
\end{Shaded}

\begin{figure}[H]

{\centering \includegraphics{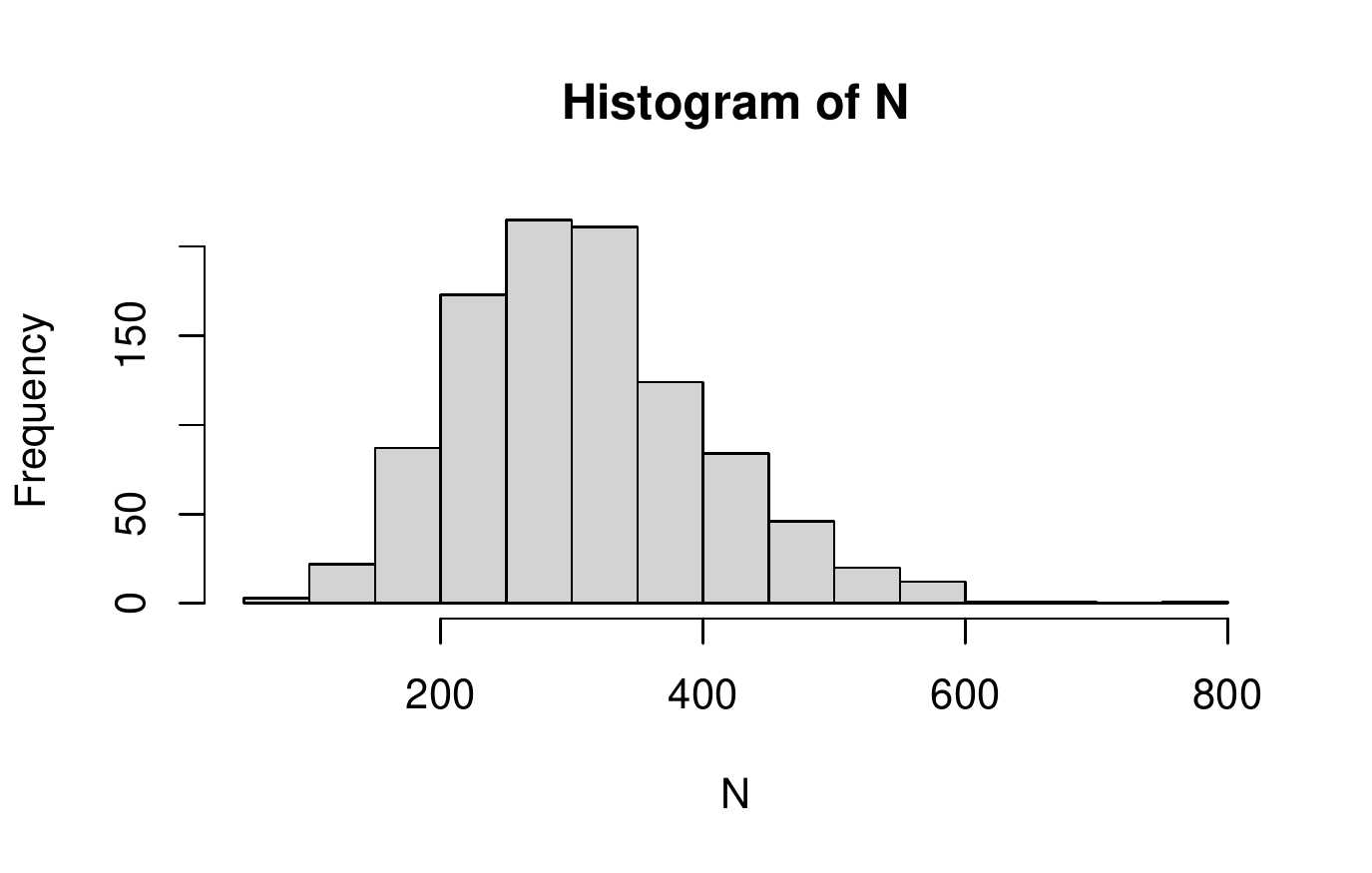}

}

\end{figure}

\hypertarget{draw-corpus}{%
\subsection{Draw corpus}\label{draw-corpus}}

Finally we draw the corpus, the observed word counts for each document.
This is the most time-consuming step in this script, much slower than
actually fitting the topic model. Experimenting with this simulation, we
found that \texttt{log1p()} scaling of the word counts produced better
results than other scaling techniques (eg, dividing by the total length
of each document, scaling words by their standard deviation) for
accounting for radical differences in document length.\footnote{Since
  \(\log(0+1) = 0\), this transformation also preserves sparsity and
  does not introduce infinite values.}

\begin{Shaded}
\begin{Highlighting}[]
\FunctionTok{tic}\NormalTok{()}
\NormalTok{corpus }\OtherTok{=} \FunctionTok{draw\_corpus}\NormalTok{(N, theta, phi)}
\FunctionTok{toc}\NormalTok{()}
\end{Highlighting}
\end{Shaded}

\begin{verbatim}
25.927 sec elapsed
\end{verbatim}

\begin{Shaded}
\begin{Highlighting}[]
\NormalTok{dtm }\OtherTok{=} \FunctionTok{mutate}\NormalTok{(corpus, }\AttributeTok{n =} \FunctionTok{log1p}\NormalTok{(n))}
\end{Highlighting}
\end{Shaded}

\hypertarget{fit-the-topic-model}{%
\subsection{Fit the topic model}\label{fit-the-topic-model}}

Fitting the topic model is extremely fast. Note that we can request
multiple values of \(k\) (numbers of topics) in a single call. Other
topic modelling packages typically fit only a single value of \(k\) at a
time.

Under the hood, we cast the document-term matrix to a sparse matrix
class if necessary. Then we extract the maximum number of desired
principal components using \texttt{irlba::prcomp\_irlba()}, centering
but not scaling the logged word counts. (Experiments with this
simulation indicated that scaling makes it more difficult to construct
probability distributions later.) Next we use the base R function
\texttt{stats:varimax()} to construct a preliminary varimax rotation of
the principal components. Because the direction of factors is arbitrary
as far as varimax is concerned, but meaningful when we convert things to
probability distributions, we check the skew of each factor's loadings
in the preliminary fit, and reverse the factors with negative skew (long
left tails with relatively large negative values).

\begin{Shaded}
\begin{Highlighting}[]
\FunctionTok{tic}\NormalTok{()}
\NormalTok{fitted }\OtherTok{=} \FunctionTok{tmfast}\NormalTok{(dtm, }\FunctionTok{c}\NormalTok{(}\DecValTok{3}\NormalTok{, }\DecValTok{5}\NormalTok{, k, }\DecValTok{2}\SpecialCharTok{*}\NormalTok{k))}
\FunctionTok{toc}\NormalTok{()}
\end{Highlighting}
\end{Shaded}

\begin{verbatim}
0.576 sec elapsed
\end{verbatim}

The object returned by \texttt{tmfast()} has a simple structure (pun
intended) and the \texttt{tmfast} S3 class. \texttt{totalvar} and
\texttt{sdev} come from the PCA step, giving the total variance across
all feature variables and the standard deviation of each extracted
principal component. (Note that these PCs do not generally correspond to
the varimax-rotated factors/topics.) \texttt{n} contains the sizes
(number of factors/topics) fitted for the models, and \texttt{varimaxes}
contains the varimax fit for each value of \texttt{n}. The varimax
objects each contain three matrices, the rotated \texttt{loadings}
(word-topics), the rotation matrix \texttt{rotmat}, and the rotated
\texttt{scores} (document-topics). Note that these are not stored as
probability distributions.

\begin{Shaded}
\begin{Highlighting}[]
\FunctionTok{str}\NormalTok{(fitted, }\AttributeTok{max.level =}\NormalTok{ 2L)}
\end{Highlighting}
\end{Shaded}

\begin{verbatim}
List of 9
 $ totalvar: num 138
 $ sdev    : num [1:20] 3.26 3.06 3.01 2.97 2.93 ...
 $ rows    : chr [1:1000] "1" "2" "3" "4" ...
 $ cols    : chr [1:999] "5" "7" "8" "11" ...
 $ center  : Named num [1:999] 0.4016 0.0943 0.4254 0.2396 0.4093 ...
  ..- attr(*, "names")= chr [1:999] "5" "7" "8" "11" ...
 $ scale   : logi FALSE
 $ rotation: num [1:999, 1:20] -0.00112 0.02725 0.01223 0.00457 -0.00883 ...
  ..- attr(*, "dimnames")=List of 2
 $ n       : num [1:4] 3 5 10 20
 $ varimax :List of 4
  ..$ 3 :List of 3
  ..$ 5 :List of 3
  ..$ 10:List of 3
  ..$ 20:List of 3
 - attr(*, "class")= chr [1:3] "tmfast" "varimaxes" "list"
\end{verbatim}

\begin{Shaded}
\begin{Highlighting}[]
\FunctionTok{str}\NormalTok{(fitted}\SpecialCharTok{$}\NormalTok{varimax[}\FunctionTok{as.character}\NormalTok{(k)])}
\end{Highlighting}
\end{Shaded}

\begin{verbatim}
List of 1
 $ 10:List of 3
  ..$ loadings: num [1:999, 1:10] -0.0479 0.1025 0.0436 -0.0414 -0.0505 ...
  .. ..- attr(*, "dimnames")=List of 2
  .. .. ..$ : chr [1:999] "5" "7" "8" "11" ...
  .. .. ..$ : NULL
  ..$ rotmat  : num [1:10, 1:10] 0.784 -0.2228 -0.23 -0.0823 0.3342 ...
  ..$ scores  : num [1:1000, 1:10] -0.384 -0.137 -0.253 -0.298 -0.388 ...
  .. ..- attr(*, "dimnames")=List of 2
  .. .. ..$ : chr [1:1000] "1" "2" "3" "4" ...
  .. .. ..$ : NULL
\end{verbatim}

Because the model contains a \texttt{sdev} component,
\texttt{screeplot()} works out of the box. Note that the first \(k\) PCs
have much higher variance than the others, and often the \(k\)th PC is
somewhat lower than the first \(k-1\). This reflects the highly
simplified structure of the simulated data. Real datasets often have a
much more gradual decline in the screeplot, likely reflecting the
complex hierarchy of topics in actual documents.

\begin{Shaded}
\begin{Highlighting}[]
\FunctionTok{screeplot}\NormalTok{(fitted, }\AttributeTok{npcs =}\NormalTok{ k }\SpecialCharTok{+} \DecValTok{5}\NormalTok{)}
\end{Highlighting}
\end{Shaded}

\begin{figure}[H]

{\centering \includegraphics{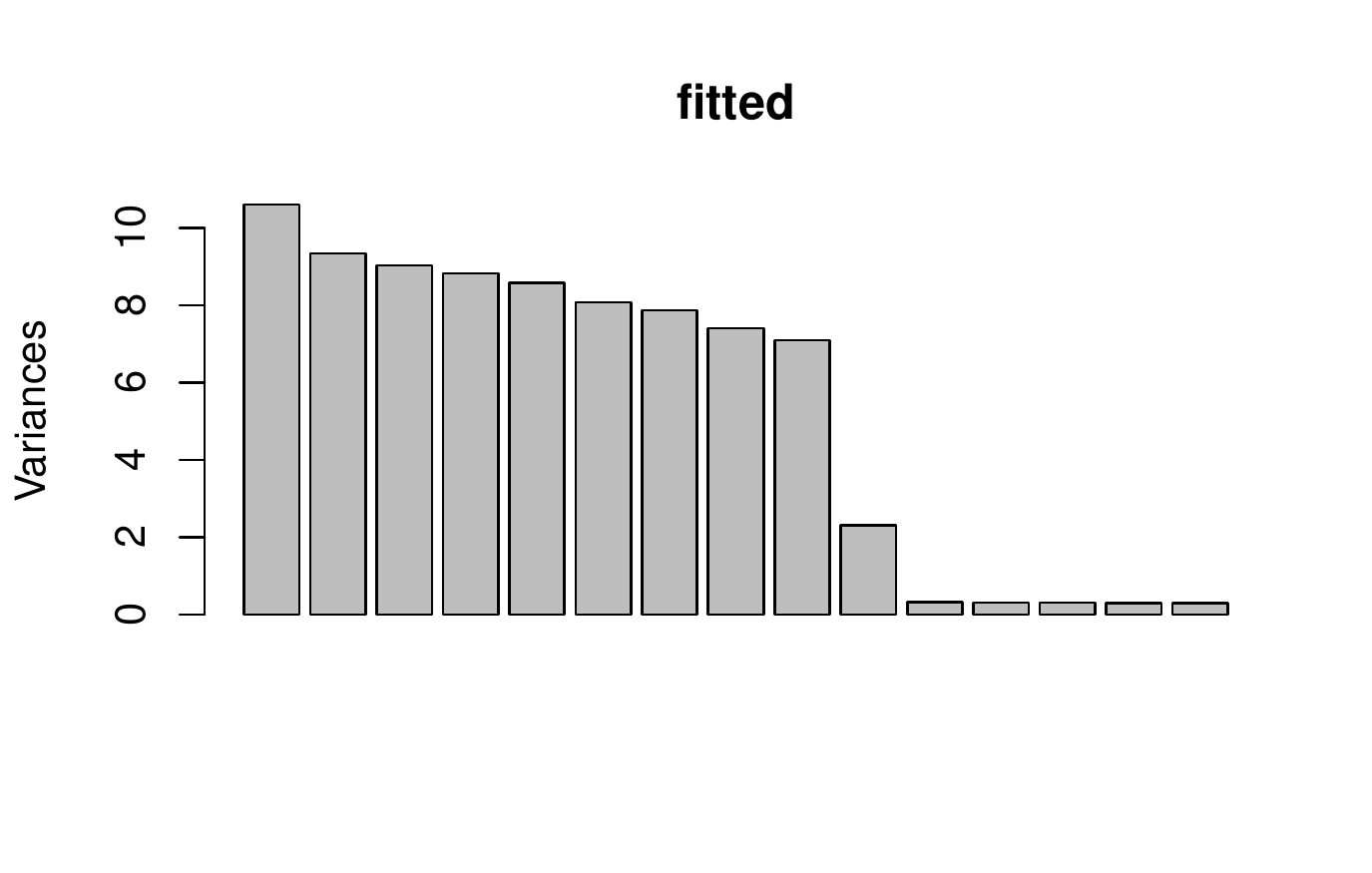}

}

\end{figure}

It's also straightforward to calculate the share of total variance
covered by successive principal components. Experimenting with this
simulation, it's common for \(k\) principal components to cover less
than half of the total variance. Again, note that the rotated varimax
factors don't correspond to the principal components, but the total
covered variance remains the same.

\begin{Shaded}
\begin{Highlighting}[]
\FunctionTok{cumsum}\NormalTok{(fitted}\SpecialCharTok{$}\NormalTok{sdev}\SpecialCharTok{\^{}}\DecValTok{2}\NormalTok{) }\SpecialCharTok{/}\NormalTok{ fitted}\SpecialCharTok{$}\NormalTok{totalvar}
\end{Highlighting}
\end{Shaded}

\begin{verbatim}
 [1] 0.07689789 0.14466433 0.21018176 0.27420309 0.33636478 0.39492291
 [7] 0.45196354 0.50563725 0.55705654 0.57380104 0.57606900 0.57828256
[13] 0.58048004 0.58264465 0.58479395 0.58691426 0.58902127 0.59107016
[19] 0.59311409 0.59514063
\end{verbatim}

\begin{Shaded}
\begin{Highlighting}[]
\FunctionTok{data.frame}\NormalTok{(}\AttributeTok{PC =} \DecValTok{1}\SpecialCharTok{:}\FunctionTok{length}\NormalTok{(fitted}\SpecialCharTok{$}\NormalTok{sdev),}
           \AttributeTok{cum\_var =} \FunctionTok{cumsum}\NormalTok{(fitted}\SpecialCharTok{$}\NormalTok{sdev}\SpecialCharTok{\^{}}\DecValTok{2}\NormalTok{) }\SpecialCharTok{/}\NormalTok{ fitted}\SpecialCharTok{$}\NormalTok{totalvar) }\SpecialCharTok{|\textgreater{}} 
    \FunctionTok{ggplot}\NormalTok{(}\FunctionTok{aes}\NormalTok{(PC, cum\_var)) }\SpecialCharTok{+}
    \FunctionTok{geom\_line}\NormalTok{() }\SpecialCharTok{+}
    \FunctionTok{geom\_point}\NormalTok{()}
\end{Highlighting}
\end{Shaded}

\begin{figure}[H]

{\centering \includegraphics{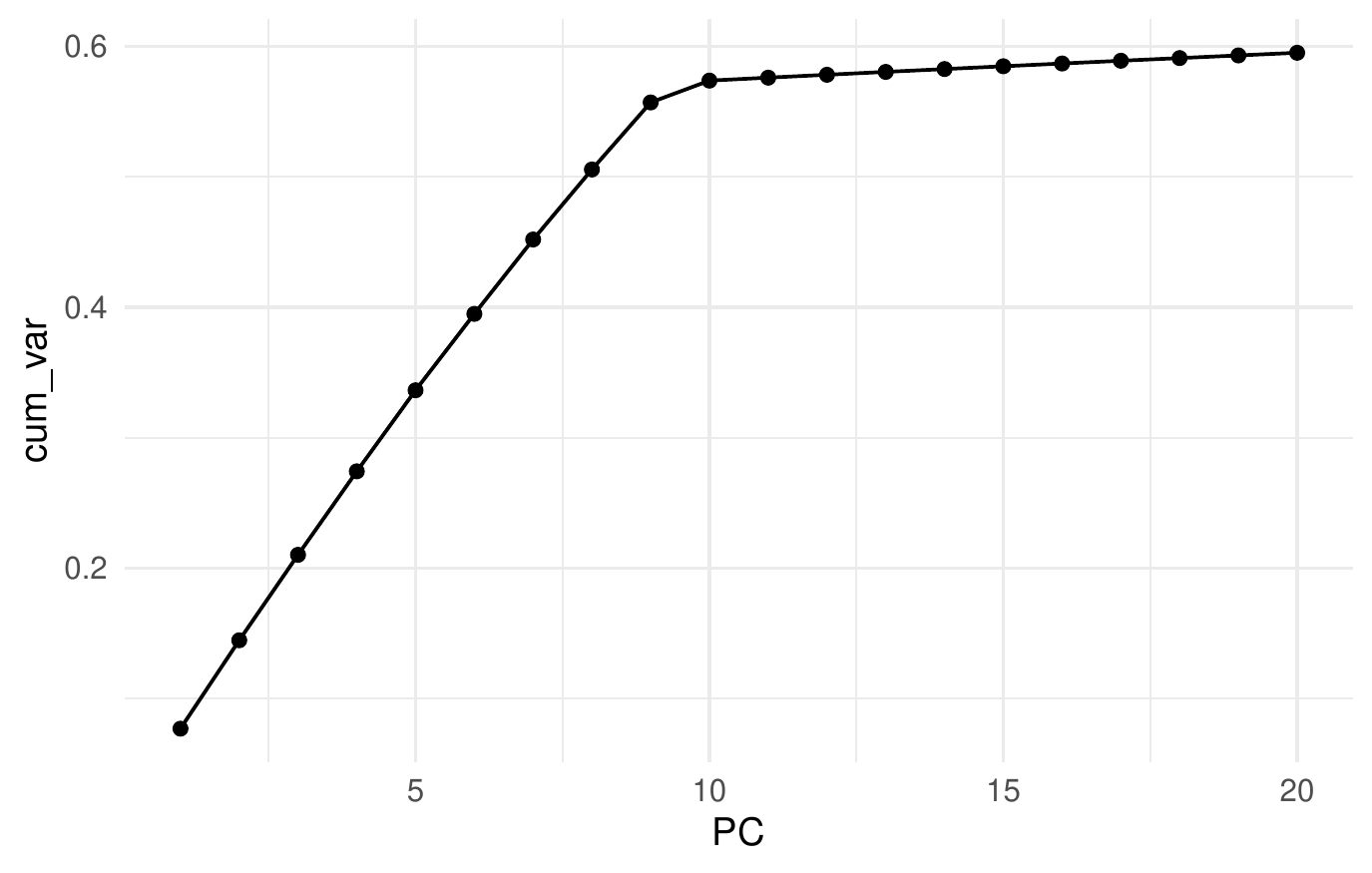}

}

\end{figure}

\hypertarget{fitting-a-conventional-topic-model-stm}{%
\subsection{Fitting a conventional topic model
(stm)}\label{fitting-a-conventional-topic-model-stm}}

For comparison, we'll also fit a conventional topic model using the
\texttt{stm} package. To address the challenge of picking a number of
topics, \texttt{stm::stm()} conducts a topic estimation process when
passed \texttt{K\ =\ 0}. With the simulation parameters and the random
seed used here, this process takes almost 12 seconds and produces a
model with 33 topics. We therefore do not run the code below.

\begin{Shaded}
\begin{Highlighting}[]
\FunctionTok{tic}\NormalTok{()}
\NormalTok{corpus }\SpecialCharTok{|\textgreater{}} 
    \FunctionTok{cast\_sparse}\NormalTok{(doc, word, n) }\SpecialCharTok{|\textgreater{}} 
    \FunctionTok{stm}\NormalTok{(}\AttributeTok{K =} \DecValTok{0}\NormalTok{, }\AttributeTok{verbose =} \ConstantTok{FALSE}\NormalTok{)}
\FunctionTok{toc}\NormalTok{()}
\end{Highlighting}
\end{Shaded}

Setting \texttt{K\ =\ k} gives us a fitted topic model in a few seconds,
about an order of magnitude slower than \texttt{tmfast()}. Profiling
experiments indicated that \texttt{tmfast()} is about 20x faster than
\texttt{stm()}.

\begin{Shaded}
\begin{Highlighting}[]
\FunctionTok{tic}\NormalTok{()}
\NormalTok{fitted\_stm }\OtherTok{=}\NormalTok{ corpus }\SpecialCharTok{|\textgreater{}} 
    \FunctionTok{cast\_sparse}\NormalTok{(doc, word, n) }\SpecialCharTok{|\textgreater{}} 
    \FunctionTok{stm}\NormalTok{(}\AttributeTok{K =}\NormalTok{ k, }\AttributeTok{verbose =} \ConstantTok{FALSE}\NormalTok{)}
\FunctionTok{toc}\NormalTok{()}
\end{Highlighting}
\end{Shaded}

\begin{verbatim}
4.732 sec elapsed
\end{verbatim}

\hypertarget{assessing-accuracy-word-topic-distributions}{%
\subsection{Assessing accuracy: Word-topic
distributions}\label{assessing-accuracy-word-topic-distributions}}

Using simulated data with true word-topic and topic-document
distributions enables us to check the accuracy of both \texttt{tmfast}
and \texttt{stm} models. Here we'll develop a method proposed by
Malaterre and Lareau
(\protect\hyperlink{ref-MalaterreEarlyDaysContemporary2022}{2022}),
comparing distributions using Hellinger distance. For discrete
probability distributions \(p, q\) over the same space \(X\), the
Hellinger distance is given by

\[ d(p,q) = \frac{1}{\sqrt{2}} \sqrt{\sum_{x \in X} (\sqrt{p(x)} - \sqrt{q(x)})^2} = \frac{1}{\sqrt{2}} \lVert \sqrt p - \sqrt q \rVert_2. \]

The last equation means that the Hellinger distance is the Euclidean
(\(L^2\)-norm) distance between the \emph{square roots} of the
distributions. Some authors working with topic models sometimes compare
distributions using the \(L^2\)-norm of the distributions themselves,
without the square root. But this approach is flawed, since probability
distributions can have different lengths in the \(L^2\) norm. (For
example, the distribution \((1, 0)\) has \(L^2\) length 1, while
\((\frac{1}{2}, \frac{1}{2} )\) has \(L^2\) length approximately 1.19.)
Cosine similarity, which is also widely used by text analysts, is
directly related to the \(L^2\)-norm and has the same problem.

Hellinger distance satisfies the equation
\[ 1 - d^2(p, q) = \sum_{x \in X} \sqrt{p(x)q(x)}. \] When working with
topic models, we're interested in pairwise sets of Hellinger distances,
either between all pairs of distributions from a single set (for
example, the topic distributions for each document, as used in
``discursive space'' analysis;
\protect\hyperlink{ref-HicksProductivityInterdisciplinaryImpacts2021}{Hicks
2021}) or two sets (such comparing fitted vs.~true word-topic
distributions as below; or word-topic distributions for two models
fitted on the same corpus but different vocabularies,
\protect\hyperlink{ref-MalaterreEarlyDaysContemporary2022}{Malaterre and
Lareau 2022}). Working with two sets of distributions
\(P = \{p_i | i \in I\}\) and \(Q = \{q_j | j \in J\}\), the right-hand
side of the last equation is equivalent to a matrix
multiplication.\footnote{For \(P\), each row corresponds to the
  elementwise square root of one distribution \(\sqrt p_i\) and each
  column to one component \(x \in X\), i.e., a cell contains the value
  \(\sqrt{p_i(x)}\). \(Q\) is the transpose, with each row corresponding
  to one component \(x \in X\) and each column corresponding to the
  square root of a distribution \(\sqrt q_j\). The product of these
  matrices is a \(i \times j\) matrix with each cell the desired sum for
  \(p\) and \(q\).} The \texttt{tmfast::hellinger()} function provides
S3 methods for calculating Hellinger pairwise distances given a single
dataframe, single matrix, or two dataframes or matrices.

First, however, we need to extract the word-topic distributions.
\texttt{tmfast} provides a \texttt{tidy()} method, following the pattern
of the topic model tidiers in the \texttt{tidytext} package. Unlike
other topic models, \texttt{tmfast} objects can contain multiple models
for different values of \(k\). So, in the second argument to
\texttt{tidy()}, we need to specify which number of topics we want. The
third argument specifies the desired set of distributions, either
word-topics (\texttt{\textquotesingle{}beta\textquotesingle{}}) or
topic-documents (\texttt{\textquotesingle{}gamma\textquotesingle{}}).

\begin{Shaded}
\begin{Highlighting}[]
\DocumentationTok{\#\# beta: fitted varimax loadings, transformed to probability distributions}
\NormalTok{beta }\OtherTok{=} \FunctionTok{tidy}\NormalTok{(fitted, k, }\StringTok{\textquotesingle{}beta\textquotesingle{}}\NormalTok{)}
\NormalTok{beta}
\end{Highlighting}
\end{Shaded}

\begin{verbatim}
# A tibble: 2,734 x 3
   token topic     beta
   <chr> <chr>    <dbl>
 1 5     V02   0.0198  
 2 5     V08   0.00454 
 3 7     V01   0.00344 
 4 7     V02   0.00276 
 5 7     V06   0.000318
 6 8     V01   0.00146 
 7 8     V02   0.00522 
 8 8     V09   0.0195  
 9 11    V02   0.0114  
10 11    V04   0.00610 
# ... with 2,724 more rows
\end{verbatim}

Word-topic distributions correspond to the varimax factor loadings.
These loadings can take any real value. To convert them to probability
distributions, within each factor (topic), we trim negative values to 0
and divide each loading by the sum of all loadings. The Zipfian plot
below compares the fitted and true word-topic distributions.
Consistently across experiments with this simulation, fitted
distributions started off a little flatter, then dropped sharply after
about 100 words. In other words, the varimax topic model highlights a
relatively long list of characteristic words for each topic --- the
actual distributions have fewer characteristic words --- and then
ignores the other words.

\begin{Shaded}
\begin{Highlighting}[]
\DocumentationTok{\#\# Compare Zipfian distributions}
\FunctionTok{bind\_rows}\NormalTok{(}\FunctionTok{mutate}\NormalTok{(beta, }\AttributeTok{type =} \StringTok{\textquotesingle{}fitted\textquotesingle{}}\NormalTok{), }
\NormalTok{          \{phi\_df }\SpecialCharTok{|\textgreater{}} 
                  \FunctionTok{rename}\NormalTok{(}\AttributeTok{beta =}\NormalTok{ phi) }\SpecialCharTok{|\textgreater{}} 
                  \FunctionTok{mutate}\NormalTok{(}\AttributeTok{type =} \StringTok{\textquotesingle{}true\textquotesingle{}}\NormalTok{)\}) }\SpecialCharTok{|\textgreater{}}
    \FunctionTok{group\_by}\NormalTok{(type, topic) }\SpecialCharTok{|\textgreater{}}
    \FunctionTok{mutate}\NormalTok{(}\AttributeTok{rank =} \FunctionTok{rank}\NormalTok{(}\FunctionTok{desc}\NormalTok{(beta))) }\SpecialCharTok{|\textgreater{}}
    \FunctionTok{arrange}\NormalTok{(type, topic, rank) }\SpecialCharTok{|\textgreater{}}
    \FunctionTok{filter}\NormalTok{(rank }\SpecialCharTok{\textless{}}\NormalTok{ vocab}\SpecialCharTok{/}\DecValTok{2}\NormalTok{) }\SpecialCharTok{|\textgreater{}}
    \FunctionTok{ggplot}\NormalTok{(}\FunctionTok{aes}\NormalTok{(rank, beta, }
               \AttributeTok{color =}\NormalTok{ type, }
               \AttributeTok{group =} \FunctionTok{interaction}\NormalTok{(topic, type))) }\SpecialCharTok{+}
    \FunctionTok{geom\_line}\NormalTok{() }\SpecialCharTok{+}
    \FunctionTok{scale\_y\_log10}\NormalTok{() }\SpecialCharTok{+}
    \FunctionTok{scale\_x\_log10}\NormalTok{()}
\end{Highlighting}
\end{Shaded}

\begin{figure}[H]

{\centering \includegraphics{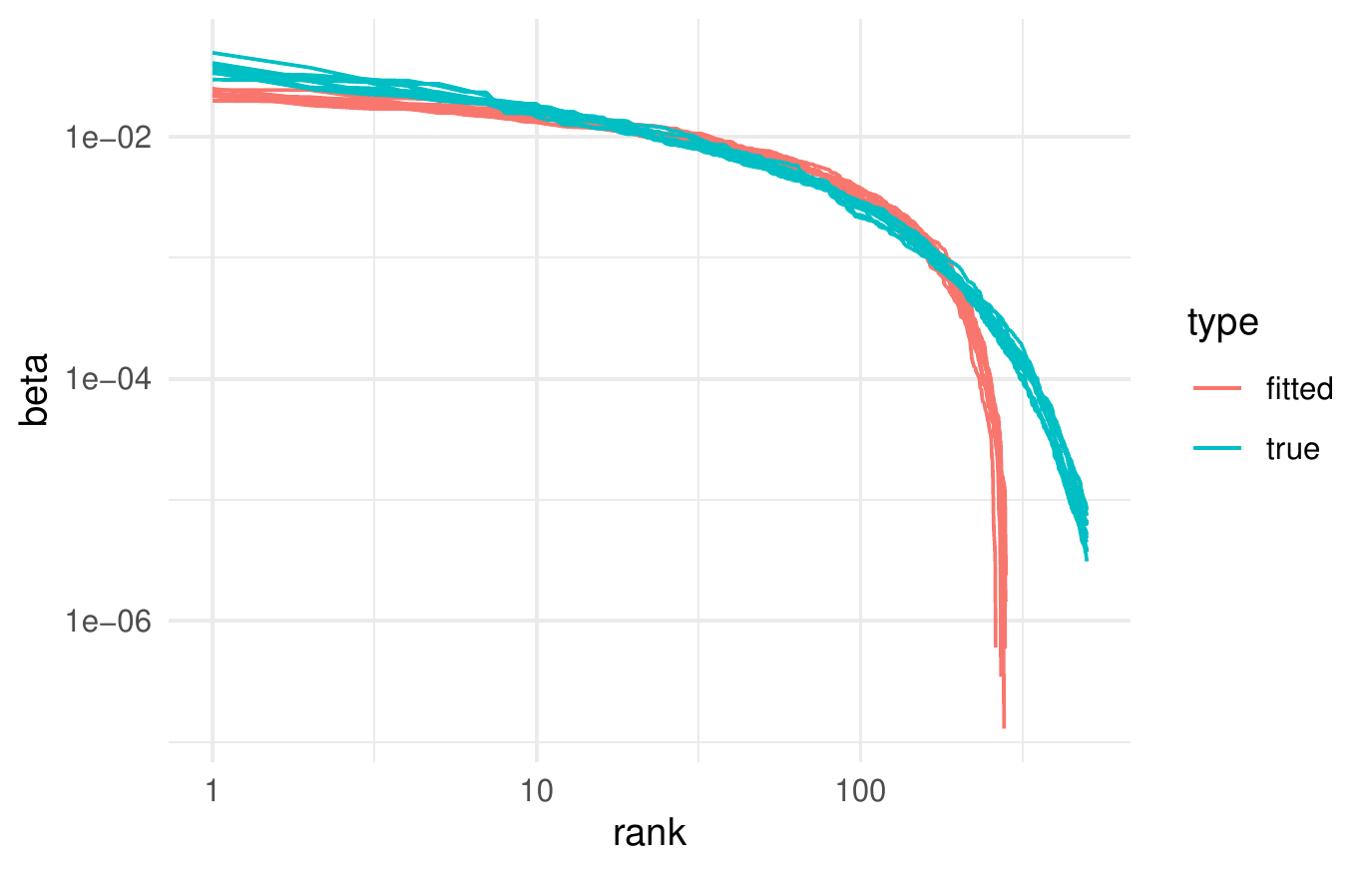}

}

\end{figure}

The Zipfian distribution doesn't tell us which fitted topics might
correspond to which true topics. For that, following Malaterre and
Lareau
(\protect\hyperlink{ref-MalaterreEarlyDaysContemporary2022}{2022}),
we'll use pairwise Hellinger distances. There's one complication,
however. The parameters chosen for this simulation typically end up not
drawing some of the words from the vocabulary, and they don't end up in
the same order as the true word-topic matrix \texttt{phi}. Fortunately
words are represented as the integers \texttt{1:vocab}, so it's
relatively painless to put them back in order and fill in the gaps
(setting the probability for the missing words to be 0 across all
topics). In the code block below, we first fix these issues with the
words, widen the long dataframe, convert it to a matrix, and then
calculate pairwise Hellinger distances with the true word-topic matrix
\texttt{phi}.

\begin{Shaded}
\begin{Highlighting}[]
\DocumentationTok{\#\# Hellinger distance of word{-}topic distributions}
\NormalTok{beta\_mx }\OtherTok{=}\NormalTok{ beta }\SpecialCharTok{|\textgreater{}}
    \DocumentationTok{\#\# Fix order of words}
    \FunctionTok{mutate}\NormalTok{(}\AttributeTok{token =} \FunctionTok{as.integer}\NormalTok{(token)) }\SpecialCharTok{|\textgreater{}}
    \FunctionTok{arrange}\NormalTok{(token) }\SpecialCharTok{|\textgreater{}}
    \DocumentationTok{\#\# And dropped words}
    \FunctionTok{complete}\NormalTok{(}\AttributeTok{token =} \DecValTok{1}\SpecialCharTok{:}\NormalTok{vocab, topic, }\AttributeTok{fill =} \FunctionTok{list}\NormalTok{(}\AttributeTok{beta =} \DecValTok{0}\NormalTok{)) }\SpecialCharTok{|\textgreater{}}
    \FunctionTok{build\_matrix}\NormalTok{(token, topic, beta, }\AttributeTok{sparse =} \ConstantTok{FALSE}\NormalTok{)}

\FunctionTok{hellinger}\NormalTok{(phi, }\FunctionTok{t}\NormalTok{(beta\_mx)) }\SpecialCharTok{|\textgreater{}} 
    \FunctionTok{print}\NormalTok{(}\AttributeTok{digits =} \DecValTok{3}\NormalTok{)}
\end{Highlighting}
\end{Shaded}

\begin{verbatim}
        V01   V02   V03   V04   V05   V06   V07   V08   V09   V10
 [1,] 0.903 0.164 0.917 0.912 0.875 0.900 0.904 0.922 0.885 0.893
 [2,] 0.906 0.878 0.910 0.899 0.157 0.915 0.891 0.904 0.887 0.903
 [3,] 0.911 0.878 0.908 0.877 0.895 0.896 0.886 0.914 0.177 0.922
 [4,] 0.936 0.918 0.167 0.907 0.907 0.878 0.912 0.908 0.911 0.914
 [5,] 0.895 0.905 0.903 0.898 0.887 0.902 0.183 0.880 0.882 0.891
 [6,] 0.923 0.912 0.878 0.893 0.910 0.181 0.901 0.896 0.893 0.906
 [7,] 0.164 0.907 0.938 0.892 0.902 0.931 0.900 0.903 0.911 0.916
 [8,] 0.915 0.888 0.916 0.898 0.900 0.907 0.887 0.880 0.925 0.171
 [9,] 0.896 0.915 0.901 0.168 0.899 0.903 0.898 0.897 0.887 0.900
[10,] 0.911 0.915 0.905 0.894 0.900 0.887 0.884 0.162 0.926 0.882
\end{verbatim}

In this distance matrix, the rows are the true topics and the columns
are the fitted topics. Low values correspond to greater similarity. It's
clear that the topics don't match up perfectly --- the minimum in each
row is about 0.17 --- but there is a clear minimum. We treat this as a
linear assignment problem, which is solved rapidly using the
\texttt{lpSolve} package. The solution --- which matches true to fitted
topics --- can then be used as a rotation with both the loadings and
scores (topic-document distributions). After rotating, the true-fitted
pairs are on the diagonal of the Hellinger distance matrix, making it
easy to extract and summarize the quality of the fit.

\begin{Shaded}
\begin{Highlighting}[]
\DocumentationTok{\#\# Use lpSolve to match fitted topics to true topics}
\NormalTok{dist }\OtherTok{=} \FunctionTok{hellinger}\NormalTok{(phi, }\FunctionTok{t}\NormalTok{(beta\_mx))}
\NormalTok{rotation }\OtherTok{=} \FunctionTok{lp.assign}\NormalTok{(dist)}\SpecialCharTok{$}\NormalTok{solution}
\NormalTok{rotation}
\end{Highlighting}
\end{Shaded}

\begin{verbatim}
      [,1] [,2] [,3] [,4] [,5] [,6] [,7] [,8] [,9] [,10]
 [1,]    0    1    0    0    0    0    0    0    0     0
 [2,]    0    0    0    0    1    0    0    0    0     0
 [3,]    0    0    0    0    0    0    0    0    1     0
 [4,]    0    0    1    0    0    0    0    0    0     0
 [5,]    0    0    0    0    0    0    1    0    0     0
 [6,]    0    0    0    0    0    1    0    0    0     0
 [7,]    1    0    0    0    0    0    0    0    0     0
 [8,]    0    0    0    0    0    0    0    0    0     1
 [9,]    0    0    0    1    0    0    0    0    0     0
[10,]    0    0    0    0    0    0    0    1    0     0
\end{verbatim}

\begin{Shaded}
\begin{Highlighting}[]
\DocumentationTok{\#\# Hellinger distance comparison using the lpSolve matching}
\FunctionTok{hellinger}\NormalTok{(phi, rotation }\SpecialCharTok{\%*\%} \FunctionTok{t}\NormalTok{(beta\_mx)) }\SpecialCharTok{|\textgreater{}} 
    \FunctionTok{print}\NormalTok{(}\AttributeTok{digits =} \DecValTok{3}\NormalTok{)}
\end{Highlighting}
\end{Shaded}

\begin{verbatim}
       [,1]  [,2]  [,3]  [,4]  [,5]  [,6]  [,7]  [,8]  [,9] [,10]
 [1,] 0.164 0.875 0.885 0.917 0.904 0.900 0.903 0.893 0.912 0.922
 [2,] 0.878 0.157 0.887 0.910 0.891 0.915 0.906 0.903 0.899 0.904
 [3,] 0.878 0.895 0.177 0.908 0.886 0.896 0.911 0.922 0.877 0.914
 [4,] 0.918 0.907 0.911 0.167 0.912 0.878 0.936 0.914 0.907 0.908
 [5,] 0.905 0.887 0.882 0.903 0.183 0.902 0.895 0.891 0.898 0.880
 [6,] 0.912 0.910 0.893 0.878 0.901 0.181 0.923 0.906 0.893 0.896
 [7,] 0.907 0.902 0.911 0.938 0.900 0.931 0.164 0.916 0.892 0.903
 [8,] 0.888 0.900 0.925 0.916 0.887 0.907 0.915 0.171 0.898 0.880
 [9,] 0.915 0.899 0.887 0.901 0.898 0.903 0.896 0.900 0.168 0.897
[10,] 0.915 0.900 0.926 0.905 0.884 0.887 0.911 0.882 0.894 0.162
\end{verbatim}

\begin{Shaded}
\begin{Highlighting}[]
\FunctionTok{hellinger}\NormalTok{(phi, rotation }\SpecialCharTok{\%*\%} \FunctionTok{t}\NormalTok{(beta\_mx)) }\SpecialCharTok{|\textgreater{}}
    \FunctionTok{diag}\NormalTok{() }\SpecialCharTok{|\textgreater{}}
    \FunctionTok{summary}\NormalTok{()}
\end{Highlighting}
\end{Shaded}

\begin{verbatim}
   Min. 1st Qu.  Median    Mean 3rd Qu.    Max. 
 0.1569  0.1639  0.1674  0.1693  0.1755  0.1829 
\end{verbatim}

And we do the same thing with the \texttt{stm} topic model.
\textbf{\texttt{stm} is somewhat more accurate than \texttt{tmfast},
with a median Hellinger distance of about 0.07 compared to 0.18. But
\texttt{stm} is significantly slower.}

\begin{Shaded}
\begin{Highlighting}[]
\NormalTok{beta\_stm\_mx }\OtherTok{=} \FunctionTok{tidy}\NormalTok{(fitted\_stm, }\AttributeTok{matrix =} \StringTok{\textquotesingle{}beta\textquotesingle{}}\NormalTok{) }\SpecialCharTok{|\textgreater{}} 
    \DocumentationTok{\#\# Fix order of words}
    \FunctionTok{mutate}\NormalTok{(}\AttributeTok{term =} \FunctionTok{as.integer}\NormalTok{(term)) }\SpecialCharTok{|\textgreater{}}
    \FunctionTok{arrange}\NormalTok{(term) }\SpecialCharTok{|\textgreater{}}
    \DocumentationTok{\#\# And dropped words}
    \FunctionTok{complete}\NormalTok{(}\AttributeTok{term =} \DecValTok{1}\SpecialCharTok{:}\NormalTok{vocab, topic, }
             \AttributeTok{fill =} \FunctionTok{list}\NormalTok{(}\AttributeTok{beta =} \DecValTok{0}\NormalTok{)) }\SpecialCharTok{|\textgreater{}}
    \FunctionTok{build\_matrix}\NormalTok{(term, topic, beta, }\AttributeTok{sparse =} \ConstantTok{FALSE}\NormalTok{)}

\FunctionTok{hellinger}\NormalTok{(phi, }\FunctionTok{t}\NormalTok{(beta\_stm\_mx)) }\SpecialCharTok{|\textgreater{}} 
    \FunctionTok{print}\NormalTok{(}\AttributeTok{digits =} \DecValTok{3}\NormalTok{)}
\end{Highlighting}
\end{Shaded}

\begin{verbatim}
           1     2      3      4      5      6      7      8     9    10
 [1,] 0.0855 0.843 0.8425 0.8821 0.8744 0.8716 0.8505 0.8654 0.879 0.868
 [2,] 0.8433 0.854 0.0823 0.8700 0.8585 0.8696 0.8616 0.8703 0.860 0.854
 [3,] 0.8481 0.085 0.8497 0.8587 0.8349 0.8773 0.8822 0.8583 0.873 0.845
 [4,] 0.8873 0.865 0.8760 0.0872 0.8614 0.9034 0.8730 0.8366 0.867 0.873
 [5,] 0.8651 0.842 0.8420 0.8669 0.8518 0.8549 0.8460 0.8605 0.838 0.079
 [6,] 0.8928 0.855 0.8715 0.8369 0.8477 0.8910 0.8628 0.0888 0.854 0.863
 [7,] 0.8760 0.875 0.8694 0.8950 0.8614 0.0843 0.8747 0.8954 0.867 0.857
 [8,] 0.8618 0.885 0.8710 0.8777 0.8576 0.8821 0.0927 0.8705 0.845 0.853
 [9,] 0.8827 0.840 0.8683 0.8580 0.0904 0.8601 0.8518 0.8531 0.853 0.858
[10,] 0.8822 0.878 0.8630 0.8688 0.8516 0.8734 0.8460 0.8559 0.091 0.846
\end{verbatim}

\begin{Shaded}
\begin{Highlighting}[]
\NormalTok{rotation\_stm }\OtherTok{=} \FunctionTok{hellinger}\NormalTok{(phi, }\FunctionTok{t}\NormalTok{(beta\_stm\_mx)) }\SpecialCharTok{|\textgreater{}} 
    \FunctionTok{lp.assign}\NormalTok{() }\SpecialCharTok{|\textgreater{}} 
\NormalTok{    magrittr}\SpecialCharTok{::}\FunctionTok{extract2}\NormalTok{(}\StringTok{\textquotesingle{}solution\textquotesingle{}}\NormalTok{)}

\FunctionTok{hellinger}\NormalTok{(phi, rotation\_stm }\SpecialCharTok{\%*\%} \FunctionTok{t}\NormalTok{(beta\_stm\_mx)) }\SpecialCharTok{|\textgreater{}}
    \FunctionTok{diag}\NormalTok{() }\SpecialCharTok{|\textgreater{}}
    \FunctionTok{summary}\NormalTok{()}
\end{Highlighting}
\end{Shaded}

\begin{verbatim}
   Min. 1st Qu.  Median    Mean 3rd Qu.    Max. 
0.07902 0.08445 0.08634 0.08661 0.08999 0.09268 
\end{verbatim}

The tidied word-topic distributions can be used in standard ways for
further analysis, such as a
\href{https://juliasilge.com/blog/2018/2018-01-25-sherlock-holmes-stm_files/figure-html/unnamed-chunk-6-1.png}{Silge
plot} of the highest probability words for each topic. But because the
``words'' in this simulation are just integers, and not semantically
meaningful, we don't construct such a plot here.

\hypertarget{renormalization-topic-document-distributions}{%
\subsection{Renormalization: Topic-document
distributions}\label{renormalization-topic-document-distributions}}

Finally, we compare fitted and true topic-document distributions. We
extract topic-document distributions using the same \texttt{tidy()}
function, specifying the matrix \texttt{gamma} and including the
rotation above to align the fitted and true topics. Tile and parallel
coordinates plots can be used to visualize all of the topic-document
distributions. These show that the \texttt{tmfast} models successfully
recover the overall association of each document's journal with a
distinctive topic.

\begin{Shaded}
\begin{Highlighting}[]
\NormalTok{gamma\_df }\OtherTok{=} \FunctionTok{tidy}\NormalTok{(fitted, k, }\StringTok{\textquotesingle{}gamma\textquotesingle{}}\NormalTok{, }
                \AttributeTok{rotation =}\NormalTok{ rotation) }\SpecialCharTok{|\textgreater{}} 
    \FunctionTok{mutate}\NormalTok{(}\AttributeTok{document =} \FunctionTok{as.integer}\NormalTok{(document),}
           \AttributeTok{journal =}\NormalTok{ (document }\SpecialCharTok{{-}} \DecValTok{1}\NormalTok{) }\SpecialCharTok{\%/\%}\NormalTok{ Mj }\SpecialCharTok{+} \DecValTok{1}\NormalTok{)}
\end{Highlighting}
\end{Shaded}

\begin{verbatim}
Warning in tidy.tmfast(fitted, k, "gamma", rotation = rotation): Rotating
scores
\end{verbatim}

\begin{Shaded}
\begin{Highlighting}[]
\FunctionTok{ggplot}\NormalTok{(gamma\_df, }\FunctionTok{aes}\NormalTok{(document, topic, }\AttributeTok{fill =}\NormalTok{ gamma)) }\SpecialCharTok{+}
    \FunctionTok{geom\_raster}\NormalTok{() }\SpecialCharTok{+}
    \FunctionTok{scale\_x\_continuous}\NormalTok{(}\AttributeTok{breaks =} \ConstantTok{NULL}\NormalTok{)}
\end{Highlighting}
\end{Shaded}

\begin{figure}[H]

{\centering \includegraphics{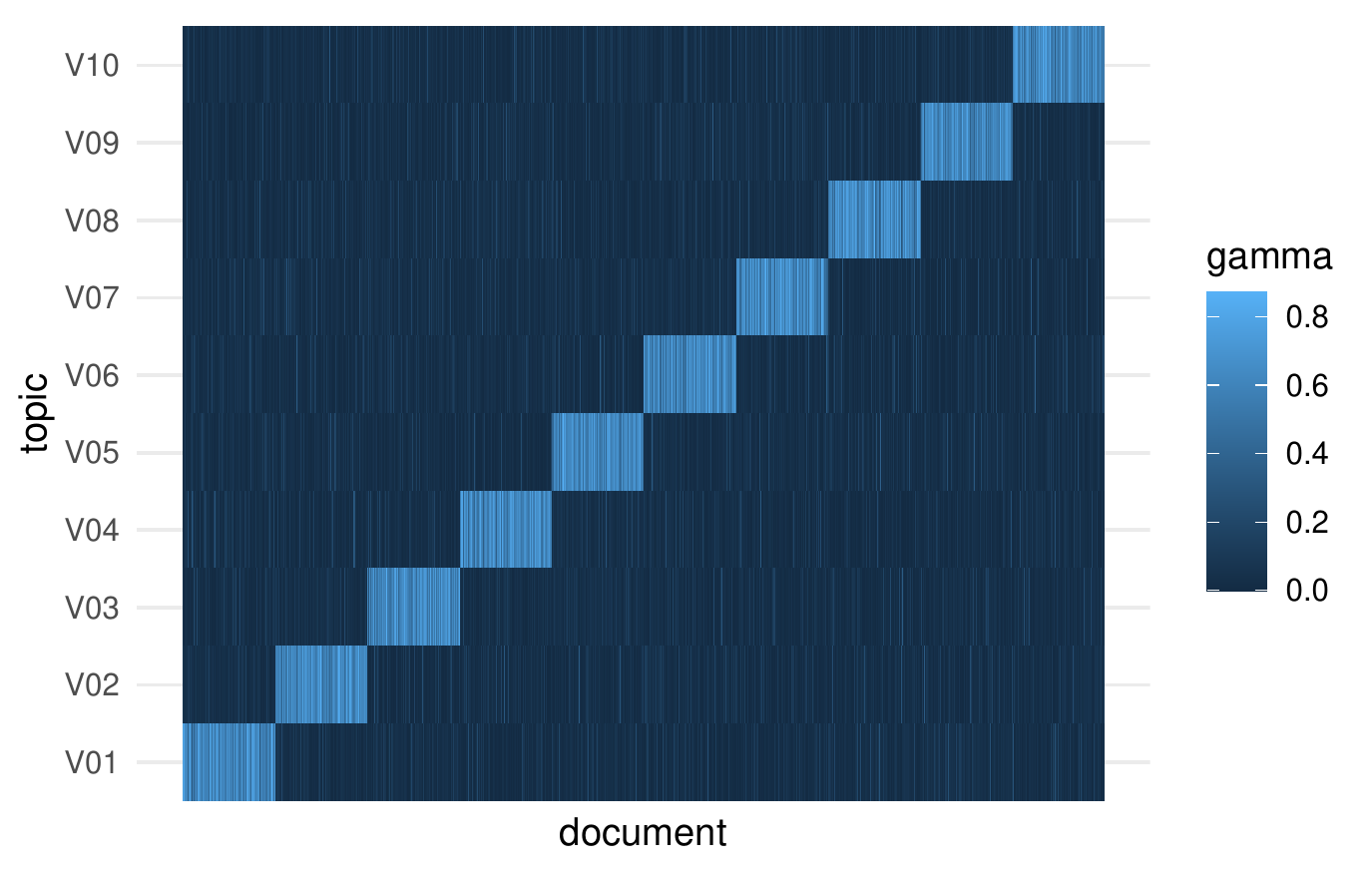}

}

\end{figure}

\begin{Shaded}
\begin{Highlighting}[]
\FunctionTok{ggplot}\NormalTok{(gamma\_df, }
       \FunctionTok{aes}\NormalTok{(topic, gamma, }
           \AttributeTok{group =}\NormalTok{ document, }
           \AttributeTok{color =} \FunctionTok{as.factor}\NormalTok{(journal))) }\SpecialCharTok{+}
    \FunctionTok{geom\_line}\NormalTok{(}\AttributeTok{alpha =}\NormalTok{ .}\DecValTok{25}\NormalTok{) }\SpecialCharTok{+}
    \FunctionTok{facet\_wrap}\NormalTok{(}\FunctionTok{vars}\NormalTok{(journal)) }\SpecialCharTok{+}
    \FunctionTok{scale\_color\_discrete}\NormalTok{(}\AttributeTok{guide =} \StringTok{\textquotesingle{}none\textquotesingle{}}\NormalTok{) }\SpecialCharTok{+}
    \FunctionTok{scale\_x\_discrete}\NormalTok{(}\AttributeTok{guide =} \StringTok{\textquotesingle{}none\textquotesingle{}}\NormalTok{)}
\end{Highlighting}
\end{Shaded}

\begin{figure}[H]

{\centering \includegraphics{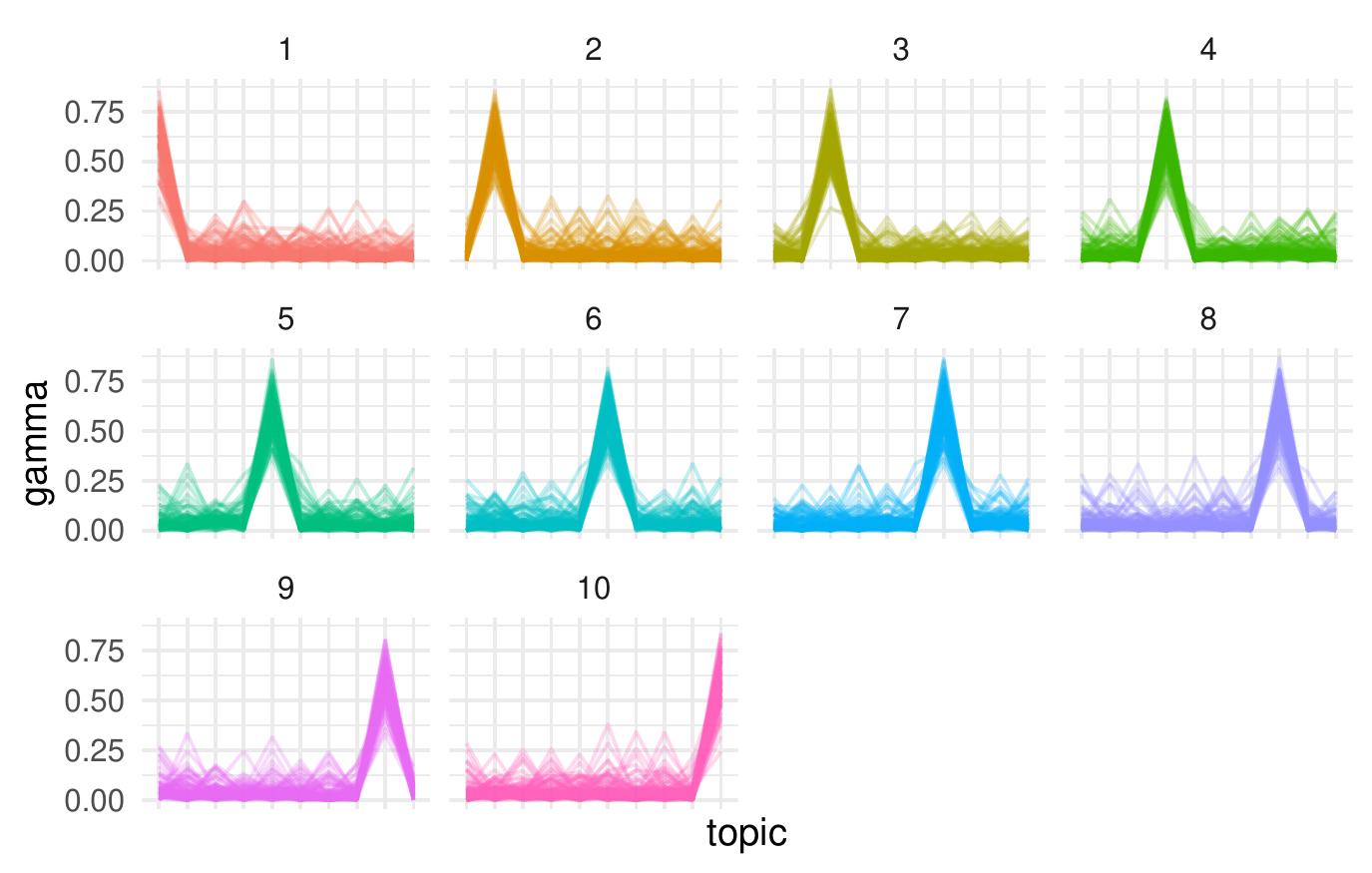}

}

\end{figure}

However, the fitted topic-document distributions are flatter than the
true ones. Consider the true and fitted distributions for document 1.
Compared to the true distribution, the fitted distribution has a
somewhat lower probability for topic \texttt{V01} and a somewhat higher
probability for the other topics.

\begin{Shaded}
\begin{Highlighting}[]
\FunctionTok{ggplot}\NormalTok{(}\AttributeTok{mapping =} \FunctionTok{aes}\NormalTok{(topic, }\AttributeTok{group =}\NormalTok{ 1L)) }\SpecialCharTok{+}
    \FunctionTok{geom\_line}\NormalTok{(}\AttributeTok{mapping =} \FunctionTok{aes}\NormalTok{(}\AttributeTok{y =}\NormalTok{ theta, }\AttributeTok{color =} \StringTok{\textquotesingle{}true\textquotesingle{}}\NormalTok{), }
              \AttributeTok{data =} \FunctionTok{filter}\NormalTok{(theta\_df, doc }\SpecialCharTok{==} \StringTok{\textquotesingle{}1\textquotesingle{}}\NormalTok{)) }\SpecialCharTok{+}
    \FunctionTok{geom\_line}\NormalTok{(}\AttributeTok{mapping =} \FunctionTok{aes}\NormalTok{(}\AttributeTok{y =}\NormalTok{ gamma, }\AttributeTok{color =} \StringTok{\textquotesingle{}fitted\textquotesingle{}}\NormalTok{), }
              \AttributeTok{data =} \FunctionTok{filter}\NormalTok{(gamma\_df, document }\SpecialCharTok{==} \StringTok{\textquotesingle{}1\textquotesingle{}}\NormalTok{))}
\end{Highlighting}
\end{Shaded}

\begin{figure}[H]

{\centering \includegraphics{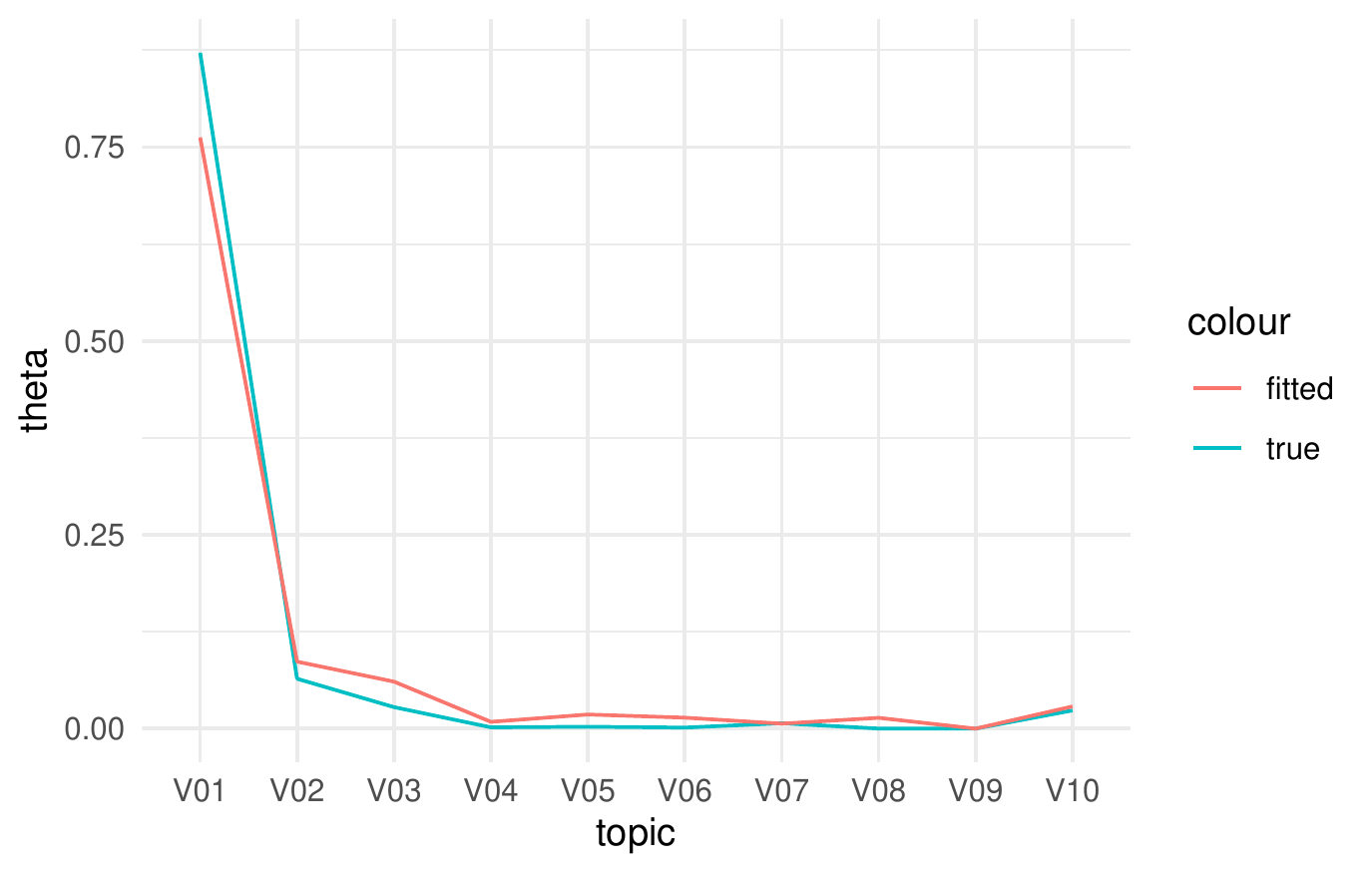}

}

\end{figure}

This flatter distribution corresponds to greater entropy. In this
simulation, the entropy of the fitted distributions are about 1 bit
greater than those of the true distributions. This discrepancy tends to
become worse with greater values of \(k\).

\begin{Shaded}
\begin{Highlighting}[]
\NormalTok{theta\_df }\SpecialCharTok{|\textgreater{}} 
    \FunctionTok{group\_by}\NormalTok{(doc) }\SpecialCharTok{|\textgreater{}} 
    \FunctionTok{summarize}\NormalTok{(}\AttributeTok{H =} \FunctionTok{entropy}\NormalTok{(theta)) }\SpecialCharTok{|\textgreater{}} 
    \FunctionTok{pull}\NormalTok{(H) }\SpecialCharTok{|\textgreater{}} 
    \FunctionTok{summary}\NormalTok{()}
\end{Highlighting}
\end{Shaded}

\begin{verbatim}
   Min. 1st Qu.  Median    Mean 3rd Qu.    Max. 
 0.1006  0.6614  0.9715  1.0100  1.3311  2.5821 
\end{verbatim}

\begin{Shaded}
\begin{Highlighting}[]
\NormalTok{gamma\_df }\SpecialCharTok{|\textgreater{}} 
    \FunctionTok{group\_by}\NormalTok{(document) }\SpecialCharTok{|\textgreater{}} 
    \FunctionTok{summarize}\NormalTok{(}\AttributeTok{H =} \FunctionTok{entropy}\NormalTok{(gamma)) }\SpecialCharTok{|\textgreater{}} 
    \FunctionTok{pull}\NormalTok{(H) }\SpecialCharTok{|\textgreater{}} 
    \FunctionTok{summary}\NormalTok{()}
\end{Highlighting}
\end{Shaded}

\begin{verbatim}
   Min. 1st Qu.  Median    Mean 3rd Qu.    Max. 
 0.8486  1.7040  1.9319  1.9129  2.1592  2.7673 
\end{verbatim}

To mitigate this problem, we add an optional renormalization step when
converting document scores to topic-document distributions. Given a
discrete probability distribution \(P\) with components \(p_i\) and
entropy \(H\), and a parameter \(\beta\), we can define a new
distribution \(P'\) with components

\[ p'_i = \frac{p_i^\beta}{\sum_i p_i^\beta} = \frac{p_i^\beta}{Z}\]

which has entropy

\[ H' = \frac{1}{Z} \sum_i [p_i^\beta \beta \log p_i] - \log Z.\]

That is, we can choose a parameter \(\beta\) that renormalizes \(P\) to
achieve a target entropy \(H'\). In LDA, the target entropy is the
expected entropy for topic-document distributions drawn from the
asymmetric Dirichlet prior. \texttt{tmfast} provides convenience
functions for calculating this expected entropy; compare this to the
mean entropy of the distributions in \texttt{theta} above. \textbf{In
actual applications, where the Dirichlet prior is an idealization,
choosing \(\alpha\) to set the target entropy is an important researcher
degree of freedom.} It is equivalent to choosing prior parameters in
other topic modeling packages.

\begin{Shaded}
\begin{Highlighting}[]
\FunctionTok{expected\_entropy}\NormalTok{(}\FunctionTok{peak\_alpha}\NormalTok{(k, }\DecValTok{1}\NormalTok{, topic\_peak, topic\_scale))}
\end{Highlighting}
\end{Shaded}

\begin{verbatim}
[1] 0.997604
\end{verbatim}

Since solving the equation for \(H'\) for \(\beta\) requires numerical
optimization, it's inefficient to do this every time we call
\texttt{tidy()}, especially with large corpora. Instead,
\texttt{tmfast::target\_power()} is used to run this optimization once,
and then return the mean value across all documents. We then use this
single value of \(\beta\) in all future calls to \texttt{tidy()}.

\begin{Shaded}
\begin{Highlighting}[]
\NormalTok{gamma\_power }\OtherTok{=} \FunctionTok{tidy}\NormalTok{(fitted, k, }\StringTok{\textquotesingle{}gamma\textquotesingle{}}\NormalTok{) }\SpecialCharTok{|\textgreater{}} 
    \FunctionTok{target\_power}\NormalTok{(document, gamma, }
                 \FunctionTok{expected\_entropy}\NormalTok{(}\FunctionTok{peak\_alpha}\NormalTok{(k, }
                                             \DecValTok{1}\NormalTok{, }
\NormalTok{                                             topic\_peak, }
\NormalTok{                                             topic\_scale)))}
\NormalTok{gamma\_power}
\end{Highlighting}
\end{Shaded}

\begin{verbatim}
[1] 1.539377
\end{verbatim}

The renormalized topic-document distributions have closer entropy to
\(\theta\). The \texttt{keep\_original} argument lets us compare the
original and renormalized distributions.

\begin{Shaded}
\begin{Highlighting}[]
\NormalTok{gamma\_df }\OtherTok{=} \FunctionTok{tidy}\NormalTok{(fitted, k, }\StringTok{\textquotesingle{}gamma\textquotesingle{}}\NormalTok{, }
                \AttributeTok{rotation =}\NormalTok{ rotation, }
                \AttributeTok{exponent =}\NormalTok{ gamma\_power, }
                \AttributeTok{keep\_original =} \ConstantTok{TRUE}\NormalTok{) }\SpecialCharTok{|\textgreater{}} 
    \FunctionTok{mutate}\NormalTok{(}\AttributeTok{document =} \FunctionTok{as.integer}\NormalTok{(document),}
           \AttributeTok{journal =}\NormalTok{ (document }\SpecialCharTok{{-}} \DecValTok{1}\NormalTok{) }\SpecialCharTok{\%/\%}\NormalTok{ Mj }\SpecialCharTok{+} \DecValTok{1}\NormalTok{)}
\end{Highlighting}
\end{Shaded}

\begin{verbatim}
Warning in tidy.tmfast(fitted, k, "gamma", rotation = rotation, exponent =
gamma_power, : Rotating scores
\end{verbatim}

\begin{Shaded}
\begin{Highlighting}[]
\NormalTok{gamma\_df }\SpecialCharTok{|\textgreater{}} 
    \FunctionTok{group\_by}\NormalTok{(document) }\SpecialCharTok{|\textgreater{}} 
    \FunctionTok{summarize}\NormalTok{(}\FunctionTok{across}\NormalTok{(}\FunctionTok{c}\NormalTok{(gamma, gamma\_rn), entropy)) }\SpecialCharTok{|\textgreater{}} 
    \FunctionTok{summarize}\NormalTok{(}\FunctionTok{across}\NormalTok{(}\FunctionTok{c}\NormalTok{(gamma, gamma\_rn), mean))}
\end{Highlighting}
\end{Shaded}

\begin{verbatim}
# A tibble: 1 x 2
  gamma gamma_rn
  <dbl>    <dbl>
1  1.91     1.03
\end{verbatim}

We can now assess accuracy of the topic-document distributions. Above we
used the \texttt{hellinger()} method for two matrices. The method for
two dataframes requires specifying the id, topic, and probability
columns. The tile plot shows that the true and fitted topics are aligned
(because we used the rotation when extracting \texttt{gamma\_df} above),
and so again we can get an overall summary from the diagonal. Without
renormalization, in the current simulation the mean Hellinger distance
is 0.24 --- not too bad, but perhaps larger than one would like. With
larger values of \(k\), this accuracy increases significantly.
Renormalization keeps the mean distance around 0.13, comparable to the
word-topic distributions.

\begin{Shaded}
\begin{Highlighting}[]
\DocumentationTok{\#\# w/o renormalization, mean distance is .24}
\FunctionTok{hellinger}\NormalTok{(theta\_df, doc, }\AttributeTok{prob1 =}\NormalTok{ theta,}
          \AttributeTok{topicsdf2 =}\NormalTok{ gamma\_df, }
          \AttributeTok{id2 =}\NormalTok{ document, }
          \AttributeTok{prob2 =}\NormalTok{ gamma, }\AttributeTok{df =} \ConstantTok{FALSE}\NormalTok{) }\SpecialCharTok{|\textgreater{}} 
    \FunctionTok{diag}\NormalTok{() }\SpecialCharTok{|\textgreater{}} 
    \FunctionTok{summary}\NormalTok{()}
\end{Highlighting}
\end{Shaded}

\begin{verbatim}
   Min. 1st Qu.  Median    Mean 3rd Qu.    Max. 
0.08499 0.20131 0.23733 0.23770 0.27244 0.37585 
\end{verbatim}

\begin{Shaded}
\begin{Highlighting}[]
\DocumentationTok{\#\# w/ renormalization, mean distance drops to .13}
\NormalTok{doc\_compare }\OtherTok{=} \FunctionTok{hellinger}\NormalTok{(theta\_df, doc, }\AttributeTok{prob1 =}\NormalTok{ theta,}
                        \AttributeTok{topicsdf2 =}\NormalTok{ gamma\_df, }
                        \AttributeTok{id2 =}\NormalTok{ document, }
                        \AttributeTok{prob2 =}\NormalTok{ gamma\_rn, }\AttributeTok{df =} \ConstantTok{TRUE}\NormalTok{)}

\NormalTok{doc\_compare }\SpecialCharTok{|\textgreater{}} 
    \FunctionTok{filter}\NormalTok{(doc }\SpecialCharTok{==}\NormalTok{ document) }\SpecialCharTok{|\textgreater{}} 
    \FunctionTok{pull}\NormalTok{(dist) }\SpecialCharTok{|\textgreater{}} 
    \FunctionTok{summary}\NormalTok{()}
\end{Highlighting}
\end{Shaded}

\begin{verbatim}
   Min. 1st Qu.  Median    Mean 3rd Qu.    Max. 
0.04808 0.10195 0.12378 0.12518 0.14564 0.24868 
\end{verbatim}

\begin{Shaded}
\begin{Highlighting}[]
\FunctionTok{ggplot}\NormalTok{(doc\_compare, }
       \FunctionTok{aes}\NormalTok{(}\FunctionTok{as.integer}\NormalTok{(doc), }
           \FunctionTok{as.integer}\NormalTok{(document), }
           \AttributeTok{fill =} \DecValTok{1} \SpecialCharTok{{-}}\NormalTok{ dist)) }\SpecialCharTok{+}
    \FunctionTok{geom\_raster}\NormalTok{() }\SpecialCharTok{+}
    \FunctionTok{scale\_x\_discrete}\NormalTok{(}\AttributeTok{breaks =} \ConstantTok{NULL}\NormalTok{, }\AttributeTok{name =} \StringTok{\textquotesingle{}true\textquotesingle{}}\NormalTok{) }\SpecialCharTok{+}
    \FunctionTok{scale\_y\_discrete}\NormalTok{(}\AttributeTok{breaks =} \ConstantTok{NULL}\NormalTok{, }\AttributeTok{name =} \StringTok{\textquotesingle{}fitted\textquotesingle{}}\NormalTok{)}
\end{Highlighting}
\end{Shaded}

\begin{figure}[H]

{\centering \includegraphics{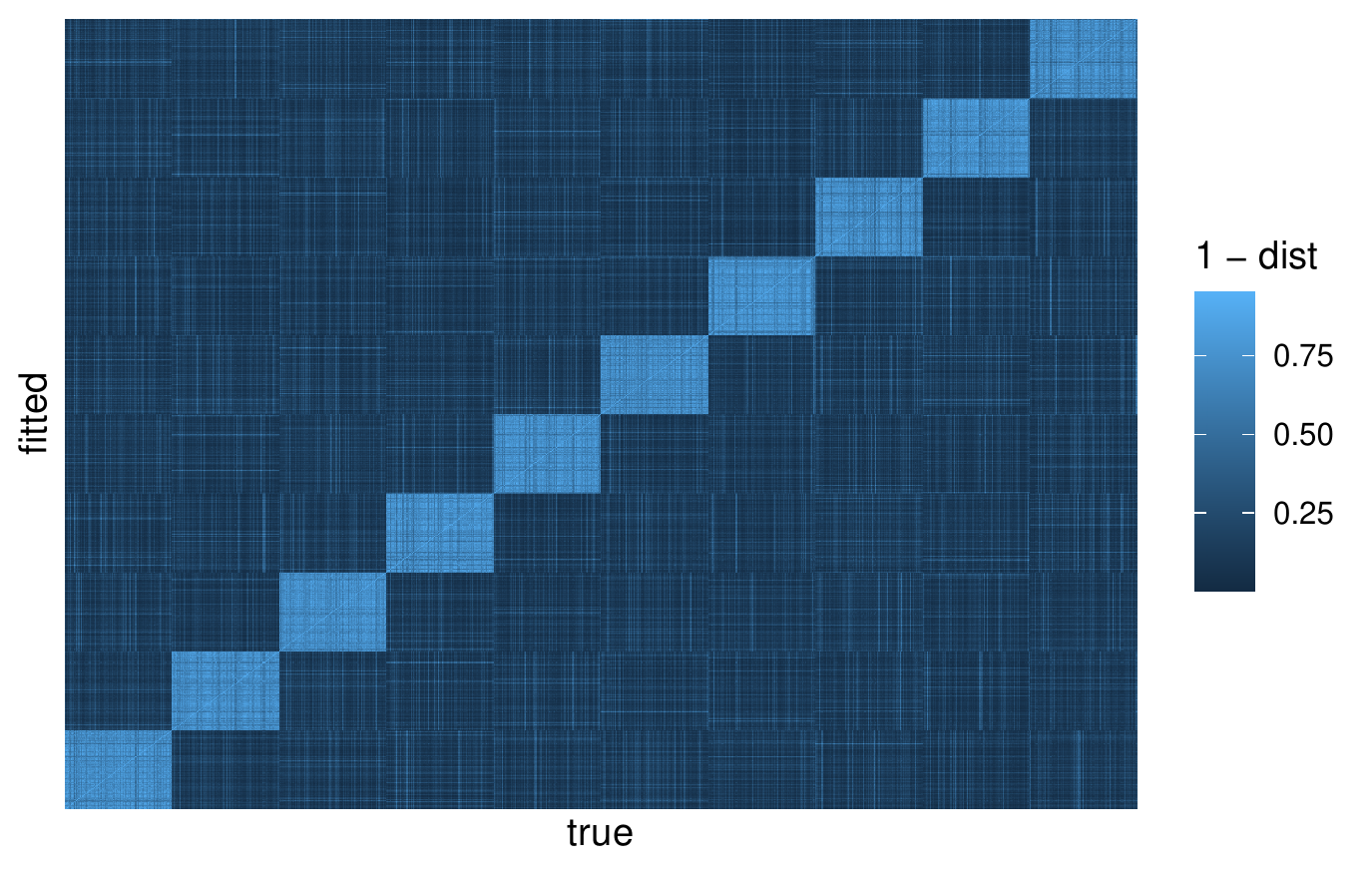}

}

\end{figure}

STM has a slightly closer fit, with a mean Hellinger distance of 0.08.

\begin{Shaded}
\begin{Highlighting}[]
\NormalTok{fitted\_stm\_gamma }\OtherTok{=} \FunctionTok{tidy}\NormalTok{(fitted\_stm, }\AttributeTok{matrix =} \StringTok{\textquotesingle{}gamma\textquotesingle{}}\NormalTok{) }\SpecialCharTok{|\textgreater{}} 
    \FunctionTok{build\_matrix}\NormalTok{(document, topic, gamma, }\AttributeTok{sparse =} \ConstantTok{FALSE}\NormalTok{)}

\FunctionTok{hellinger}\NormalTok{(theta, fitted\_stm\_gamma }\SpecialCharTok{\%*\%} \FunctionTok{t}\NormalTok{(rotation\_stm)) }\SpecialCharTok{|\textgreater{}} 
    \FunctionTok{diag}\NormalTok{() }\SpecialCharTok{|\textgreater{}} 
    \FunctionTok{summary}\NormalTok{()}
\end{Highlighting}
\end{Shaded}

\begin{verbatim}
   Min. 1st Qu.  Median    Mean 3rd Qu.    Max. 
0.03216 0.07148 0.08638 0.08823 0.10260 0.19884 
\end{verbatim}

\hypertarget{sec-realbooks}{%
\section{Example 2: Literature from the long nineteenth
century}\label{sec-realbooks}}

Our second example will analyze a literary corpus from the long
nineteenth century, attempting to recover the author of each document.
In order to reduce the number of requests sent to Project Gutenberg, we
construct a convenience function to identify and retrieve fulltext works
given an author's Gutenberg ID, and wrap this in
\texttt{memoise::memoise()} to create a local cache.

\begin{Shaded}
\begin{Highlighting}[]
\FunctionTok{library}\NormalTok{(tidyverse)            }\CommentTok{\# infrastructure}
\FunctionTok{theme\_set}\NormalTok{(}\FunctionTok{theme\_minimal}\NormalTok{())}
\FunctionTok{library}\NormalTok{(ggbeeswarm)}
\FunctionTok{library}\NormalTok{(memoise)}
\FunctionTok{library}\NormalTok{(tictoc)}
\FunctionTok{library}\NormalTok{(glue)}

\FunctionTok{library}\NormalTok{(gutenbergr)           }\CommentTok{\# text retrieval and manipulation}
\FunctionTok{library}\NormalTok{(tidytext)}
\FunctionTok{library}\NormalTok{(tmfast)               }\CommentTok{\# topic modeling}
\FunctionTok{library}\NormalTok{(stm)                  }\CommentTok{\# topic modeling}

\NormalTok{get\_author }\OtherTok{=} \ControlFlowTok{function}\NormalTok{(author\_id) \{}
    \FunctionTok{gutenberg\_works}\NormalTok{(gutenberg\_author\_id }\SpecialCharTok{==}\NormalTok{ author\_id, }
\NormalTok{                    has\_text) }\SpecialCharTok{|\textgreater{}} 
        \FunctionTok{gutenberg\_download}\NormalTok{(}\AttributeTok{meta\_fields =} \FunctionTok{c}\NormalTok{(}\StringTok{\textquotesingle{}author\textquotesingle{}}\NormalTok{, }\StringTok{\textquotesingle{}title\textquotesingle{}}\NormalTok{), }
                           \AttributeTok{mirror =} \StringTok{\textquotesingle{}http://aleph.gutenberg.org\textquotesingle{}}\NormalTok{)}
\NormalTok{\}}
\NormalTok{get\_author }\OtherTok{=} \FunctionTok{memoise}\NormalTok{(get\_author, }
                     \AttributeTok{cache =} \FunctionTok{cache\_filesystem}\NormalTok{(}\StringTok{\textquotesingle{}realbooks\textquotesingle{}}\NormalTok{))}
\end{Highlighting}
\end{Shaded}

\hypertarget{corpus-assembly}{%
\subsection{Corpus assembly}\label{corpus-assembly}}

We first retrieve all works in Project Gutenberg by our target authors:
Jane Austen, Charlotte and Emily Brontë, Louisa May Alcott, George
Eliot, Charles Dickens, HG Wells, and HP Lovecraft. For these authors,
the \texttt{memoise} local cache ends up at about 286 MB.

\begin{Shaded}
\begin{Highlighting}[]
\DocumentationTok{\#\# Jane Austen is author 68}
\CommentTok{\# gutenberg\_authors |\textgreater{} }
\CommentTok{\#     filter(str\_detect(author, \textquotesingle{}Austen\textquotesingle{}))}
\NormalTok{austen\_df }\OtherTok{=} \FunctionTok{get\_author}\NormalTok{(}\DecValTok{68}\NormalTok{)}
\end{Highlighting}
\end{Shaded}

\begin{Shaded}
\begin{Highlighting}[]
\DocumentationTok{\#\# Anne Brontë is 404}
\CommentTok{\# filter(gutenberg\_authors, str\_detect(author, \textquotesingle{}Brontë\textquotesingle{}))}
\NormalTok{a\_bronte\_df }\OtherTok{=} \FunctionTok{get\_author}\NormalTok{(}\DecValTok{404}\NormalTok{)}
\end{Highlighting}
\end{Shaded}

\begin{Shaded}
\begin{Highlighting}[]
\DocumentationTok{\#\# Charlotte Brontë is 408}
\CommentTok{\# filter(gutenberg\_authors, str\_detect(author, \textquotesingle{}Brontë\textquotesingle{}))}
\NormalTok{c\_bronte\_df }\OtherTok{=} \FunctionTok{get\_author}\NormalTok{(}\DecValTok{408}\NormalTok{)}
\end{Highlighting}
\end{Shaded}

\begin{Shaded}
\begin{Highlighting}[]
\DocumentationTok{\#\# Emily Brontë is 405}
\CommentTok{\# filter(gutenberg\_authors, str\_detect(author, \textquotesingle{}Brontë\textquotesingle{}))}
\NormalTok{e\_bronte\_df }\OtherTok{=} \FunctionTok{get\_author}\NormalTok{(}\DecValTok{405}\NormalTok{)}
\end{Highlighting}
\end{Shaded}

\begin{Shaded}
\begin{Highlighting}[]
\DocumentationTok{\#\# Louisa May Alcott is 102}
\CommentTok{\# filter(gutenberg\_authors, str\_detect(author, \textquotesingle{}Alcott\textquotesingle{}))}
\NormalTok{alcott\_df }\OtherTok{=} \FunctionTok{get\_author}\NormalTok{(}\DecValTok{102}\NormalTok{)}
\end{Highlighting}
\end{Shaded}

\begin{Shaded}
\begin{Highlighting}[]
\DocumentationTok{\#\# George Eliot is 90}
\CommentTok{\# filter(gutenberg\_authors, str\_detect(author, \textquotesingle{}Eliot\textquotesingle{}))}
\NormalTok{eliot\_df }\OtherTok{=} \FunctionTok{get\_author}\NormalTok{(}\DecValTok{90}\NormalTok{)}
\end{Highlighting}
\end{Shaded}

\begin{Shaded}
\begin{Highlighting}[]
\DocumentationTok{\#\# Mary Wollstonecraft Shelley is 61}
\CommentTok{\# filter(gutenberg\_authors, str\_detect(author, \textquotesingle{}Shelley\textquotesingle{}))}
\NormalTok{shelley\_df }\OtherTok{=} \FunctionTok{get\_author}\NormalTok{(}\DecValTok{61}\NormalTok{)}
\end{Highlighting}
\end{Shaded}

\begin{Shaded}
\begin{Highlighting}[]
\DocumentationTok{\#\# Charles Dickens is 37}
\CommentTok{\# filter(gutenberg\_authors, str\_detect(author, \textquotesingle{}Dickens\textquotesingle{}))}
\NormalTok{dickens\_df }\OtherTok{=} \FunctionTok{get\_author}\NormalTok{(}\DecValTok{37}\NormalTok{)}
\end{Highlighting}
\end{Shaded}

\begin{Shaded}
\begin{Highlighting}[]
\DocumentationTok{\#\# HG Wells is 30}
\CommentTok{\# filter(gutenberg\_authors, str\_detect(author, \textquotesingle{}Wells\textquotesingle{}))}
\NormalTok{wells\_df }\OtherTok{=} \FunctionTok{get\_author}\NormalTok{(}\DecValTok{30}\NormalTok{)}
\end{Highlighting}
\end{Shaded}

\begin{Shaded}
\begin{Highlighting}[]
\DocumentationTok{\#\# HP Lovecraft is 34724}
\CommentTok{\# filter(gutenberg\_authors, str\_detect(author, \textquotesingle{}Lovecraft\textquotesingle{}))}
\NormalTok{lovecraft\_df }\OtherTok{=} \FunctionTok{get\_author}\NormalTok{(}\DecValTok{34724}\NormalTok{)}
\end{Highlighting}
\end{Shaded}

We combine these results, and use \texttt{tidytext::unnest\_tokens()} to
convert the result into a long-format document-term matrix. Note that
token extraction can take a long moment. We also construct a dataframe
to link titles to authors in the topic model output.

\begin{Shaded}
\begin{Highlighting}[]
\NormalTok{dataf }\OtherTok{=} \FunctionTok{bind\_rows}\NormalTok{(austen\_df, }
\NormalTok{                  a\_bronte\_df, }
\NormalTok{                  c\_bronte\_df, }
\NormalTok{                  e\_bronte\_df, }
\NormalTok{                  alcott\_df, }
\NormalTok{                  eliot\_df,}
\NormalTok{                  shelley\_df,}
\NormalTok{                  dickens\_df, }
\NormalTok{                  wells\_df, }
\NormalTok{                  lovecraft\_df) }\SpecialCharTok{|\textgreater{}} 
    \FunctionTok{unnest\_tokens}\NormalTok{(term, text, }\AttributeTok{token =} \StringTok{\textquotesingle{}words\textquotesingle{}}\NormalTok{) }\SpecialCharTok{|\textgreater{}} 
    \FunctionTok{count}\NormalTok{(gutenberg\_id, author, title, term)}

\NormalTok{dataf}
\end{Highlighting}
\end{Shaded}

\begin{verbatim}
# A tibble: 1,812,904 x 5
   gutenberg_id author                        title            term            n
          <int> <chr>                         <chr>            <chr>       <int>
 1           35 Wells, H. G. (Herbert George) The Time Machine _can_           1
 2           35 Wells, H. G. (Herbert George) The Time Machine _cancan_        1
 3           35 Wells, H. G. (Herbert George) The Time Machine _down_          1
 4           35 Wells, H. G. (Herbert George) The Time Machine _four_          1
 5           35 Wells, H. G. (Herbert George) The Time Machine _him_           1
 6           35 Wells, H. G. (Herbert George) The Time Machine _how_           1
 7           35 Wells, H. G. (Herbert George) The Time Machine _i_             1
 8           35 Wells, H. G. (Herbert George) The Time Machine _instantan~     1
 9           35 Wells, H. G. (Herbert George) The Time Machine _minus_         1
10           35 Wells, H. G. (Herbert George) The Time Machine _nil_           1
# ... with 1,812,894 more rows
\end{verbatim}

\begin{Shaded}
\begin{Highlighting}[]
\NormalTok{meta\_df }\OtherTok{=} \FunctionTok{distinct}\NormalTok{(dataf, author, title)}
\end{Highlighting}
\end{Shaded}

The number of works by each author varies widely, as does the total
token count.

\begin{Shaded}
\begin{Highlighting}[]
\FunctionTok{distinct}\NormalTok{(dataf, author, title) }\SpecialCharTok{|\textgreater{}} 
    \FunctionTok{count}\NormalTok{(author)}
\end{Highlighting}
\end{Shaded}

\begin{verbatim}
# A tibble: 10 x 2
   author                                 n
   <chr>                              <int>
 1 Alcott, Louisa May                    45
 2 Austen, Jane                          10
 3 Brontë, Anne                           2
 4 Brontë, Charlotte                      6
 5 Brontë, Emily                          1
 6 Dickens, Charles                      77
 7 Eliot, George                         18
 8 Lovecraft, H. P. (Howard Phillips)     7
 9 Shelley, Mary Wollstonecraft          17
10 Wells, H. G. (Herbert George)         70
\end{verbatim}

\begin{Shaded}
\begin{Highlighting}[]
\FunctionTok{with}\NormalTok{(dataf, }\FunctionTok{n\_distinct}\NormalTok{(author, title))}
\end{Highlighting}
\end{Shaded}

\begin{verbatim}
[1] 253
\end{verbatim}

\begin{Shaded}
\begin{Highlighting}[]
\NormalTok{dataf }\SpecialCharTok{|\textgreater{}} 
    \FunctionTok{group\_by}\NormalTok{(author, title) }\SpecialCharTok{|\textgreater{}} 
    \FunctionTok{summarize}\NormalTok{(}\AttributeTok{n =} \FunctionTok{sum}\NormalTok{(n)) }\SpecialCharTok{|\textgreater{}} 
    \FunctionTok{summarize}\NormalTok{(}\AttributeTok{min =} \FunctionTok{min}\NormalTok{(n), }
              \AttributeTok{median =} \FunctionTok{median}\NormalTok{(n), }
              \AttributeTok{max =} \FunctionTok{max}\NormalTok{(n), }
              \AttributeTok{total =} \FunctionTok{sum}\NormalTok{(n)) }\SpecialCharTok{|\textgreater{}} 
    \FunctionTok{arrange}\NormalTok{(}\FunctionTok{desc}\NormalTok{(total))}
\end{Highlighting}
\end{Shaded}

\begin{verbatim}
`summarise()` has grouped output by 'author'. You can override using the
`.groups` argument.
\end{verbatim}

\begin{verbatim}
# A tibble: 10 x 5
   author                                min  median    max   total
   <chr>                               <int>   <dbl>  <int>   <int>
 1 Dickens, Charles                     1364  31226  360502 6785632
 2 Wells, H. G. (Herbert George)        3958  64936. 470557 5224147
 3 Alcott, Louisa May                   2660  55483  194549 2977676
 4 Eliot, George                        1871 108236. 320413 2247001
 5 Austen, Jane                        23192 101879  784790 1652092
 6 Shelley, Mary Wollstonecraft        12514  53643  183856 1434844
 7 Brontë, Charlotte                    1416 138921  219783  699938
 8 Brontë, Anne                        68716 119946. 171177  239893
 9 Lovecraft, H. P. (Howard Phillips)   3654  12073   99008  160200
10 Brontë, Emily                      117082 117082  117082  117082
\end{verbatim}

\begin{Shaded}
\begin{Highlighting}[]
\NormalTok{dataf }\SpecialCharTok{|\textgreater{}} 
    \FunctionTok{group\_by}\NormalTok{(author, title) }\SpecialCharTok{|\textgreater{}} 
    \FunctionTok{summarize}\NormalTok{(}\AttributeTok{n =} \FunctionTok{sum}\NormalTok{(n)) }\SpecialCharTok{|\textgreater{}} 
    \FunctionTok{ggplot}\NormalTok{(}\FunctionTok{aes}\NormalTok{(author, n, }\AttributeTok{color =}\NormalTok{ author)) }\SpecialCharTok{+}
    \FunctionTok{geom\_boxplot}\NormalTok{() }\SpecialCharTok{+}
    \FunctionTok{geom\_beeswarm}\NormalTok{() }\SpecialCharTok{+}
    \FunctionTok{scale\_color\_discrete}\NormalTok{(}\AttributeTok{guide =} \StringTok{\textquotesingle{}none\textquotesingle{}}\NormalTok{) }\SpecialCharTok{+}
    \FunctionTok{coord\_flip}\NormalTok{()}
\end{Highlighting}
\end{Shaded}

\begin{verbatim}
`summarise()` has grouped output by 'author'. You can override using the
`.groups` argument.
\end{verbatim}

\begin{figure}[H]

{\centering \includegraphics[width=6in,height=4in]{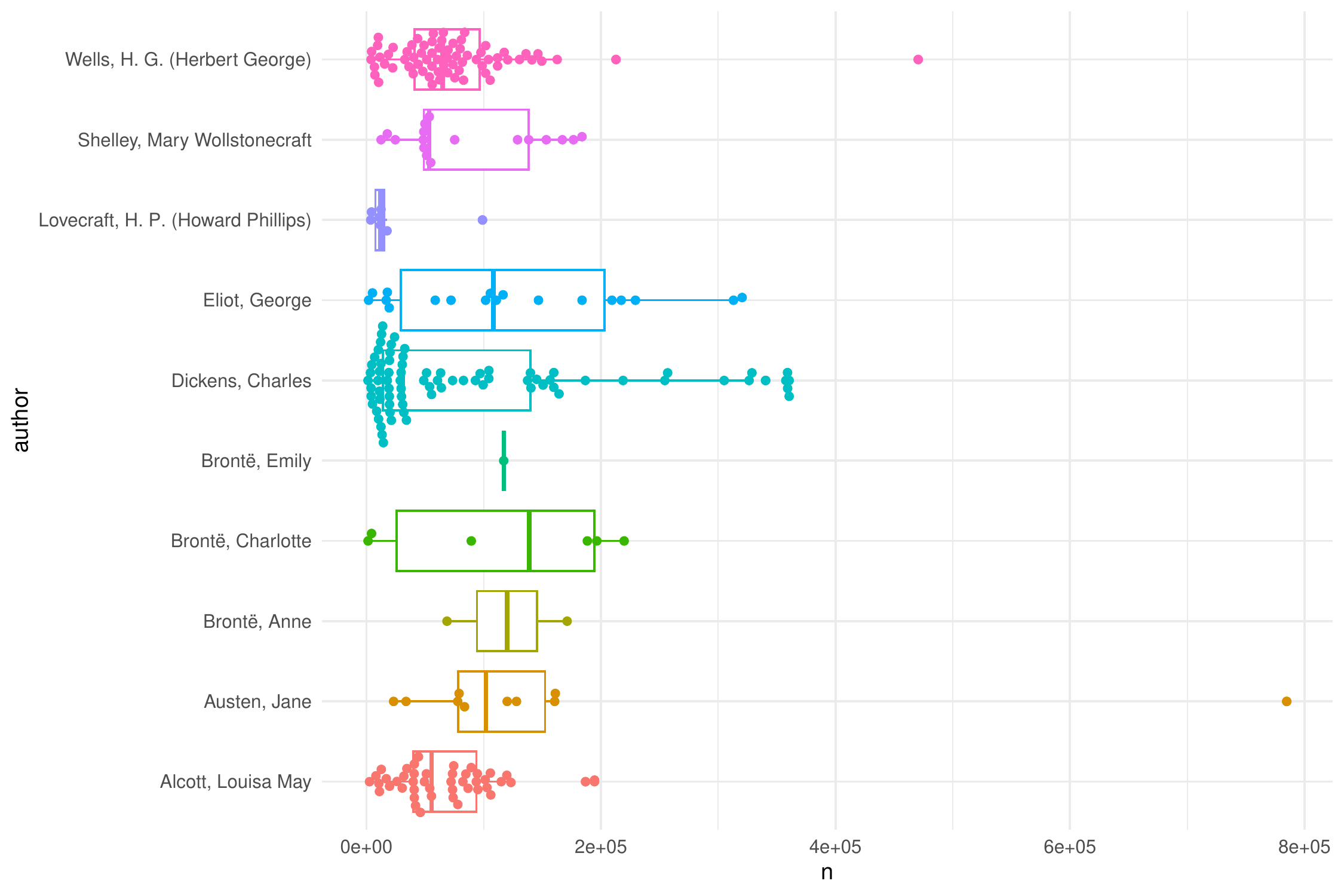}

}

\end{figure}

\hypertarget{vocabulary-selection}{%
\subsection{Vocabulary selection}\label{vocabulary-selection}}

In line with a common rule of thumb in topic modeling, we aim for a
vocabulary of about 10 times as many terms as documents in the corpus.

\begin{Shaded}
\begin{Highlighting}[]
\NormalTok{vocab\_size }\OtherTok{=} \FunctionTok{n\_distinct}\NormalTok{(dataf}\SpecialCharTok{$}\NormalTok{author, dataf}\SpecialCharTok{$}\NormalTok{title) }\SpecialCharTok{*} \DecValTok{10}
\NormalTok{vocab\_size}
\end{Highlighting}
\end{Shaded}

\begin{verbatim}
[1] 2530
\end{verbatim}

\texttt{tmfast} provides two information-theoretic methods for
vocabulary selection. Both are based on the idea of a two-player
guessing game. I pick one of the documents from the corpus, then one of
the terms from the document. I tell you the term, and you have to guess
which document I picked. More informative terms have greater information
gain (calculated as the Kullback-Leibler divergence) relative to a
``baseline'' distribution based purely on the process used to pick the
document. The difference between the two methods is in the
document-picking process. The \texttt{ndH} method assumes the document
was picked uniformly at random from the corpus, so that no document is
more likely to be picked than any other. The \texttt{ndR} method assumes
document probability is proportional to the document length, so that
shorter documents are less likely to be picked. This method implies that
terms that are distinctive of shorter documents have high information
gain, since they indicate ``surprising'' short documents.

On either method, the most informative terms are often typographical or
OCR errors, since these only occur in a single document. To balance
this, we multiply the information gain (\(\Delta H\) for the uniform
process, \(\Delta R\) for the length-weighted process) by the log
frequency of the term across the entire corpus (\(\log n\)). So
\texttt{ndH} is shorthand for \(\log(n) \Delta H\) while \texttt{ndR} is
shorthand for \(\log(n) \Delta R\).

\begin{Shaded}
\begin{Highlighting}[]
\FunctionTok{tic}\NormalTok{()}
\NormalTok{H\_df }\OtherTok{=} \FunctionTok{ndH}\NormalTok{(dataf, title, term, n)}
\NormalTok{R\_df }\OtherTok{=} \FunctionTok{ndR}\NormalTok{(dataf, title, term, n) }\SpecialCharTok{|\textgreater{}}
    \FunctionTok{mutate}\NormalTok{(}\AttributeTok{in\_vocab =} \FunctionTok{rank}\NormalTok{(}\FunctionTok{desc}\NormalTok{(ndR)) }\SpecialCharTok{\textless{}=}\NormalTok{ vocab\_size)}
\FunctionTok{toc}\NormalTok{()}
\end{Highlighting}
\end{Shaded}

\begin{verbatim}
17.518 sec elapsed
\end{verbatim}

\begin{Shaded}
\begin{Highlighting}[]
\NormalTok{H\_df}
\end{Highlighting}
\end{Shaded}

\begin{verbatim}
# A tibble: 116,449 x 5
   term           H    dH     n   ndH
   <chr>      <dbl> <dbl> <int> <dbl>
 1 kipps     0.125   7.86  1454  82.6
 2 dombey    0.386   7.60  1618  81.0
 3 boffin    0.108   7.87  1127  79.8
 4 pecksniff 0.366   7.62  1320  79.0
 5 gwendolen 0.239   7.74  1048  77.7
 6 lydgate   0.0235  7.96   867  77.7
 7 deronda   0.404   7.58  1155  77.1
 8 nicholas  0.977   7.01  1931  76.5
 9 tito      0.0835  7.90   811  76.3
10 squeers   0.253   7.73   895  75.8
# ... with 116,439 more rows
\end{verbatim}

\begin{Shaded}
\begin{Highlighting}[]
\NormalTok{R\_df}
\end{Highlighting}
\end{Shaded}

\begin{verbatim}
# A tibble: 116,449 x 5
   term           n    dR   ndR in_vocab
   <chr>      <int> <dbl> <dbl> <lgl>   
 1 kipps       1454  7.48  78.6 TRUE    
 2 hoopdriver   469  8.50  75.4 TRUE    
 3 scrooge     1007  7.54  75.3 TRUE    
 4 lewisham     575  8.03  73.6 TRUE    
 5 benham       689  7.56  71.3 TRUE    
 6 melville     243  8.97  71.1 TRUE    
 7 bealby       458  8.04  71.0 TRUE    
 8 veronica     659  7.52  70.4 TRUE    
 9 bert         556  7.71  70.3 TRUE    
10 christie    1797  6.43  69.5 TRUE    
# ... with 116,439 more rows
\end{verbatim}

The resulting term ranking of the two methods tend to be similar, but
\texttt{ndR} is preferable in the current case because of the additional
weight it gives to distinctive terms from shorter documents.

\begin{Shaded}
\begin{Highlighting}[]
\FunctionTok{inner\_join}\NormalTok{(H\_df, R\_df, }\AttributeTok{by =} \StringTok{\textquotesingle{}term\textquotesingle{}}\NormalTok{) }\SpecialCharTok{|\textgreater{}} 
    \FunctionTok{ggplot}\NormalTok{(}\FunctionTok{aes}\NormalTok{(ndH, ndR, }\AttributeTok{color =}\NormalTok{ in\_vocab)) }\SpecialCharTok{+}
    \FunctionTok{geom\_point}\NormalTok{(}\FunctionTok{aes}\NormalTok{(}\AttributeTok{alpha =} \FunctionTok{rank}\NormalTok{(}\FunctionTok{desc}\NormalTok{(ndH)) }\SpecialCharTok{\textless{}=}\NormalTok{ vocab\_size))}
\end{Highlighting}
\end{Shaded}

\begin{verbatim}
Warning: Using alpha for a discrete variable is not advised.
\end{verbatim}

\begin{figure}[H]

{\centering \includegraphics[width=6in,height=4in]{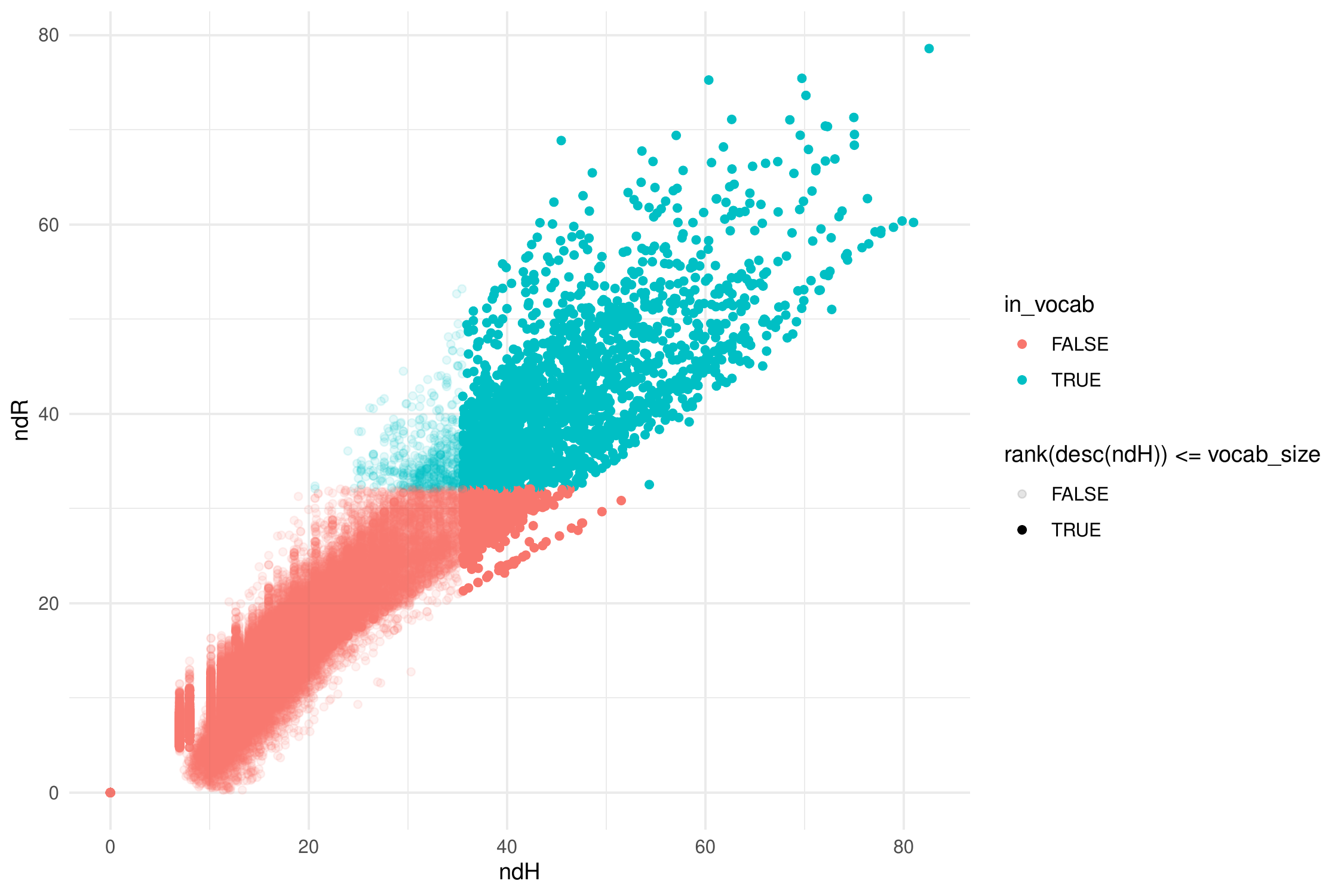}

}

\end{figure}

\begin{Shaded}
\begin{Highlighting}[]
\FunctionTok{inner\_join}\NormalTok{(H\_df, R\_df, }\AttributeTok{by =} \StringTok{\textquotesingle{}term\textquotesingle{}}\NormalTok{) }\SpecialCharTok{|\textgreater{}} 
    \FunctionTok{mutate}\NormalTok{(}\AttributeTok{ndH\_rank =} \FunctionTok{rank}\NormalTok{(}\FunctionTok{desc}\NormalTok{(ndH)), }
           \AttributeTok{ndR\_rank =} \FunctionTok{rank}\NormalTok{(}\FunctionTok{desc}\NormalTok{(ndR))) }\SpecialCharTok{|\textgreater{}} 
    \FunctionTok{ggplot}\NormalTok{(}\FunctionTok{aes}\NormalTok{(ndH\_rank, ndR\_rank, }\AttributeTok{color =}\NormalTok{ in\_vocab)) }\SpecialCharTok{+}
    \FunctionTok{geom\_point}\NormalTok{(}\FunctionTok{aes}\NormalTok{(}\AttributeTok{alpha =}\NormalTok{ ndH\_rank }\SpecialCharTok{\textless{}=}\NormalTok{ vocab\_size)) }\SpecialCharTok{+}
    \FunctionTok{scale\_x\_log10}\NormalTok{() }\SpecialCharTok{+} 
    \FunctionTok{scale\_y\_log10}\NormalTok{()}
\end{Highlighting}
\end{Shaded}

\begin{verbatim}
Warning: Using alpha for a discrete variable is not advised.
\end{verbatim}

\begin{figure}[H]

{\centering \includegraphics[width=6in,height=4in]{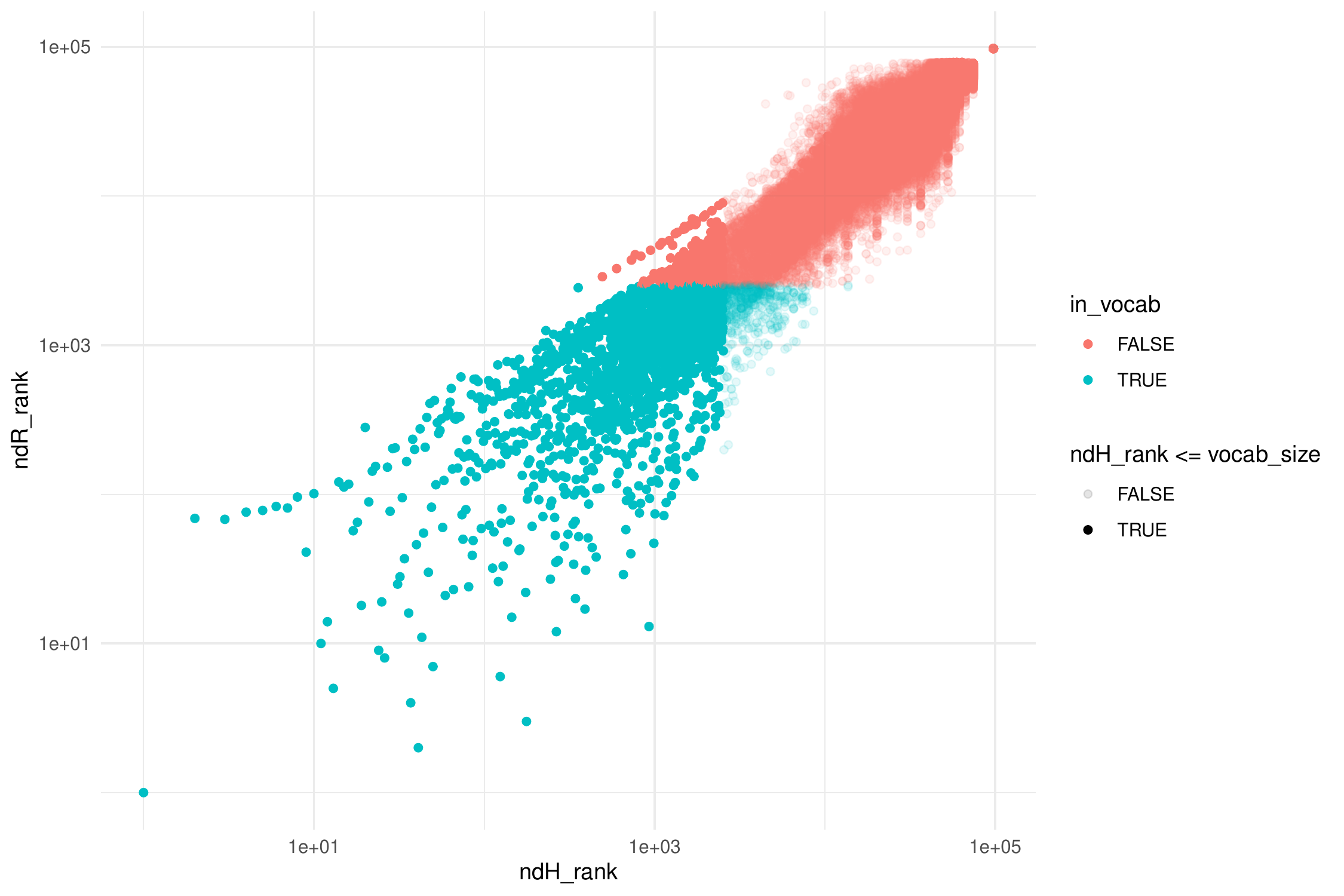}

}

\end{figure}

\begin{Shaded}
\begin{Highlighting}[]
\NormalTok{vocab }\OtherTok{=}\NormalTok{ R\_df }\SpecialCharTok{|\textgreater{}} 
    \FunctionTok{filter}\NormalTok{(in\_vocab) }\SpecialCharTok{|\textgreater{}} 
    \FunctionTok{pull}\NormalTok{(term)}
\FunctionTok{head}\NormalTok{(vocab, }\DecValTok{50}\NormalTok{)}
\end{Highlighting}
\end{Shaded}

\begin{verbatim}
 [1] "kipps"      "hoopdriver" "scrooge"    "lewisham"   "benham"    
 [6] "melville"   "bealby"     "veronica"   "bert"       "christie"  
[11] "sylvia"     "snitchey"   "boldheart"  "britling"   "castruccio"
[16] "bounderby"  "lillian"    "maggie"     "marjorie"   "craggs"    
[21] "n't"        "kemp"       "bab"        "redwood"    "harman"    
[26] "cavor"      "chatteris"  "brumley"    "ammi"       "heathcliff"
[31] "tackleton"  "gladys"     "helwyze"    "tetterby"   "montgomery"
[36] "lodore"     "trafford"   "treherne"   "jill"       "ludovico"  
[41] "tito"       "lomi"       "canaris"    "trotty"     "villiers"  
[46] "falkner"    "doubledick" "amanda"     "gradgrind"  "linton"    
\end{verbatim}

\begin{Shaded}
\begin{Highlighting}[]
\NormalTok{dataf }\SpecialCharTok{|\textgreater{}} 
    \FunctionTok{filter}\NormalTok{(term }\SpecialCharTok{\%in\%}\NormalTok{ vocab) }\SpecialCharTok{|\textgreater{}} 
    \FunctionTok{group\_by}\NormalTok{(author, title) }\SpecialCharTok{|\textgreater{}} 
    \FunctionTok{summarize}\NormalTok{(}\AttributeTok{n =} \FunctionTok{sum}\NormalTok{(n)) }\SpecialCharTok{|\textgreater{}} 
    \FunctionTok{ggplot}\NormalTok{(}\FunctionTok{aes}\NormalTok{(author, n, }\AttributeTok{color =}\NormalTok{ author)) }\SpecialCharTok{+}
    \FunctionTok{geom\_boxplot}\NormalTok{() }\SpecialCharTok{+}
    \FunctionTok{geom\_beeswarm}\NormalTok{() }\SpecialCharTok{+}
    \FunctionTok{scale\_color\_discrete}\NormalTok{(}\AttributeTok{guide =} \StringTok{\textquotesingle{}none\textquotesingle{}}\NormalTok{) }\SpecialCharTok{+}
    \FunctionTok{coord\_flip}\NormalTok{()}
\end{Highlighting}
\end{Shaded}

\begin{verbatim}
`summarise()` has grouped output by 'author'. You can override using the
`.groups` argument.
\end{verbatim}

\begin{figure}[H]

{\centering \includegraphics[width=6in,height=4in]{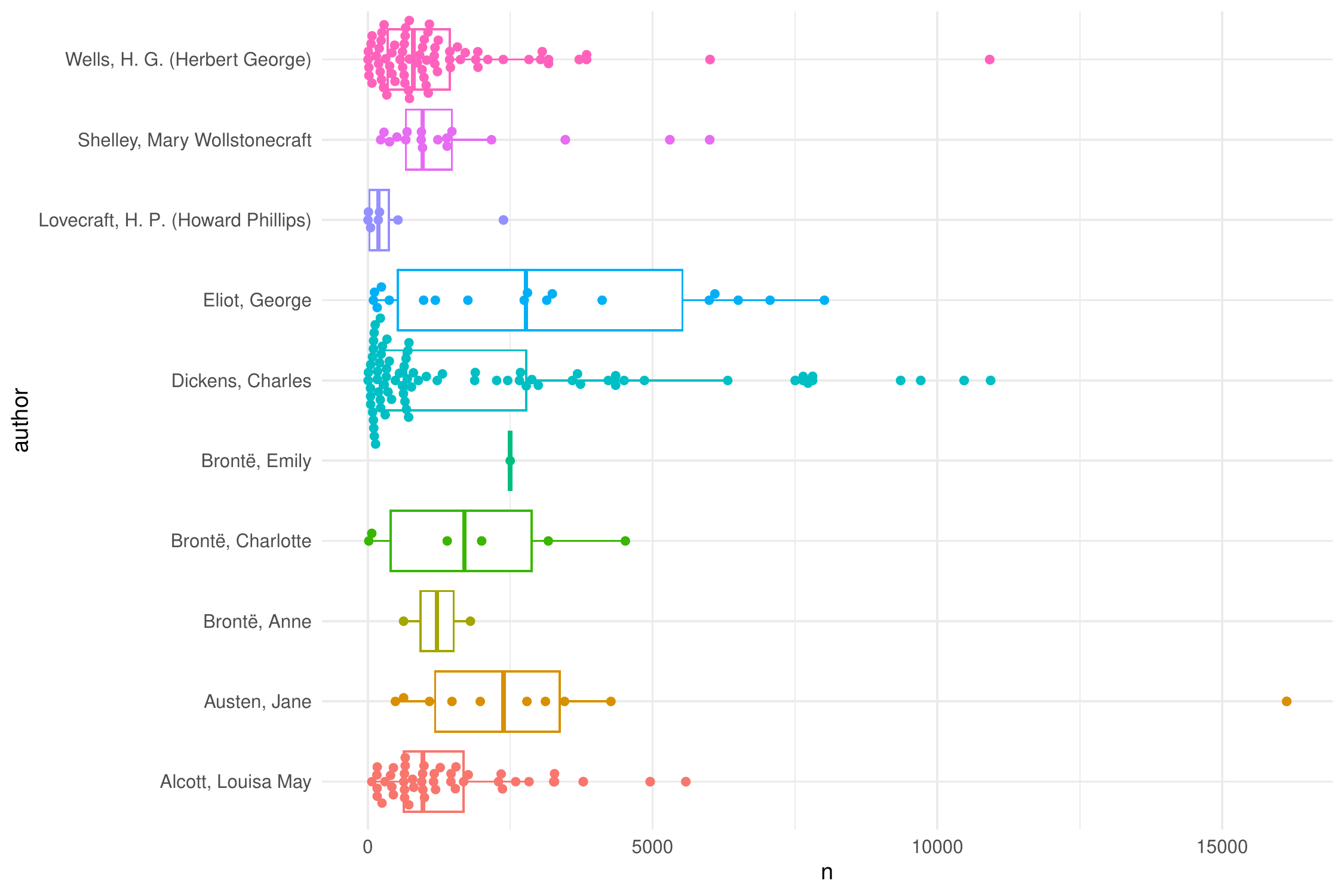}

}

\end{figure}

\hypertarget{fit-topic-models}{%
\subsection{Fit topic models}\label{fit-topic-models}}

\begin{Shaded}
\begin{Highlighting}[]
\NormalTok{dtm }\OtherTok{=}\NormalTok{ dataf }\SpecialCharTok{|\textgreater{}} 
    \FunctionTok{filter}\NormalTok{(term }\SpecialCharTok{\%in\%}\NormalTok{ vocab) }\SpecialCharTok{|\textgreater{}} 
    \FunctionTok{mutate}\NormalTok{(}\AttributeTok{n =} \FunctionTok{log1p}\NormalTok{(n))}

\NormalTok{n\_authors }\OtherTok{=} \FunctionTok{n\_distinct}\NormalTok{(dataf}\SpecialCharTok{$}\NormalTok{author)}

\FunctionTok{tic}\NormalTok{()}
\NormalTok{fitted\_tmf }\OtherTok{=} \FunctionTok{tmfast}\NormalTok{(dtm, }\AttributeTok{n =} \FunctionTok{c}\NormalTok{(}\DecValTok{5}\NormalTok{,}
\NormalTok{                               n\_authors, }
\NormalTok{                               n\_authors }\SpecialCharTok{+} \DecValTok{5}\NormalTok{),}
                    \AttributeTok{row =}\NormalTok{ title, }\AttributeTok{column =}\NormalTok{ term, }\AttributeTok{value =}\NormalTok{ n)}
\FunctionTok{toc}\NormalTok{()}
\end{Highlighting}
\end{Shaded}

\begin{verbatim}
0.801 sec elapsed
\end{verbatim}

\begin{Shaded}
\begin{Highlighting}[]
\FunctionTok{screeplot}\NormalTok{(fitted\_tmf, }\AttributeTok{npcs =}\NormalTok{ n\_authors }\SpecialCharTok{+} \DecValTok{5}\NormalTok{)}
\end{Highlighting}
\end{Shaded}

\begin{figure}[H]

{\centering \includegraphics[width=6in,height=4in]{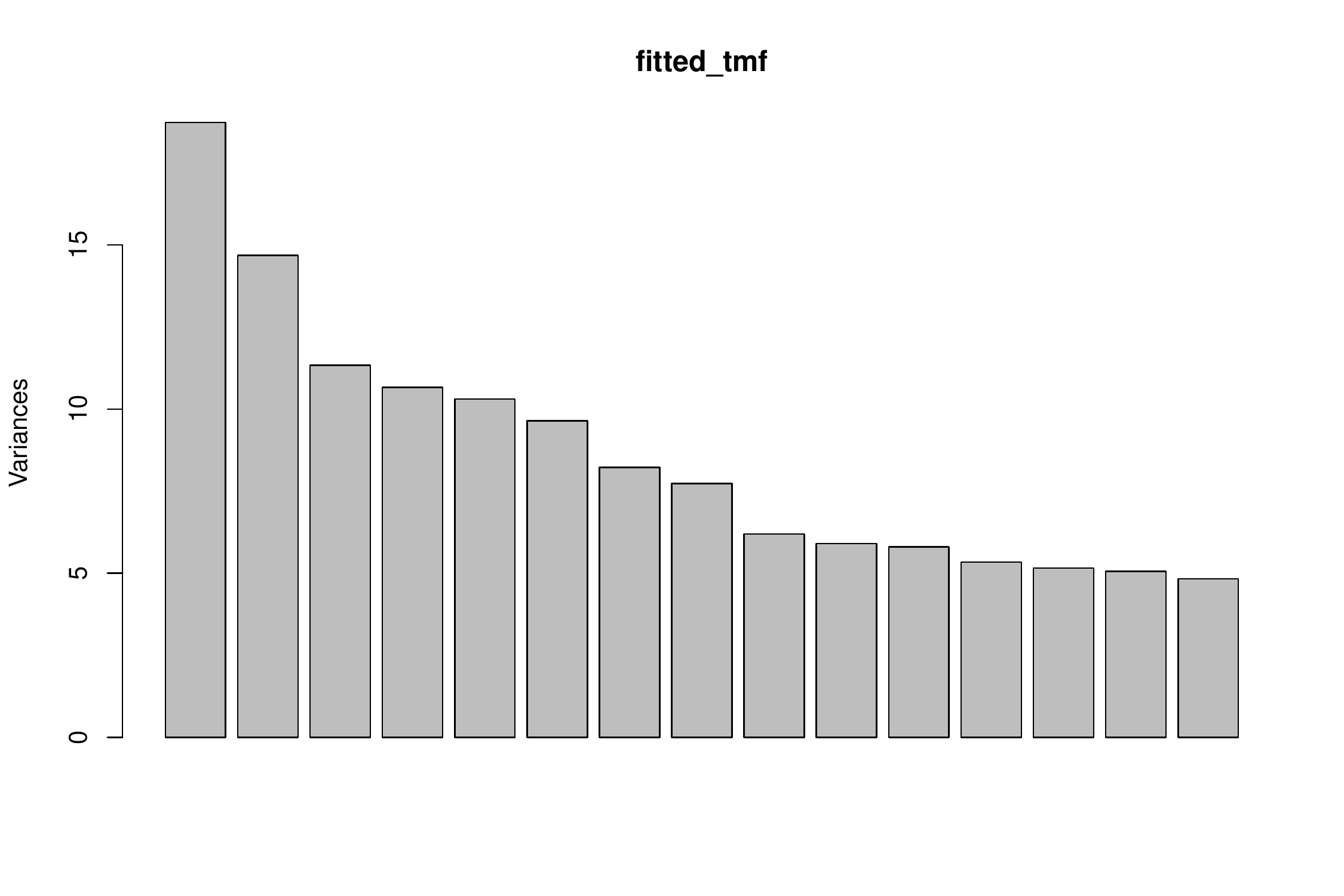}

}

\end{figure}

\hypertarget{topic-exploration}{%
\subsection{Topic exploration}\label{topic-exploration}}

Without renormalization, most of the works are spread across a few
topics, and the topics don't clearly correspond to authors.

\begin{Shaded}
\begin{Highlighting}[]
\FunctionTok{tidy}\NormalTok{(fitted\_tmf, n\_authors, }\StringTok{\textquotesingle{}gamma\textquotesingle{}}\NormalTok{) }\SpecialCharTok{|\textgreater{}} 
    \FunctionTok{left\_join}\NormalTok{(meta\_df, }\AttributeTok{by =} \FunctionTok{c}\NormalTok{(}\StringTok{\textquotesingle{}document\textquotesingle{}} \OtherTok{=} \StringTok{\textquotesingle{}title\textquotesingle{}}\NormalTok{)) }\SpecialCharTok{|\textgreater{}} 
    \FunctionTok{ggplot}\NormalTok{(}\FunctionTok{aes}\NormalTok{(document, gamma, }\AttributeTok{fill =}\NormalTok{ topic)) }\SpecialCharTok{+}
    \FunctionTok{geom\_col}\NormalTok{() }\SpecialCharTok{+}
    \FunctionTok{facet\_wrap}\NormalTok{(}\FunctionTok{vars}\NormalTok{(author), }\AttributeTok{scales =} \StringTok{\textquotesingle{}free\_x\textquotesingle{}}\NormalTok{) }\SpecialCharTok{+}
    \FunctionTok{scale\_x\_discrete}\NormalTok{(}\AttributeTok{guide =} \StringTok{\textquotesingle{}none\textquotesingle{}}\NormalTok{) }\SpecialCharTok{+}
    \FunctionTok{scale\_fill\_viridis\_d}\NormalTok{()}
\end{Highlighting}
\end{Shaded}

\begin{figure}[H]

{\centering \includegraphics[width=6in,height=4in]{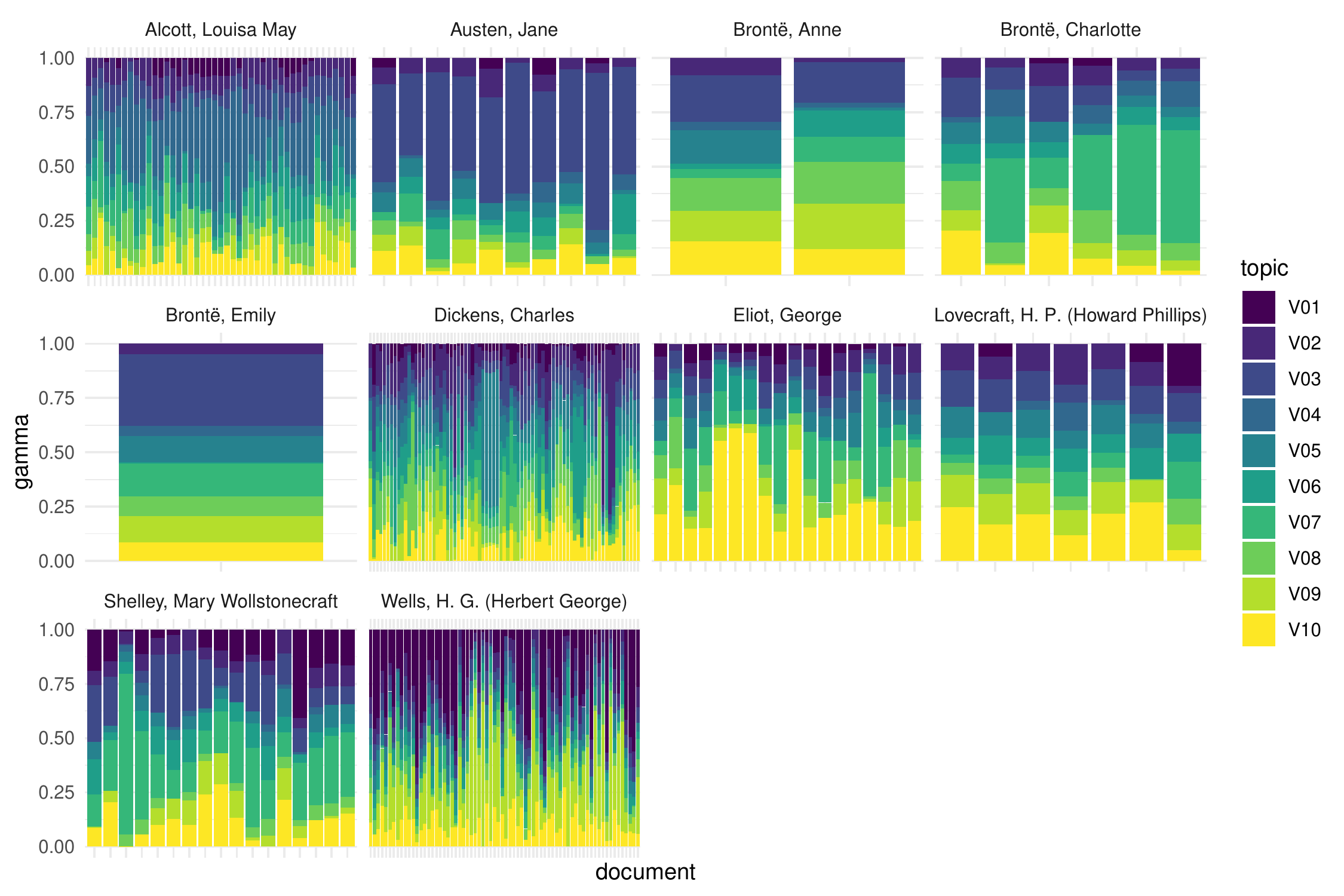}

}

\end{figure}

To renormalize, we need to choose a theoretical Dirichlet distribution.

\begin{Shaded}
\begin{Highlighting}[]
\NormalTok{alpha }\OtherTok{=} \FunctionTok{peak\_alpha}\NormalTok{(n\_authors, }\DecValTok{1}\NormalTok{, }\AttributeTok{peak =}\NormalTok{ .}\DecValTok{8}\NormalTok{, }\AttributeTok{scale =} \DecValTok{10}\NormalTok{)}
\NormalTok{target\_entropy }\OtherTok{=} \FunctionTok{expected\_entropy}\NormalTok{(alpha)}
\NormalTok{target\_entropy}
\end{Highlighting}
\end{Shaded}

\begin{verbatim}
[1] 0.997604
\end{verbatim}

\begin{Shaded}
\begin{Highlighting}[]
\NormalTok{exponent }\OtherTok{=} \FunctionTok{tidy}\NormalTok{(fitted\_tmf, n\_authors, }\StringTok{\textquotesingle{}gamma\textquotesingle{}}\NormalTok{) }\SpecialCharTok{|\textgreater{}} 
    \FunctionTok{target\_power}\NormalTok{(document, gamma, target\_entropy)}
\NormalTok{exponent}
\end{Highlighting}
\end{Shaded}

\begin{verbatim}
[1] 4.064884
\end{verbatim}

\begin{Shaded}
\begin{Highlighting}[]
\FunctionTok{tidy}\NormalTok{(fitted\_tmf, n\_authors, }\StringTok{\textquotesingle{}gamma\textquotesingle{}}\NormalTok{, }\AttributeTok{exponent =}\NormalTok{ exponent) }\SpecialCharTok{|\textgreater{}} 
    \FunctionTok{left\_join}\NormalTok{(meta\_df, }\AttributeTok{by =} \FunctionTok{c}\NormalTok{(}\StringTok{\textquotesingle{}document\textquotesingle{}} \OtherTok{=} \StringTok{\textquotesingle{}title\textquotesingle{}}\NormalTok{)) }\SpecialCharTok{|\textgreater{}} 
    \FunctionTok{ggplot}\NormalTok{(}\FunctionTok{aes}\NormalTok{(document, gamma, }\AttributeTok{fill =}\NormalTok{ topic)) }\SpecialCharTok{+}
    \FunctionTok{geom\_col}\NormalTok{() }\SpecialCharTok{+}
    \FunctionTok{facet\_wrap}\NormalTok{(}\FunctionTok{vars}\NormalTok{(author), }\AttributeTok{scales =} \StringTok{\textquotesingle{}free\_x\textquotesingle{}}\NormalTok{) }\SpecialCharTok{+}
    \FunctionTok{scale\_x\_discrete}\NormalTok{(}\AttributeTok{guide =} \StringTok{\textquotesingle{}none\textquotesingle{}}\NormalTok{) }\SpecialCharTok{+}
    \FunctionTok{scale\_fill\_viridis\_d}\NormalTok{()}
\end{Highlighting}
\end{Shaded}

\begin{figure}[H]

{\centering \includegraphics[width=6in,height=4in]{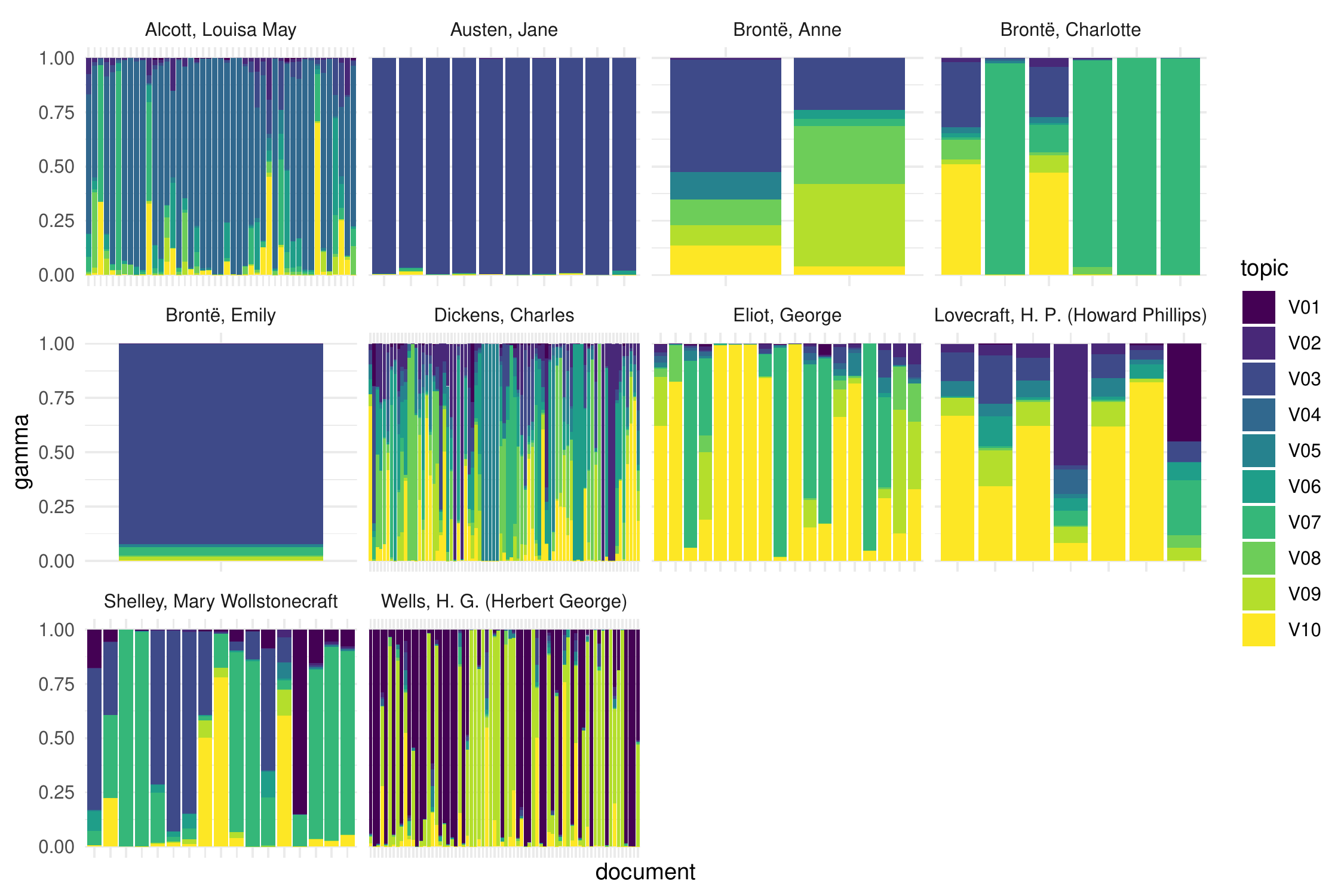}

}

\end{figure}

\begin{Shaded}
\begin{Highlighting}[]
\FunctionTok{tidy}\NormalTok{(fitted\_tmf, n\_authors, }\StringTok{\textquotesingle{}gamma\textquotesingle{}}\NormalTok{, }\AttributeTok{exponent =}\NormalTok{ exponent) }\SpecialCharTok{|\textgreater{}} 
    \FunctionTok{left\_join}\NormalTok{(meta\_df, }\AttributeTok{by =} \FunctionTok{c}\NormalTok{(}\StringTok{\textquotesingle{}document\textquotesingle{}} \OtherTok{=} \StringTok{\textquotesingle{}title\textquotesingle{}}\NormalTok{)) }\SpecialCharTok{|\textgreater{}} 
    \FunctionTok{ggplot}\NormalTok{(}\FunctionTok{aes}\NormalTok{(document, topic, }\AttributeTok{fill =}\NormalTok{ gamma)) }\SpecialCharTok{+}
    \FunctionTok{geom\_raster}\NormalTok{() }\SpecialCharTok{+}
    \FunctionTok{facet\_grid}\NormalTok{(}\AttributeTok{cols =} \FunctionTok{vars}\NormalTok{(}\FunctionTok{str\_wrap}\NormalTok{(author, }
                                    \AttributeTok{width =} \DecValTok{20}\NormalTok{)), }
               \AttributeTok{scales =} \StringTok{\textquotesingle{}free\_x\textquotesingle{}}\NormalTok{, }
               \AttributeTok{switch =} \StringTok{\textquotesingle{}x\textquotesingle{}}\NormalTok{) }\SpecialCharTok{+}
    \FunctionTok{scale\_x\_discrete}\NormalTok{(}\AttributeTok{guide =} \StringTok{\textquotesingle{}none\textquotesingle{}}\NormalTok{)}
\end{Highlighting}
\end{Shaded}

\begin{figure}[H]

{\centering \includegraphics[width=6in,height=4in]{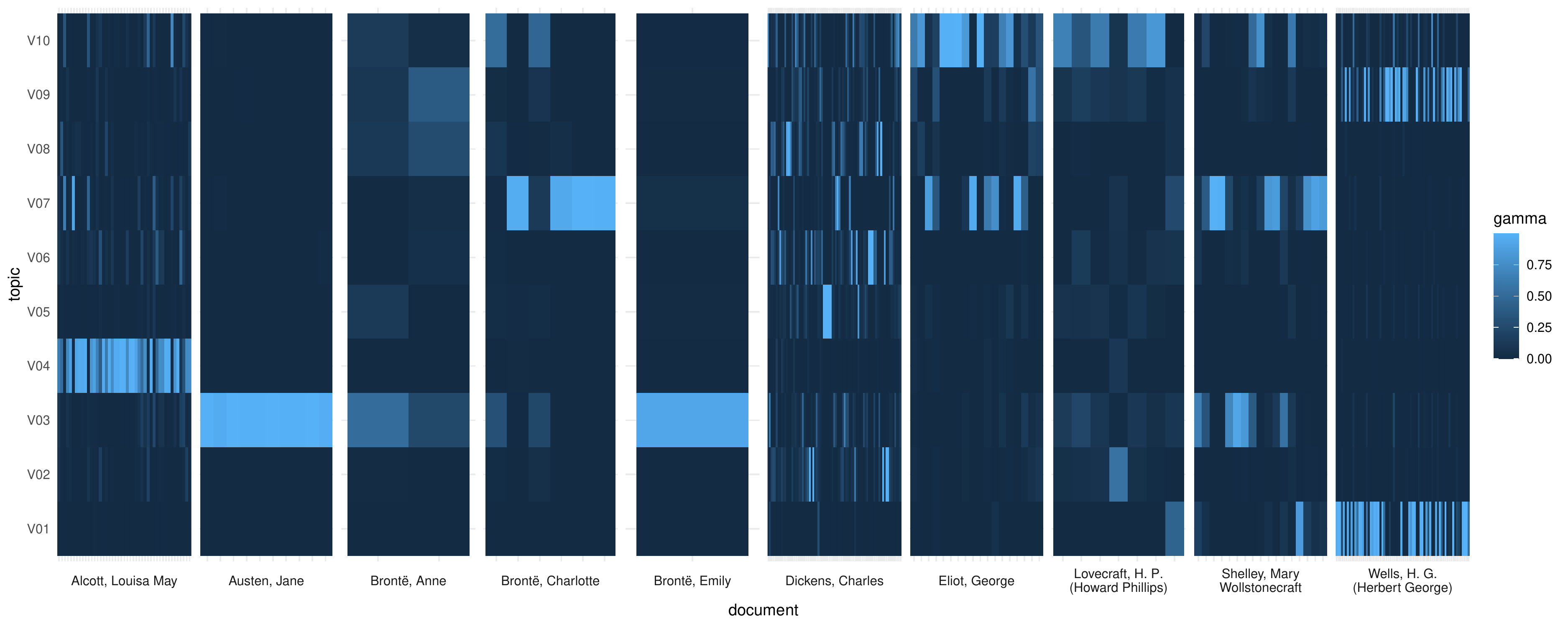}

}

\end{figure}

After renormalization, there are distinctive topics for Alcott (4) and
Wells (1 and 9). Austen, Anne Brontë, Emily Brontë, and some of
Shelley's works appear together in topic 3. Charlotte Brontë and some of
Eliot's and Shelley's works split topic 5. Eliot and Lovecraft share
topic 10. And Dickens' works are spread across multiple topics, with 2,
6, and 8 appearing to be distinctive to him.

To aid interpretation, we create a crosswalk dataframe connecting topics
to authors.

\begin{Shaded}
\begin{Highlighting}[]
\NormalTok{topic\_author }\OtherTok{=} \FunctionTok{tribble}\NormalTok{(}
    \SpecialCharTok{\textasciitilde{}}\NormalTok{ topic, }\SpecialCharTok{\textasciitilde{}}\NormalTok{ authors,}
    \StringTok{\textquotesingle{}V01\textquotesingle{}}\NormalTok{, }\StringTok{\textquotesingle{}Wells\textquotesingle{}}\NormalTok{, }
    \StringTok{\textquotesingle{}V02\textquotesingle{}}\NormalTok{, }\StringTok{\textquotesingle{}Dickens\textquotesingle{}}\NormalTok{, }
    \StringTok{\textquotesingle{}V03\textquotesingle{}}\NormalTok{, }\StringTok{\textquotesingle{}Austin, A \& E Brontë\textquotesingle{}}\NormalTok{, }
    \StringTok{\textquotesingle{}V04\textquotesingle{}}\NormalTok{, }\StringTok{\textquotesingle{}Alcott\textquotesingle{}}\NormalTok{, }
    \StringTok{\textquotesingle{}V05\textquotesingle{}}\NormalTok{, }\StringTok{\textquotesingle{}Dickens\textquotesingle{}}\NormalTok{, }
    \StringTok{\textquotesingle{}V06\textquotesingle{}}\NormalTok{, }\StringTok{\textquotesingle{}Dickens\textquotesingle{}}\NormalTok{, }
    \StringTok{\textquotesingle{}V07\textquotesingle{}}\NormalTok{, }\StringTok{\textquotesingle{}C Brontë, Eliot, Shelley\textquotesingle{}}\NormalTok{, }
    \StringTok{\textquotesingle{}V08\textquotesingle{}}\NormalTok{, }\StringTok{\textquotesingle{}Dickens\textquotesingle{}}\NormalTok{, }
    \StringTok{\textquotesingle{}V09\textquotesingle{}}\NormalTok{, }\StringTok{\textquotesingle{}Wells\textquotesingle{}}\NormalTok{, }
    \StringTok{\textquotesingle{}V10\textquotesingle{}}\NormalTok{, }\StringTok{\textquotesingle{}Eliot, Lovecraft\textquotesingle{}}
\NormalTok{)}
\end{Highlighting}
\end{Shaded}

To explore these topics further, we turn to the word-topic distribution.
These distributions could be renormalized, as with the topic-doc
distributions. But the exponent for the word-topic distributions is
usually quite close to 1, meaning renormalization doesn't change these
distributions very much.

\begin{Shaded}
\begin{Highlighting}[]
\NormalTok{target\_entropy\_term }\OtherTok{=} \FunctionTok{expected\_entropy}\NormalTok{(.}\DecValTok{1}\NormalTok{, }\AttributeTok{k =}\NormalTok{ vocab\_size)}
\NormalTok{target\_entropy\_term}
\end{Highlighting}
\end{Shaded}

\begin{verbatim}
[1] 8.597192
\end{verbatim}

\begin{Shaded}
\begin{Highlighting}[]
\NormalTok{exponent\_term }\OtherTok{=} \FunctionTok{tidy}\NormalTok{(fitted\_tmf, n\_authors, }\StringTok{\textquotesingle{}beta\textquotesingle{}}\NormalTok{) }\SpecialCharTok{|\textgreater{}} 
    \FunctionTok{target\_power}\NormalTok{(topic, beta, target\_entropy\_term)}
\NormalTok{exponent\_term}
\end{Highlighting}
\end{Shaded}

\begin{verbatim}
[1] 1.066448
\end{verbatim}

We therefore skip renormalization and move directly to a Silge plot,
showing the top 10 terms for each topic.
\texttt{tidytext::reorder\_within()} and
\texttt{tidytext::scale\_x\_reordered()} are useful for constructing
this plot.

\begin{Shaded}
\begin{Highlighting}[]
\NormalTok{beta\_df }\OtherTok{=} \FunctionTok{tidy}\NormalTok{(fitted\_tmf, n\_authors, }\StringTok{\textquotesingle{}beta\textquotesingle{}}\NormalTok{)}

\NormalTok{top\_terms }\OtherTok{=}\NormalTok{ beta\_df }\SpecialCharTok{|\textgreater{}} 
    \FunctionTok{group\_by}\NormalTok{(topic) }\SpecialCharTok{|\textgreater{}} 
    \FunctionTok{arrange}\NormalTok{(topic, }\FunctionTok{desc}\NormalTok{(beta)) }\SpecialCharTok{|\textgreater{}} 
    \FunctionTok{top\_n}\NormalTok{(}\DecValTok{15}\NormalTok{, beta) }\SpecialCharTok{|\textgreater{}} 
    \FunctionTok{left\_join}\NormalTok{(topic\_author, }\AttributeTok{by =} \StringTok{\textquotesingle{}topic\textquotesingle{}}\NormalTok{)}
\NormalTok{top\_terms}
\end{Highlighting}
\end{Shaded}

\begin{verbatim}
# A tibble: 150 x 4
# Groups:   topic [10]
   token     topic    beta authors
   <chr>     <chr>   <dbl> <chr>  
 1 empire    V01   0.0162  Wells  
 2 britain   V01   0.0124  Wells  
 3 peoples   V01   0.0122  Wells  
 4 russia    V01   0.0117  Wells  
 5 king      V01   0.0111  Wells  
 6 asia      V01   0.0104  Wells  
 7 socialism V01   0.00995 Wells  
 8 section   V01   0.00971 Wells  
 9 egypt     V01   0.00926 Wells  
10 ii        V01   0.00892 Wells  
# ... with 140 more rows
\end{verbatim}

\begin{Shaded}
\begin{Highlighting}[]
\NormalTok{top\_terms }\SpecialCharTok{|\textgreater{}} 
    \FunctionTok{mutate}\NormalTok{(}\AttributeTok{token =} \FunctionTok{reorder\_within}\NormalTok{(token, }
                                  \AttributeTok{by =}\NormalTok{ beta, }
                                  \AttributeTok{within =}\NormalTok{ topic)) }\SpecialCharTok{|\textgreater{}} 
    \FunctionTok{ggplot}\NormalTok{(}\FunctionTok{aes}\NormalTok{(token, beta)) }\SpecialCharTok{+}
    \FunctionTok{geom\_point}\NormalTok{() }\SpecialCharTok{+}
    \FunctionTok{geom\_segment}\NormalTok{(}\FunctionTok{aes}\NormalTok{(}\AttributeTok{xend =}\NormalTok{ token), }\AttributeTok{yend =} \DecValTok{0}\NormalTok{) }\SpecialCharTok{+}
    \FunctionTok{facet\_wrap}\NormalTok{(}\FunctionTok{vars}\NormalTok{(topic, authors), }\AttributeTok{scales =} \StringTok{\textquotesingle{}free\_y\textquotesingle{}}\NormalTok{) }\SpecialCharTok{+}
    \FunctionTok{coord\_flip}\NormalTok{() }\SpecialCharTok{+}
    \FunctionTok{scale\_x\_reordered}\NormalTok{()}
\end{Highlighting}
\end{Shaded}

\begin{figure}[H]

{\centering \includegraphics[width=6in,height=4in]{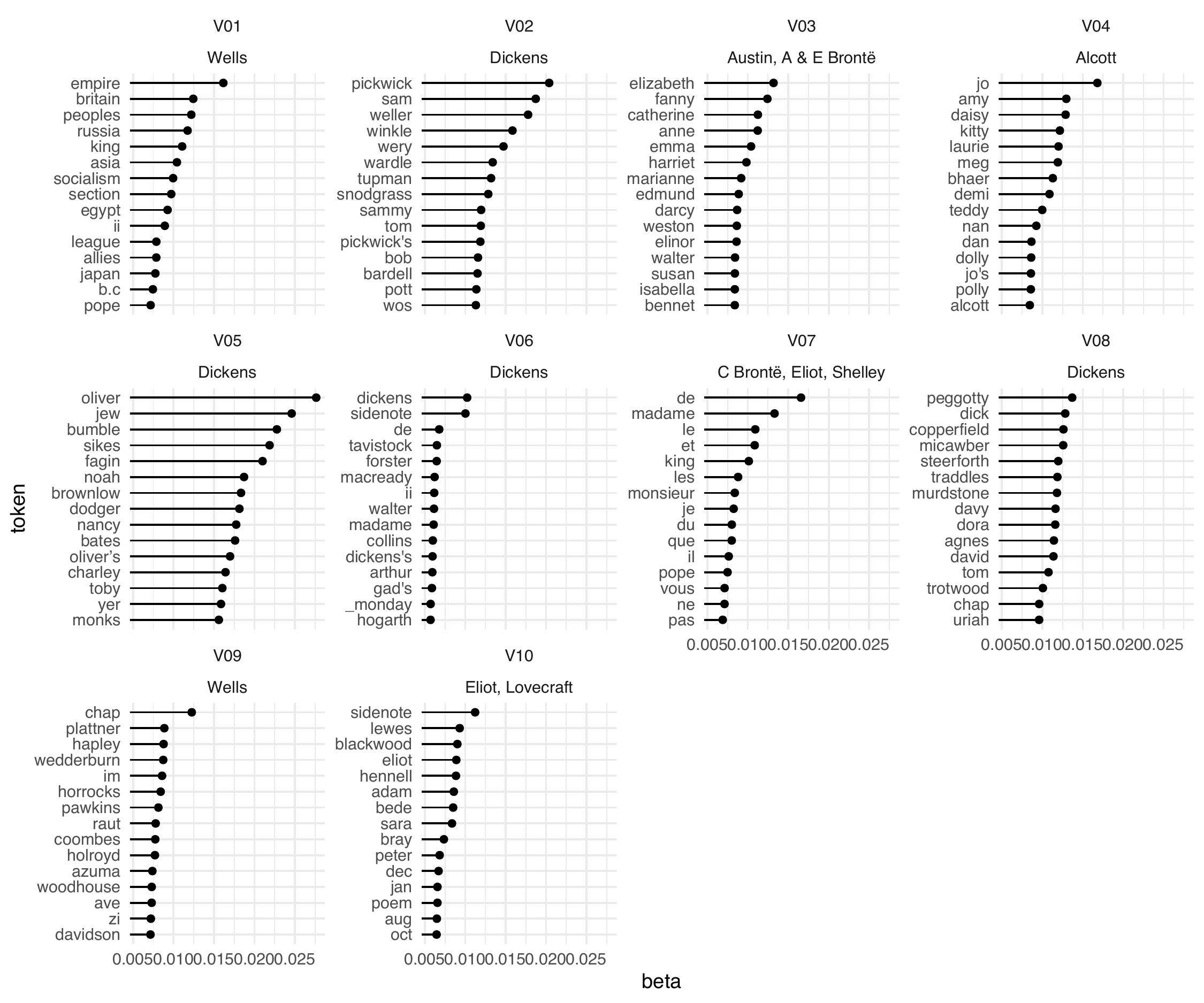}

}

\end{figure}

Most topics (2, 3, 4, 5, 6, 8, 9) focus on character names, with three
of the four Dickens topics corresponding to \emph{The Pickwick Papers}
(topic 2), \emph{Oliver Twist} (5), and \emph{David Copperfield} (8).
Wells' topics appear to distinguish non-fiction essays (topic 1) from
fiction (9). Topic 7 groups together Charlotte Brontë, Eliot, and
Shelley based on the use of French. Topic 10 has a mix of character
names with months of the year; it appears to be a ``miscellaneous''
topic, often created by topic models to accommodate documents that don't
fit elsewhere.

\hypertarget{reproducibility}{%
\section{Reproducibility}\label{reproducibility}}

\begin{Shaded}
\begin{Highlighting}[]
\NormalTok{sessioninfo}\SpecialCharTok{::}\FunctionTok{session\_info}\NormalTok{()}
\end{Highlighting}
\end{Shaded}

\begin{verbatim}
- Session info ---------------------------------------------------------------
 setting  value
 version  R version 4.1.2 (2021-11-01)
 os       macOS Big Sur 10.16
 system   x86_64, darwin17.0
 ui       X11
 language (EN)
 collate  en_US.UTF-8
 ctype    en_US.UTF-8
 tz       America/Los_Angeles
 date     2023-05-02
 pandoc   2.16.2 @ /usr/local/bin/ (via rmarkdown)

- Packages -------------------------------------------------------------------
 package     * version          date (UTC) lib source
 assertthat    0.2.1            2019-03-21 [1] CRAN (R 4.1.0)
 backports     1.4.1            2021-12-13 [1] CRAN (R 4.1.0)
 beeswarm      0.4.0            2021-06-01 [1] CRAN (R 4.1.0)
 broom         1.0.2            2022-12-15 [1] CRAN (R 4.1.2)
 cachem        1.0.7            2023-02-24 [1] CRAN (R 4.1.2)
 cellranger    1.1.0            2016-07-27 [1] CRAN (R 4.1.0)
 cli           3.6.0            2023-01-09 [1] CRAN (R 4.1.2)
 colorspace    2.0-3            2022-02-21 [1] CRAN (R 4.1.2)
 crayon        1.5.1            2022-03-26 [1] CRAN (R 4.1.2)
 data.table    1.14.2           2021-09-27 [1] CRAN (R 4.1.0)
 DBI           1.1.2            2021-12-20 [1] CRAN (R 4.1.0)
 dbplyr        2.2.1            2022-06-27 [1] CRAN (R 4.1.2)
 digest        0.6.31           2022-12-11 [1] CRAN (R 4.1.2)
 dplyr       * 1.0.10           2022-09-01 [1] CRAN (R 4.1.2)
 ellipsis      0.3.2            2021-04-29 [1] CRAN (R 4.1.0)
 evaluate      0.20             2023-01-17 [1] CRAN (R 4.1.2)
 fansi         1.0.3            2022-03-24 [1] CRAN (R 4.1.2)
 farver        2.1.1            2022-07-06 [1] CRAN (R 4.1.2)
 fastmap       1.1.1            2023-02-24 [1] CRAN (R 4.1.2)
 forcats     * 0.5.2            2022-08-19 [1] CRAN (R 4.1.2)
 fs            1.6.1            2023-02-06 [1] CRAN (R 4.1.2)
 generics      0.1.3            2022-07-05 [1] CRAN (R 4.1.2)
 ggbeeswarm  * 0.6.0            2017-08-07 [1] CRAN (R 4.1.0)
 ggplot2     * 3.4.0            2022-11-04 [1] CRAN (R 4.1.2)
 glue        * 1.6.2            2022-02-24 [1] CRAN (R 4.1.2)
 gtable        0.3.0            2019-03-25 [1] CRAN (R 4.1.0)
 gutenbergr  * 0.2.3            2022-12-14 [1] CRAN (R 4.1.2)
 haven         2.5.1            2022-08-22 [1] CRAN (R 4.1.2)
 hms           1.1.2            2022-08-19 [1] CRAN (R 4.1.2)
 htmltools     0.5.4            2022-12-07 [1] CRAN (R 4.1.2)
 httr          1.4.4            2022-08-17 [1] CRAN (R 4.1.2)
 irlba         2.3.5            2021-12-06 [1] CRAN (R 4.1.0)
 janeaustenr   0.1.5            2017-06-10 [1] CRAN (R 4.1.0)
 jsonlite      1.8.4            2022-12-06 [1] CRAN (R 4.1.2)
 knitr         1.42             2023-01-25 [1] CRAN (R 4.1.2)
 labeling      0.4.2            2020-10-20 [1] CRAN (R 4.1.0)
 lattice       0.20-45          2021-09-22 [1] CRAN (R 4.1.2)
 lifecycle     1.0.3            2022-10-07 [1] CRAN (R 4.1.2)
 lpSolve     * 5.6.15           2020-01-24 [1] CRAN (R 4.1.0)
 lubridate     1.9.0            2022-11-06 [1] CRAN (R 4.1.2)
 magrittr      2.0.3            2022-03-30 [1] CRAN (R 4.1.2)
 Matrix        1.3-4            2021-06-01 [1] CRAN (R 4.1.2)
 memoise     * 2.0.1            2021-11-26 [1] CRAN (R 4.1.0)
 mnormt        2.0.2            2020-09-01 [1] CRAN (R 4.1.0)
 modelr        0.1.10           2022-11-11 [1] CRAN (R 4.1.2)
 munsell       0.5.0            2018-06-12 [1] CRAN (R 4.1.0)
 nlme          3.1-153          2021-09-07 [1] CRAN (R 4.1.2)
 pillar        1.8.1            2022-08-19 [1] CRAN (R 4.1.2)
 pkgconfig     2.0.3            2019-09-22 [1] CRAN (R 4.1.0)
 plyr          1.8.7            2022-03-24 [1] CRAN (R 4.1.2)
 psych         2.1.9            2021-09-22 [1] CRAN (R 4.1.0)
 purrr       * 1.0.0            2022-12-20 [1] CRAN (R 4.1.2)
 R6            2.5.1            2021-08-19 [1] CRAN (R 4.1.0)
 Rcpp          1.0.9            2022-07-08 [1] CRAN (R 4.1.2)
 readr       * 2.1.3            2022-10-01 [1] CRAN (R 4.1.2)
 readxl        1.4.1            2022-08-17 [1] CRAN (R 4.1.2)
 reprex        2.0.2            2022-08-17 [1] CRAN (R 4.1.2)
 reshape2      1.4.4            2020-04-09 [1] CRAN (R 4.1.0)
 rlang         1.1.0            2023-03-14 [1] CRAN (R 4.1.2)
 rmarkdown     2.14             2022-04-25 [1] CRAN (R 4.1.2)
 rstudioapi    0.13             2020-11-12 [1] CRAN (R 4.1.0)
 rvest         1.0.3            2022-08-19 [1] CRAN (R 4.1.2)
 scales        1.2.0            2022-04-13 [1] CRAN (R 4.1.2)
 sessioninfo   1.2.2            2021-12-06 [1] CRAN (R 4.1.0)
 SnowballC     0.7.0            2020-04-01 [1] CRAN (R 4.1.0)
 stm         * 1.3.6            2020-09-18 [1] CRAN (R 4.1.0)
 stringi       1.7.12           2023-01-11 [1] CRAN (R 4.1.2)
 stringr     * 1.5.0            2022-12-02 [1] CRAN (R 4.1.2)
 tibble      * 3.1.8            2022-07-22 [1] CRAN (R 4.1.2)
 tictoc      * 1.0.1            2021-04-19 [1] CRAN (R 4.1.0)
 tidyr       * 1.2.1            2022-09-08 [1] CRAN (R 4.1.2)
 tidyselect    1.1.2            2022-02-21 [1] CRAN (R 4.1.2)
 tidytext    * 0.3.2            2021-09-30 [1] CRAN (R 4.1.0)
 tidyverse   * 1.3.1            2021-04-15 [1] CRAN (R 4.1.0)
 timechange    0.1.1            2022-11-04 [1] CRAN (R 4.1.2)
 tmfast      * 0.0.0.2023-04-15 2023-04-15 [1] local
 tmvnsim       1.0-2            2016-12-15 [1] CRAN (R 4.1.0)
 tokenizers    0.2.1            2018-03-29 [1] CRAN (R 4.1.0)
 tzdb          0.3.0            2022-03-28 [1] CRAN (R 4.1.2)
 utf8          1.2.2            2021-07-24 [1] CRAN (R 4.1.0)
 vctrs         0.6.0            2023-03-16 [1] CRAN (R 4.1.2)
 vipor         0.4.5            2017-03-22 [1] CRAN (R 4.1.0)
 viridisLite   0.4.0            2021-04-13 [1] CRAN (R 4.1.0)
 withr         2.5.0            2022-03-03 [1] CRAN (R 4.1.2)
 xfun          0.37             2023-01-31 [1] CRAN (R 4.1.2)
 xml2          1.3.3            2021-11-30 [1] CRAN (R 4.1.0)
 yaml          2.3.7            2023-01-23 [1] CRAN (R 4.1.2)

 [1] /Library/Frameworks/R.framework/Versions/4.1/Resources/library

------------------------------------------------------------------------------
\end{verbatim}

\hypertarget{references}{%
\section*{References}\label{references}}
\addcontentsline{toc}{section}{References}

\hypertarget{refs}{}
\begin{CSLReferences}{1}{0}
\leavevmode\vadjust pre{\hypertarget{ref-BaglamaAugmentedImplicitlyRestarted2005}{}}%
Baglama, James, and Lothar Reichel. 2005. {``Augmented Implicitly
Restarted Lanczos Bidiagonalization Methods.''} \emph{SIAM Journal on
Scientific Computing} 27 (1): 19--42.
\url{https://doi.org/10.1137/04060593X}.

\leavevmode\vadjust pre{\hypertarget{ref-BaglamaIrlbaFastTruncated2022}{}}%
Baglama, Jim, Lothar Reichel, and B. W. Lewis. 2022. {``Irlba: Fast
Truncated Singular Value Decomposition and Principal Components Analysis
for Large Dense and Sparse Matrices.''}
\url{https://cran.r-project.org/web/packages/irlba/index.html}.

\leavevmode\vadjust pre{\hypertarget{ref-BleiLatentDirichletAllocation2003}{}}%
Blei, David M., Andrew Y. Ng, and Michael I. Jordan. 2003. {``Latent
Dirichlet Allocation.''} \emph{The Journal of Machine Learning Research}
3: 993--1022. \url{http://dl.acm.org/citation.cfm?id=944937}.

\leavevmode\vadjust pre{\hypertarget{ref-GelmanGardenForkingPaths2013}{}}%
Gelman, Andrew, and Eric Loken. 2013. {``The Garden of Forking Paths:
Why Multiple Comparisons Can Be a Problem, Even When There Is No
{`Fishing Expedition'} or {`p-Hacking'} and the Research Hypothesis Was
Posited Ahead of Time.''} \emph{Downloaded January} 30: 2014.
\url{http://www.stat.columbia.edu/~gelman/research/unpublished/p_hacking.pdf}.

\leavevmode\vadjust pre{\hypertarget{ref-HicksProductivityInterdisciplinaryImpacts2021}{}}%
Hicks, Daniel J. 2021. {``Productivity and Interdisciplinary Impacts of
Organized Research Units.''} \emph{Quantitative Science Studies} 2 (3):
990--1022. \url{https://doi.org/10.1162/qss_a_00150}.

\leavevmode\vadjust pre{\hypertarget{ref-MalaterreEarlyDaysContemporary2022}{}}%
Malaterre, Christophe, and Francis Lareau. 2022. {``The Early Days of
Contemporary Philosophy of Science: Novel Insights from Machine
Translation and Topic-Modeling of Non-Parallel Multilingual Corpora.''}
\emph{Synthese} 200 (3): 242.
\url{https://doi.org/10.1007/s11229-022-03722-x}.

\leavevmode\vadjust pre{\hypertarget{ref-RobertsStmPackageStructural2019}{}}%
Roberts, Margaret E, Brandon M Stewart, and Dustin Tingley. 2019.
{``Stm: An R Package for Structural Topic Models.''} \emph{Journal of
Statistical Software} 91 (October): 1--40.
\url{https://doi.org/10.18637/jss.v091.i02}.

\leavevmode\vadjust pre{\hypertarget{ref-RoheVintageFactorAnalysis2020}{}}%
Rohe, Karl, and Muzhe Zeng. 2020. {``Vintage Factor Analysis with
Varimax Performs Statistical Inference.''} arXiv.
\url{https://doi.org/10.48550/arXiv.2004.05387}.

\leavevmode\vadjust pre{\hypertarget{ref-SteegenIncreasingTransparencyMultiverse2016}{}}%
Steegen, Sara, Francis Tuerlinckx, Andrew Gelman, and Wolf Vanpaemel.
2016. {``Increasing Transparency Through a Multiverse Analysis.''}
\emph{Perspectives on Psychological Science} 11 (5): 702--12.
\url{https://doi.org/10.1177/1745691616658637}.

\end{CSLReferences}

\end{document}